\documentclass[
 reprint,
nofootinbib,
pra,
 amsmath,amssymb,
 aps,
]{revtex4-2}

\usepackage{graphicx}
\usepackage{dcolumn}
\usepackage{bm}
\usepackage[colorlinks]{hyperref}

\usepackage{ragged2e}

\usepackage{caption}
\usepackage{subcaption}

\usepackage{physics}
\usepackage{dsfont}
\usepackage{mathtools}
\usepackage{amsthm}

\usepackage{dsfont}
\usepackage{faktor}

\usepackage{xcolor}

\definecolor{capBlue}{RGB}{93,93,235}
\definecolor{capGreen}{RGB}{93,215,93}
\definecolor{capRed}{RGB}{235,93,93}
\usepackage{tikz,tikz-3dplot}
\usetikzlibrary{shapes,arrows,cd,chains,decorations.markings,decorations.pathmorphing,calc}
\usetikzlibrary{decorations.pathreplacing}
\tikzset{
->-/.style args={#1rotate#2}{decoration={markings, mark=at position #1 with {\arrow[scale=1.5,rotate = #2 ]{stealth}}}, postaction={decorate}}
}

\tikzstyle{WhiteNode}=[circle, draw=black, fill=white, inner sep=0pt, minimum size=10pt]
\tikzstyle{BlackNode}=[circle, draw=black, fill=black, inner sep=0pt, minimum size=10pt]
\tikzstyle{YellowNode}=[circle, draw=black, fill=yellow, inner sep=0pt, minimum size=10pt]
\tikzstyle{OrangeNode}=[circle, draw=black, fill=orange, inner sep=0pt, minimum size=10pt]

\newcommand{\xangle}{0}
\newcommand{\yangle}{90}
\newcommand{\zangle}{90}

\newcommand{\xlength}{1}
\newcommand{\ylength}{1}
\newcommand{\zlength}{0.5}

\pgfmathsetmacro{\xx}{\xlength*cos(\xangle)}
\pgfmathsetmacro{\xy}{\xlength*sin(\xangle)}
\pgfmathsetmacro{\yx}{\ylength*cos(\yangle)}
\pgfmathsetmacro{\yy}{\ylength*sin(\yangle)}
\pgfmathsetmacro{\zx}{\zlength*cos(\zangle)}
\pgfmathsetmacro{\zy}{\zlength*sin(\zangle)}

\pgfmathsetmacro{\scaleOne}{0.75}

\pgfmathsetmacro\a{2}
\pgfmathsetmacro{\phi}{\a*(1+sqrt(5))/2}
\pgfmathsetmacro{\gr}{\a*(1+sqrt(5))/2}
\pgfmathsetmacro{\igr}{\a*2/(1+sqrt(5))}

\DeclareMathOperator{\Aut}{Aut}

\DeclareMathOperator{\bbR}{\mathbb{R}}

\DeclareMathOperator{\bbE}{\mathbb{E}}
\DeclareMathOperator{\bbS}{\mathbb{S}}
\DeclareMathOperator{\bbH}{\mathbb{H}}

\DeclareMathOperator{\AdS}{\mathrm{AdS}}

\DeclareMathOperator{\Mink}{\mathrm{Mink}}
\DeclareMathOperator{\dS}{\mathrm{dS}}

\DeclareMathOperator{\Iso}{\mathrm{Iso}}

\usepackage{empheq}
\usepackage[most]{tcolorbox}
\newtcbox{\mymath}[1][]{%
    nobeforeafter, math upper, tcbox raise base,
    enhanced, colframe=blue!30!black,
    colback=blue!30, boxrule=1pt,
    #1}

\newmuskip\pFqskip
\mathchardef\pFcomma=\mathcode`, 

\newcommand{\Mod}[1]{ \, \mathrm{mod} \, #1}
\newcommand{\tile}[1]{{\bf{{#1}}}}
\newcommand{\btile}[1]{{\bf{\bar{{#1}}}}}

\graphicspath{ {./images/} }

\newtheorem{conjecture}{Conjecture}

\begin{document}

\preprint{hYp3R-Qu451}

\title{Holographic Foliations: Self-Similar Quasicrystals from Hyperbolic Honeycombs}


\renewcommand{\andname}{\ignorespaces}

\author{Latham Boyle}
\affiliation{Higgs Centre for Theoretical Physics, University of Edinburgh, UK}
\affiliation{Perimeter Institute for Theoretical Physics, Waterloo, Ontario, Canada}
\author{Justin Kulp}
\affiliation{Simons Center for Geometry and Physics, Stony Brook University, Stony Brook, NY, USA}
\affiliation{Yang Institute for Theoretical Physics, Stony Brook University, Stony Brook, NY, USA}
\date{\today}

\begin{abstract}
    Discrete geometries in hyperbolic space are of longstanding interest in pure mathematics and have come to recent attention in holography, quantum information, and condensed matter physics. Working at a purely geometric level, we describe how any regular tessellation of ($d+1$)-dimensional hyperbolic space naturally admits a $d$-dimensional boundary geometry with self-similar ``quasicrystalline'' properties. In particular, the boundary geometry is described by a local, invertible, self-similar substitution tiling, that discretizes conformal geometry. We greatly refine an earlier description of these local substitution rules that appear in the 1D/2D example and use the refinement to give the first extension to higher dimensional bulks; including a detailed account for all regular 3D hyperbolic tessellations. We comment on global issues, including the reconstruction of bulk geometries from boundary data, and introduce the notion of a ``holographic foliation'': a foliation by a stack of self-similar quasicrystals, where the full geometry of the bulk (and of the foliation itself) is encoded in any single leaf in a local, invertible way. In the $\{3,5,3\}$ tessellation of 3D hyperbolic space by regular icosahedra, we find a 2D boundary quasicrystal admitting points of 5-fold symmetry which is not the Penrose tiling, and record and comment on a related conjecture of William Thurston. We end with a large list of open questions for future analytic and numerical studies.
\end{abstract}


\maketitle

\tableofcontents

\section{Introduction}\label{sec:Introduction}
For more than 25 years, physicists have been enormously interested in hyperbolic space because of its central role in the AdS/CFT correspondence \cite{Maldacena:1997re, Witten:1998qj} and holography more broadly \cite{tHooft:1993dmi, Susskind:1994vu, Bousso:2002ju}.
    
More recently, in the past decade, physicists have been increasingly interested in \textit{discrete} models of holography, in an effort to clarify the connections between quantum information, tensor networks, condensed matter physics, and the ``it from qubit'' perspective \cite{Swingle:2009bg, VanRaamsdonk:2010pw, Swingle:2012wq, MolinaVilaplana:2012rmg, Qi:2013caa, Orus:2014poa, Czech:2015kbp, Pastawski:2015qua, Hayden:2016cfa, Evenbly:2017hyg, Osborne:2017woa, Milsted:2018san}. They have thus been led to study regular tilings in hyperbolic space, since these naturally discretize the geometry of hyperbolic space in a way that preserves an infinite discrete subgroup of its original symmetries. Some recent work emphasizing this particular connection includes: numerical studies on hyperbolic lattices \cite{Brower:2019kyh, Asaduzzaman:2020hjl, Asaduzzaman:2021ufo,Asaduzzaman:2021bcw, Brower:2022atv, Basteiro:2022zur, Basteiro:2022pyp, Basteiro:2022xvu,gluscevich2023dynamic, Dey:2024jno, Erdmenger:2024jsb, Okunishi:2024amt,Huang:2024dux}, topoelectric circuits and ``circuit QED'' \cite{kollar2019hyperbolic, boettcher2020quantum, bienias2022circuit, saa2021higher}, hyperbolic band theory \cite{maciejko2021hyperbolic, ikeda2021hyperbolic, boettcher2022crystallography, maciejko2021automorphic, kienzle2022hyperbolic, NonAbHypBand, Shankar:2023tsw,Sun:2024rxr}, tensor networks \cite{jahn2020central, jahn2022tensor, steinberg2022conformal, gesteau2022holographic} and error-correcting codes \cite{2015arXiv150604029B, 2017QS&T....2c5007B, Taylor:2021hsx, Li:2023ihz, Steinberg:2024ack}.
    
In this paper, we point out that: \textit{every regular tessellation of hyperbolic space naturally foliates into collections of self-similar quasicrystalline layers.} Moreover, each layer is related to the next by a self-similarity transformation that is encoded in a {\it local} and {\it invertible} ``inflation/deflation" rule.  Using the local inflation/deflation rule one can thus reconstruct the full tiling and its foliation starting from a single layer (in a sense
which would not be possible if the individual layers were ordinary periodic crystals). In this kinematic sense, some ``holographic'' character of hyperbolic space is already visible at the level of its discrete geometry, even before any physical/dynamical fields are included. We call a foliation with this special property a {\it holographic foliation.}

As a necessary ingredient in this story, we introduce a new characterization of crystals and (self-similar) quasicrystals which naturally extends from flat space to hyperbolic space. Traditional self-similar quasicrystals (which live in Euclidean space), like the famous Penrose tiling \cite{gardner1977extraordinary, penrose1979pentaplexity, gardner1997penrose}, are special objects that are of interest to both physicists \cite{shechtman1984metallic, levine1984quasicrystals, socolar1986quasicrystals, duncan2020topological, else2021quantum, manna2024noncrystalline} and mathematicians \cite{senechal1996quasicrystals, bombieri1986distributions, bombieri1987quasicrystals,  grunbaum1987tilings, baake2002guide, baake2013aperiodic}. The new self-similar quasicrystals introduced here (which live in hyperbolic space) are similarly special and interesting. In particular, we will find that the quasicrystals arising as layers of a holographic foliation of $\mathbb{H}^{d>2}$ include genuinely new examples of quasicrystalline patterns.

The fact that regular tilings of hyperbolic space $\mathbb{H}^{d+1}$ induce $d$-dimensional self-similar boundary quasicrystals was first pointed out in \cite{Boyle:2018uiv}, and has been studied in a number of subsequent papers (see references above); but the understanding that these quasicrystals naturally foliate the space is new to the present paper and allow us to better define the class of all boundary quasicrystals. Moreover, all previous work on this topic has focused on the 1D/2D case (i.e. tilings of 2D hyperbolic space $\mathbb{H}^{2}$, and the corresponding 1D boundary 
quasicrystals), which is simpler in a number of aspects. As a result, generalizing the story requires a number of new conceptual ingredients. The present paper is the first to show how this story generalizes to higher dimensions.

The layout of the paper is as follows.  In Section \ref{sec:Section2} we provide a lightning fast review of the geometry of hyperbolic space and its regular tessellations, then introduce a new characterization of crystals and self-similar quasicrystals that naturally generalizes from Euclidean to hyperbolic space.  In Section \ref{sec:1DQCfrom2D}, we warm-up for our higher-dimensional examples by reviewing the 1D/2D story; taking the opportunity to recast it in a new way which improves the original formalism developed in \cite{Boyle:2018uiv} by the introduction of so-called ``half-step rules.'' In particular, the half-step rules reflect a more refined discretization of hyperbolic space by lattices preserving the same subgroup of isometries.

After these refinements, we move on to higher dimensions. Based on our improved story, in Section \ref{sec:H3Tiling} we consider an extended example: the self-dual $\{3,5,3\}$ tiling of $\mathbb{H}^{3}$ by icosahedra and its corresponding 2D quasicrystal. It had been conjectured in \cite{Boyle:2018uiv}, and apparently also earlier by William Thurston \cite{PenroseCommunication}, that in this case the 2D quasicrystal would be the famous Penrose tiling. In a sense we explain, the conjecture turns out to be false, and we obtain a manifestly new quasicrystalline pattern with points of 5-fold rotational symmetry. In Section \ref{sec:GeneralStory} we sketch a general scheme for producing holographic foliations in $\mathbb{H}^{d}$ and analyze the remaining 2D/3D examples.

Finally, in Section \ref{sec:Conclusion}, we end by discussing some interesting open questions and speculations for future directions. Several more technical points and reference formulae are included in the appendices.

\section{Preliminaries}\label{sec:Section2}

\subsection{Models of Hyperbolic Space}\label{sec:modelsOfH}

The maximally symmetric $d$-dimensional Riemannian manifolds are the $d$-dimensional sphere $\mathbb{S}^{d}$, $d$-dimensional hyperbolic space $\mathbb{H}^{d}$, and $d$-dimensional Euclidean space $\mathbb{E}^{d}$.  These spaces have constant positive, negative, or vanishing Riemann curvature respectively.
The analogous maximally-symmetric constant curvature $d$-dimensional manifolds of Lorentzian signature are de Sitter space $\dS_d$, anti-de Sitter space $\AdS_d$ and Minkowski space $\Mink_d$, with positive, negative and zero curvature respectively.

\begin{figure*}
	\begin{minipage}{0.28\textwidth}
        \centering
        \includegraphics[width=.95\linewidth]{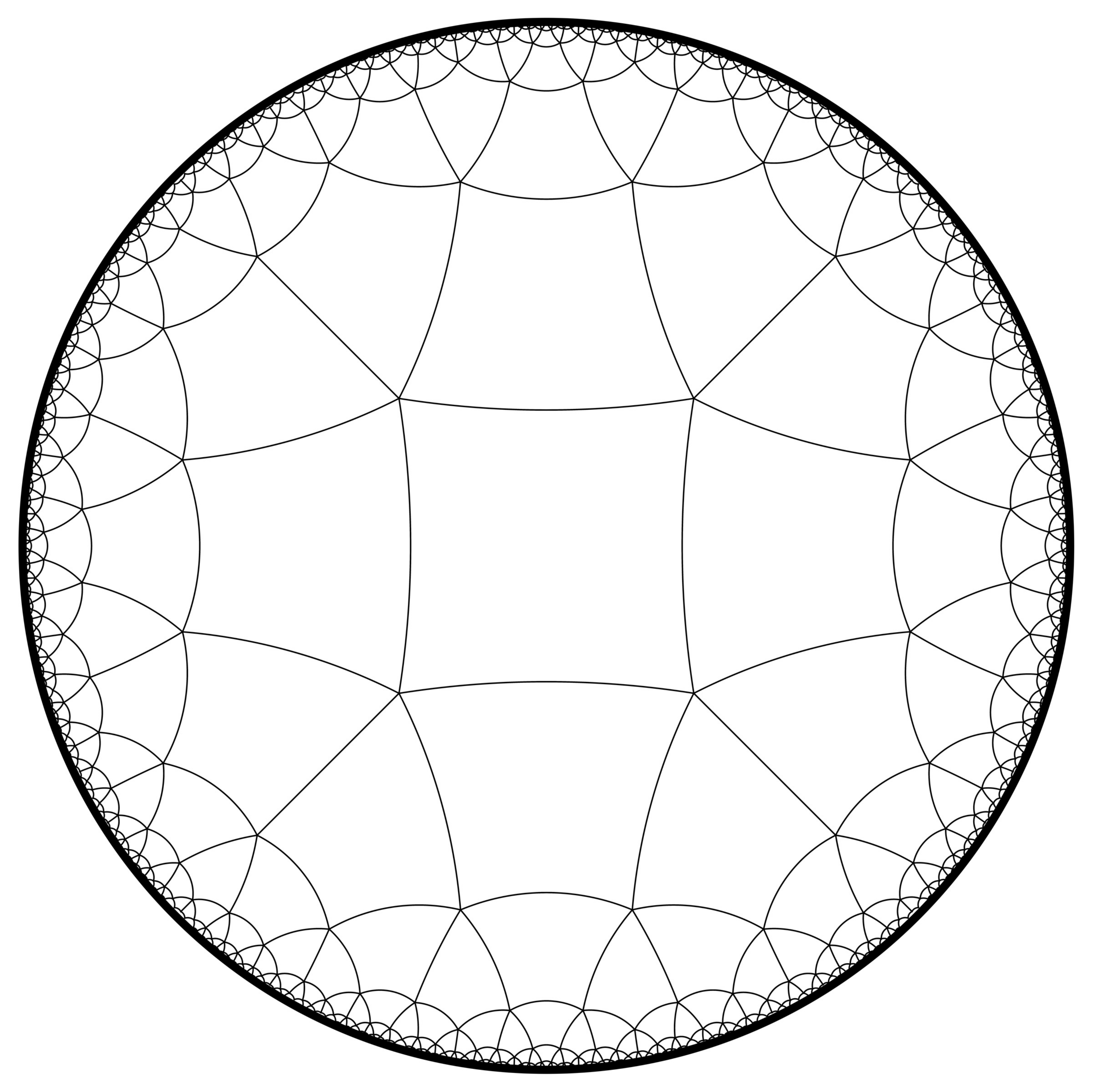}
	\end{minipage}
    \begin{minipage}{0.32\textwidth}
        \centering
        \includegraphics[width=.95\linewidth]{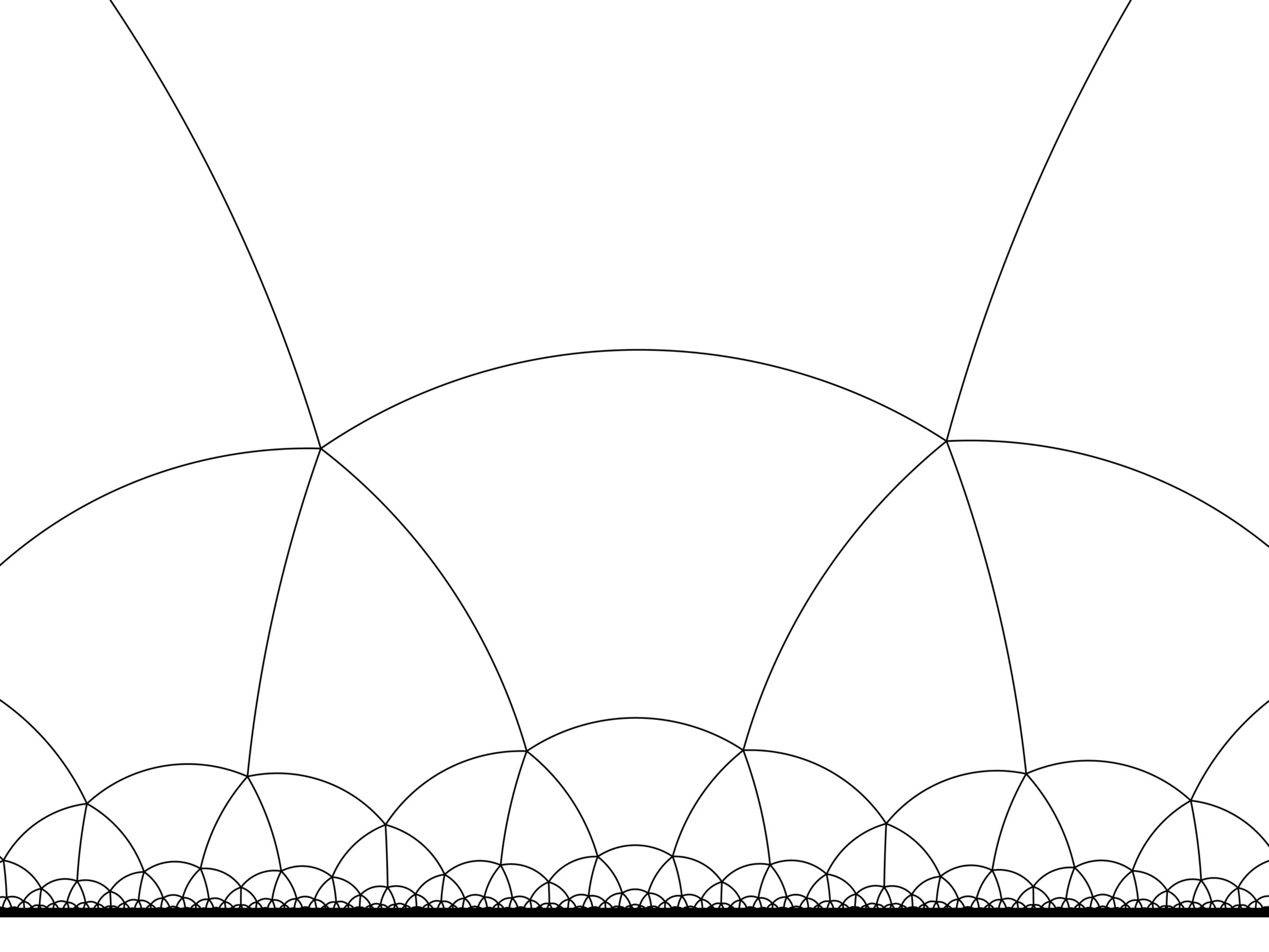}
	\end{minipage}
	\begin{minipage}{0.35\textwidth}
        \centering
        \includegraphics[trim={6cm 0cm 3cm 1cm}, clip, width=1.05\linewidth]{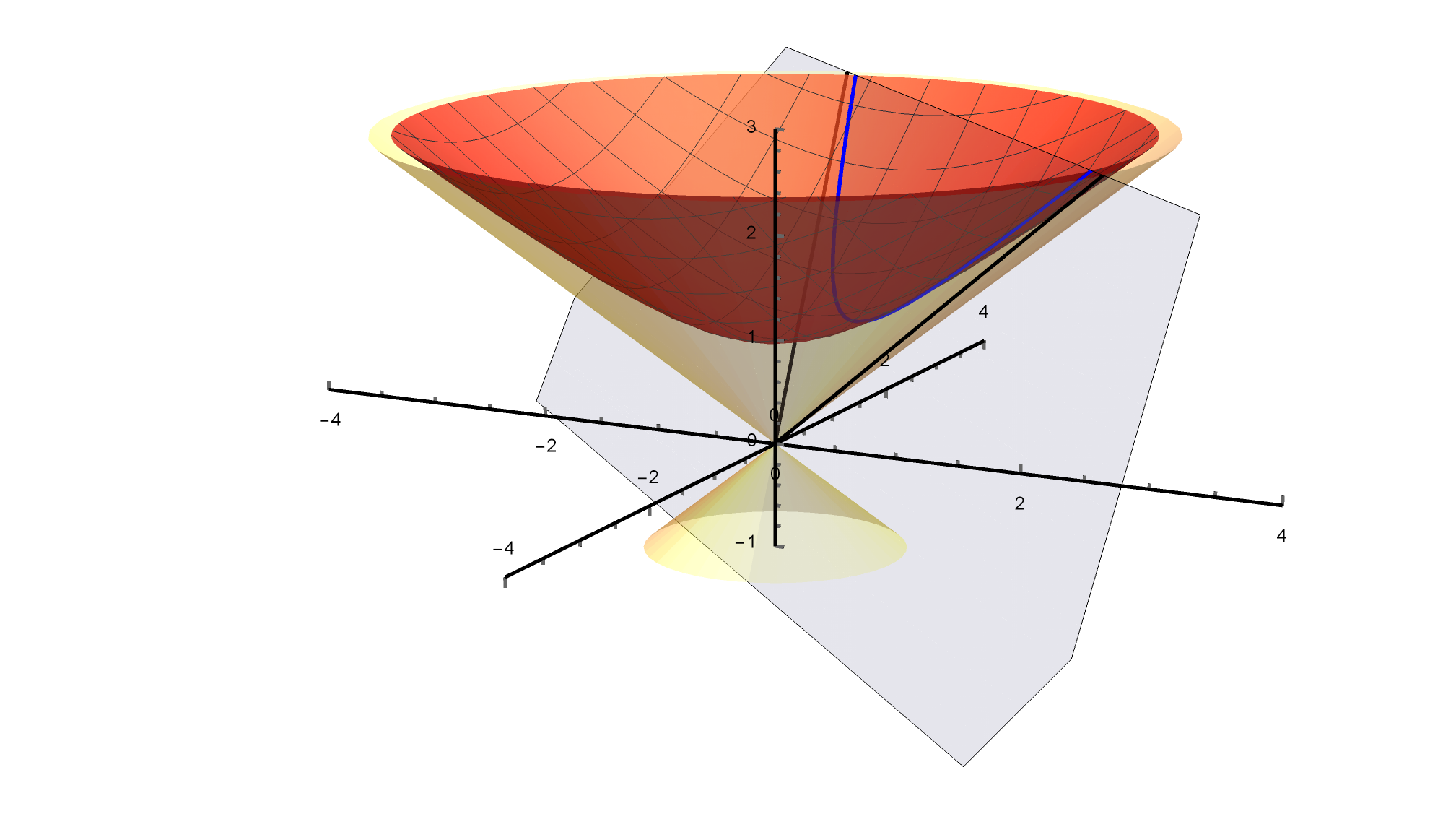}
	\end{minipage}
	\caption{\RaggedRight{Left, the Poincar\'e Ball presentation of $\bbH^2$ with a $\{4,5\}$ tessellation drawn on top. In the picture, an $O(2)$ subgroup is manifest (but broken by the tessellation) and the boundary of $\bbH^2$ is clear. Center, the same tessellation of $\bbH^2$ is drawn in the Upper Half-Space model. Right, $\bbH^2$ is depicted as a hyperboloid (\textcolor[RGB]{232,70,52}{\textbf{red}}) in $\Mink_{2+1}$ (with lightcone depicted in \textcolor{yellow!80!orange}{\textbf{yellow}}). The Lorentz group acts naturally on the hyperboloid, and the boundary is the ``celestial circle'' which lives at future timelike infinity. A \textcolor{capBlue}{\textbf{blue plane}} through the origin in Minkowski space cuts the hyperboloid and appears as a geodesic (\textcolor{capBlue!80!black}{\textbf{dark blue curve}}) in $\bbH^2$. Drawn in units of $L$.}}
	\label{fig:threeModels}
\end{figure*}

To understand the geometry of hyperbolic space,\footnote{For an introduction to classical results in hyperbolic geometry, see e.g. \cite{thurstonNotes, thurston1997three, elstrodt2013groups, coxeter1998non}.} it is useful to keep three different models (coordinate systems) in mind: the Poincar\'e Ball, the Upper Half-Space, and the Hyperboloid Model. See Figure \ref{fig:threeModels}.

In the Poincar\'e Ball model, we consider $\bbH^d$ as the open unit ball $B^d$ inside an ambient $\bbR^d$, with Cartesian coordinates $\{x^1,\ldots,x^d\}$, equipped with metric
\begin{equation}
   ds^2 = 4L^2\frac{(dx^1)^2+\dots+(dx^d)^2}{(1-\abs{x}^2{)}^2}\,.
\end{equation}
This model makes manifest the existence of a boundary at hyperbolic infinity, $\abs{x}^2 = 1$.  Famously, the metric does not extend to the boundary of the ball $\bar{B}^{d}$, which is only equipped with a conformal metric, defined up to an arbitrary Weyl rescaling $g_{\mu\nu}(x)\to\Omega^{2}(x)g_{\mu\nu}$ (see e.g. \cite{Witten:1998qj}). In this model, ``straight lines" (geodesics) in $\mathbb{H}^{d}$ look like circles in the ambient $\mathbb{R}^{d}$ that perpendicularly intersect the boundary $\bar{B}^{d}$; 2-planes in $\mathbb{H}^{d}$ correspond to 2-spheres in $\mathbb{R}^{d}$ that perpendicularly intersect $\bar{B}^{d}$; and so on.

In the Upper Half-Space model we again consider $\mathbb{R}^{d}$, with 
Cartesian coordinates $\{y^1,\ldots,y^d\}$, and take $\bbH^d$ to be
the ``upper half space'' ($y^d>0$) equipped with the metric
\begin{equation}
    ds^2 = L^2 \frac{(dy^1)^2+\dots+(dy^d)^2}{(y^d)^2}\,.
\end{equation}
The boundary is the hyperplane $y^d = 0$ (compactified by adding the point at $y^d=\infty$). In this model, $n$-planes in $\bbH^d$ again correspond to $n$-spheres in $\mathbb{R}^{d}$ that perpendicularly intersect the boundary.

Finally, in the Hyperboloid Model, we start with a $d+1$ dimensional Minkowski spacetime $\Mink_{d+1}$ with coordinates $\{X^0,\ldots,X^d\}$ and consider the two-sheeted hyperboloid $H$ defined by
\begin{equation}
    -(X^0)^2 + (X^1)^2 + \dots + (X^d)^2 = - L^2\,. \label{eq:Hyperboloid}
\end{equation}
This hyperboloid's ``upper sheet'' $H_+$ (with $X^0 > 0$) has a positive-definite induced metric, and is a copy of $\bbH^d$.  To be more explicit, if we take the usual Minkowski metric and switch to standard spherical coordinates:
\begin{align}
    ds_{\Mink_{d+1}}^2 
        &= -(dX^0)^2+d\vec{X}^2\\
        &= -(dt)^2 + dr^2 + r^2 d\Omega_{d-1}^2\,,
\end{align}
then the line element restricted to the upper hyperboloid $t=\sqrt{L^2+r^2}$ is
\begin{align}
    ds^2 
        &= \frac{dr^2}{1+(r/L)^2} + r^2 d\Omega_{d-1}^2\\
        &= d\rho^2 + (\sinh^2\rho)\,d\Omega_{d-1}^2\,,\label{eq:hyperboloidMetric}
\end{align}
where we have defined $r=\sinh\rho$. From the final line element, we see that $\rho$ is the intrinsic radial coordinate in the hyperboloid itself, i.e. it measures the proper distance along a radial geodesic.  In this model, the intersection between $H_+$ and an $(n+1)$-plane through the origin of $\Mink_{d+1}$ is an $n$-plane in $\mathbb{H}^{d}$.\footnote{The Hyperboloid Model is familiar to those who do e.g. conformal bootstrap \cite{Costa:2011mg, Costa:2011dw, simmonsDuffin:shadows} under the name ``embedding space formalism'' (see also \cite{Dirac:1936fq, Mack:1969rr, Weinberg:2010fx}). The boundary of the hyperbolic space is the ``celestial sphere'' of Minkowski space.}

The Hyperboloid Model makes it straightforward to understand $\Iso(\bbH^d)$, the isometry group of $\bbH^d$. In particular, all isometries of the two-sheeted hyperboloid $H$ come from Lorentz transformations of the ambient Minkowski space $\Mink_{d+1}$, so the isometry group of $H$ is just the ``full Lorentz group''  $\Iso(H)=O(1,d)$.  Hence, the isometry group of the one-sheeted hyperboloid $H_{+}=\bbH^d$ is the orthochronous Lorentz group $\Iso(\mathbb{H}^{d}) \cong O_{+}(1,d)$, and the orientation preserving isometries are the connected component $SO_+(1,d)$. The beauty of this approach is that it lets us see the whole group of isometries through the boosts and rotations of the ambient Minkowski space, and also encodes their action in a linear way on the ambient coordinates.

\subsection{Regular Tessellations and Their Symmetries}\label{sec:tessellations}

The study of tessellations (or ``tilings" or ``honeycombs") built from regular shapes (regular polygons, polyhedra, polytopes, $\dots$ ) is a beautiful area of classical geometry going back -- at minumum -- to Ancient Greece \cite{Senechal_whichTetrahedra}, with other later pioneering contributions by Kepler and others \cite{keplerBook}. The regular tessellations in flat space $\bbE^d$ and positively curved space $\bbS^d$ were enumerated by Schl\"afli, and those in negatively curved space $\bbH^d$ by Schlegel. The whole subject was given a beautiful unified formulation based on the theory of ``reflection groups" (or ``Coxeter groups") by Coxeter \cite{schlegel1883theorie, coxeter1954regular, coxeterHyperbolic, coxeter1973regular, coxeter2013generators}.

In order to proceed, we remind readers of the Schl\"afli symbol for presenting regular polytopes or regular tessellations by polytopes: $\{p\}$ denotes a (convex) regular $p$-gon; $\{p,q\}$ denotes a regular polyhedron or tessellation in which $q$ regular $p$-gons meet at each vertex; $\{p,q,r\}$ denotes an arrangement where $r$ regular $\{p,q\}$'s meet around each edge; and so on. For example, $\{4,3\}$ describes a cube ({i.e.} a tiling of $\bbS^2$ by squares); $\{4,4\}$ describes a tessellation of $\bbE^2$ by squares; and $\{4,3,4\}$ describes a regular honeycomb of cubes tessellating $\bbE^3$.

In 2D Euclidean space $\bbE^2$, the internal angles of a $\{p\}$ are $(1-2/p)\pi$, so to fit $q$ around a vertex they must satisfy $(p-2)(q-2)=4$.  In positively or negatively curved space ({i.e.}\ $\mathbb{S}^{2}$ or $\mathbb{H}^{2}$), the internal angles of a $\{p\}$ are larger or smaller than $(1-2/p)\pi$ respectively (as can be inferred from the Gauss-Bonnet formula, see also \cite{thurstonNotes, thurston1997three, elstrodt2013groups, coxeter1998non}), so $(p-2)(q-2)<4$ or $>4$ respectively.  Indeed, if we fix the overall curvature radius $L$, then by making a $\{p\}$ in $\bbH^2$ arbitrarily large, we can make its internal angles arbitrarily small, and hence can fit an arbitrarily large number $q$ of such $\{p\}$'s around a vertex.

\begin{figure*}
	\begin{minipage}{0.48\textwidth}
    \includegraphics[width=.95\linewidth]{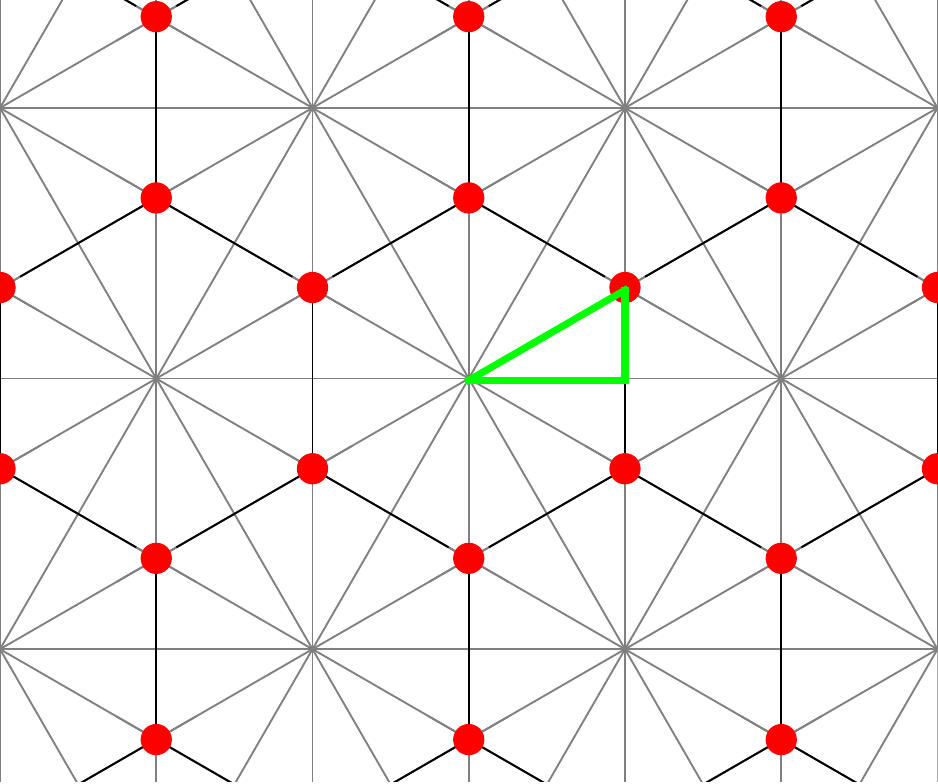}
	\end{minipage}
    \begin{minipage}{0.48\textwidth}
        \includegraphics[width=.95\linewidth]{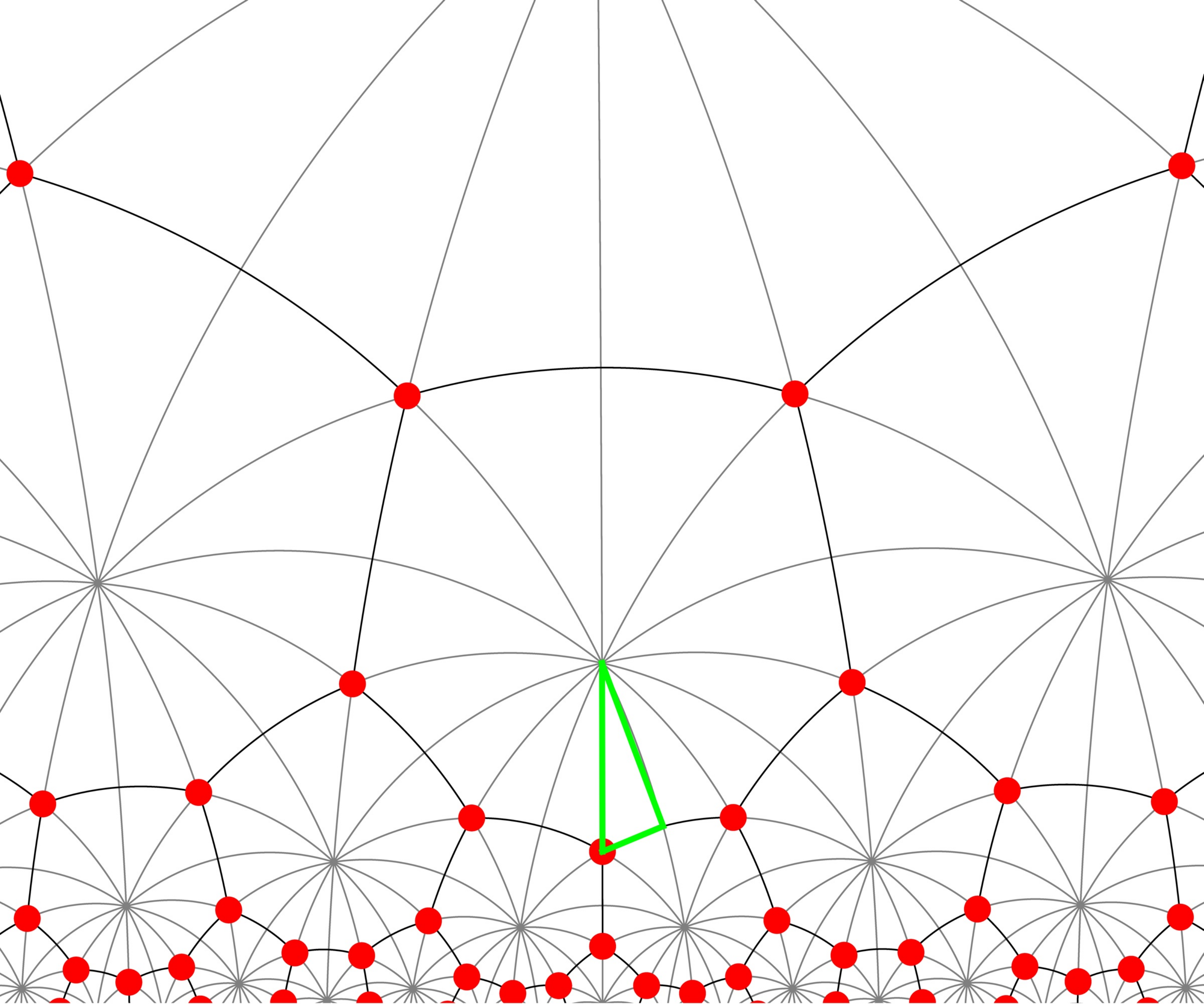}
	\end{minipage}
	\caption{\RaggedRight{Left, a \textcolor[RGB]{0,220,0}{\textbf{green triangle}} with interior angles $(\tfrac{\pi}{2},\tfrac{\pi}{3},\tfrac{\pi}{6})$ forms the fundamental domain for mirror planes corresponding to the Coxeter-Dynkin diagram in Equation \eqref{eq:CDD236}. Reflections of this fundamental domain in the mirrors triangulates $\bbE^2$. Tracking the node (\textcolor[RGB]{232,70,52}{\textbf{red}}) opposite the ``6-mirror'' under these reflections generates the $\{6,3\}$ tiling. Right, a $(\tfrac{\pi}{2},\tfrac{\pi}{3},\tfrac{\pi}{7})$ triangle (\textcolor[RGB]{0,220,0}{\textbf{green}}) forms the fundamental domain for the mirrors in Equation \eqref{eq:CDD237}. Reflections of this fundamental domain in the mirrors triangulates $\bbH^2$. Tracking the node (\textcolor[RGB]{232,70,52}{\textbf{red}}) opposite the ``7-mirror'' generates the $\{7,3\}$ tiling. In both cases, one can see the dual $\{3,6\}$ and $\{3,7\}$ triangle honeycombs whose vertices are the centers of the hexagons and heptagons respectively.}}
	\label{fig:tessKaleidoscope}
\end{figure*}

Altogether, if
\begin{equation}
    \frac{1}{p}+\frac{1}{q} \quad 
        \begin{cases}
            =\frac{1}{2} & \text{$\{p,q\}$ tessellates $\bbE^2$} \\
            >\frac{1}{2} & \text{$\{p,q\}$ tessellates $\bbS^2$} \\
            <\frac{1}{2} & \text{$\{p,q\}$ tessellates $\bbH^2$} 
        \end{cases}\,.
\end{equation}
Thus there are only three regular tilings of $\bbE^2$: by squares $\{4,4\}$, triangles $\{3,6\}$, and hexagons $\{6,3\}$; only five regular tilings of $\bbS^2$ corresponding to (the projections of) the five Platonic solids: tetrahedron $\{3,3\}$, octahedron $\{3,4\}$,  cube $\{4,3\}$, icosahedron $\{3,5\}$, and dodecahedron $\{5,3\}$; and an infinite number of regular tilings of $\bbH^2$: $\{3,7\}$, $\{7,3\}$, $\{4,5\}$, $\{5,4\}$, and so on.  We compare two examples in Figure \ref{fig:tessKaleidoscope}.

If we restrict ourselves to only tessellations with finite cells and finitely many cells meeting at a vertex, then the only regular tessellations of hyperbolic space are:
\begin{enumerate}
    \item[$\mathbb{H}^2$.] $\{p,q\}$ with $(p-2)(q-2)>4$,
    \item[$\mathbb{H}^3$.] $\{4,3,5\}$, $\{5,3,4\}$, $\{5,3,5\}$, and $\{3,5,3\}$,
    \item[$\mathbb{H}^4$.] $\{3,3,3,5\}$, $\{5,3,3,3\}$, $\{4,3,3,5\}$, $\{5,3,3,4\}$, and $\{5,3,3,5\}$,
\end{enumerate}
with no further examples in $\bbH^d$ for $d>4$ \cite{schlegel1883theorie, coxeterHyperbolic}.

Given a regular tessellation $\{p,q,...\,,r\}$ of a space $\mathbb{M}$ ($\bbE^d$, $\bbS^d$, or $\bbH^d$), its isometry group $[p,q,...\,,r]$ is the discrete subgroup of $\Iso(\mathbb{M})$ that maps $\{p,q,...\,,r\}$ to itself.  

To understand the isometry group $[p,q,\ldots,r]$, first imagine all of the tessellation's ``mirrors" -- {i.e.}\ all of the codimension-one hyperplanes across which the tessellation $\{p,q,\ldots,r\}$ reflects into itself.  These mirrors (the light gray lines in Figure \ref{fig:tessKaleidoscope}) split the space into a collection of congruent connected regions (the triangular regions in Figure \ref{fig:tessKaleidoscope}), and we can choose one such region as the ``fundamental domain" (the green triangle in Figure \ref{fig:tessKaleidoscope}).  
 
The fundamental domain is bounded by $n$ ``fundamental mirrors" $(M_1,\ldots,M_n)$.  Any two of these mirrors, $M_i$ and $M_j$, meet at an angle $\pi/p_{ij}$, where $p_{ij}$ are integers, with $p_{ii}=1$ and all other $p_{ij}\geq 2$.  The $n$ fundamental mirrors define $n$ ``fundamental reflections" $(R_1,\ldots,R_n)$.  Algebraically, $[p,q,...\,,r]$ is generated by these $n$ $R_{i}$'s, modulo the relations $(R_i R_j)^{p_{ij}}=1$.  Geometrically, the space is tessellated by congruent copies of the fundamental domain which may all be reached from the fundamental domain by repeated reflections through the fundamental mirrors, so that the copies of the fundamental domain and the elements of $[p,q,...\,,r]$ are in one-to-one correspondence.

More generally, a Coxeter group is any group with a presentation of the form
\begin{equation}
    \langle R_1,\dots, R_n \,\vert\, (R_i R_j)^{p_{ij}}=1 \rangle\,,
\end{equation}
where $p_{ij}$ are integers, with $p_{ii}=1$ and all other $p_{ij}\geq 2$.  

A Coxeter group is summarized by its ``Coxeter diagram" -- a graph in which each node corresponds to one of the fundamental mirrors, and each pair of distinct nodes $i$ and $j$ is connected by an edge labelled by the integer $p_{ij}$ (unless $p_{ij}=3$, in which case the label is omitted, or $p_{ij}=2$, in which case the edge is omitted altogether).  Thus, from an algebraic standpoint, the Coxeter diagram summarizes the group generators and relations; and from a geometric standpoint, it summarizes the shape of a fundamental domain bounded by mirrors.

The group $[p,q,...\,,r]$ is a special case of a Coxeter group in which the Coxeter diagram has the linear form
\begin{equation}
\begin{tikzpicture}[baseline={(current bounding box.center)}, scale = 1]
	\tikzstyle{vertex}=[circle, fill=black, minimum size=2pt,inner sep=2pt];
	\def\r{1};
	\node[vertex] (T1) at (\r*1,\r*0) {};
	\node[vertex] (T2) at (\r*2,\r*0) {};
	\node[vertex] (T3) at (\r*3,\r*0) {};
	\node[vertex] (T4) at (\r*4,\r*0) {};
	\node[vertex] (T5) at (\r*5,\r*0) {};
	
    \draw[-] (T1) -- (T2) node[midway, above] {$p$};	
	\draw[-] (T2) -- (T3) node[midway, above] {$q$};
	\node[] (dots) at (\r*3.5,\r*0) {$\dots$};
	\draw[-] (T4) -- (T5) node[midway, above] {$r$};
\end{tikzpicture}\label{eq:CoxeterDynkinpqr}
\end{equation}
as opposed to something more general like a tree with multiple branches or a loop.  

For example, on the left side of Figure \ref{fig:tessKaleidoscope}, the fundamental domain is a triangle with internal angles $(\frac{\pi}{2}, \frac{\pi}{3}, \frac{\pi}{6})$, corresponding to the diagram
\begin{equation}\label{eq:CDD236}
\begin{tikzpicture}[scale = 1]
	\tikzstyle{vertex}=[circle, fill=black, minimum size=2pt,inner sep=2pt];
	\def\r{1};
	\node[vertex] (T1) at (\r*1,\r*0) {};
	\node[vertex] (T2) at (\r*2,\r*0) {};
	\node[vertex] (T3) at (\r*3,\r*0) {};
	
    \draw[-] (T1) -- (T2) node[midway, below] {$\hphantom{6}$};	
	\draw[-] (T2) -- (T3) node[midway, above] {$6$};
\end{tikzpicture}
\end{equation}
On the right side of Figure \ref{fig:tessKaleidoscope}, the fundamental domain is a triangle with angle $(\frac{\pi}{2}, \frac{\pi}{3}, \frac{\pi}{7})$, corresponding to
\begin{equation}\label{eq:CDD237}
\begin{tikzpicture}[baseline={(current bounding box.center)}, scale = 1]
	\tikzstyle{vertex}=[circle, fill=black, minimum size=2pt,inner sep=2pt];
	\def\r{1};
	\node[vertex] (T1) at (\r*1,\r*0) {};
	\node[vertex] (T2) at (\r*2,\r*0) {};
	\node[vertex] (T3) at (\r*3,\r*0) {};
	
    \draw[-] (T1) -- (T2) node[midway, below] {$\vphantom{7}$};	
	\draw[-] (T2) -- (T3) node[midway, above] {$7$};
\end{tikzpicture}
\end{equation}

In the hyperbolic case we can give an elegant description of the mirrors/reflections in $[p,q,\ldots,r]$ using the Hyperboloid Model. On one hand, just as a 2D plane in $\bbE^3$ may be specified by giving three points in $\mathbb{E}^{3}$, a hyperbolic mirror (a codimension-1 hyperplane in $\bbH^d$) may be specified by giving $d$ points in $\bbH^d$. In the Hyperboloid Model, the $i$th such point has coordinates $X_i^{\mu}$ where $-(X_i^0)^2+(X_i^1)^2+\ldots+(X_i^d)^2=-L^2$.  On the other hand, we have seen that the same hyperbolic mirror may be regarded as the intersection of $\bbH^d$ with a Minkowski mirror (a codimension-1 hyperplane through the origin in $\Mink_{d+1}$ and the same $d$ points in $\mathbb{H}^d$, see Figure \ref{fig:threeModels}). This Minkowski mirror defines the Minkowski reflection:
\begin{equation}
    M^{\mu}_{\nu} = \delta^{\mu}_{\nu} - 2\frac{n^\mu n_\nu}{n^2}\,,\label{eq:reflectionMatrix}
\end{equation}
where $n^{\mu}$ is a vector orthogonal to the mirror
\begin{equation}
    n_{\mu} = \epsilon_{\mu\nu_1\cdots\nu_{d}} X_1^{\nu_1}\cdots X_d^{\nu_d}\,.
\end{equation}
Altogether, any point $Z$ is reflected through the mirror to a new point
\begin{equation}
    (Z^\prime)^\mu = Z^\mu - 2 \frac{Z^\alpha n_\alpha}{n^\beta n_\beta} n^\mu\,.
\end{equation}
In particular, if $Z$ is a point in the upper hyperboloid $H^{+}$, then so is $Z'$, and the resulting map $\bbH^d\to\bbH^d$ is precisely the desired reflection through the original mirror in $\bbH^d$. This perspective is useful for explicitly constructing the generators/action of the discrete isometry group $[p,q,\ldots,r]$.


\subsection{Substitution Tilings and Quasicrystals}\label{sec:SubTilingsAndQuasi}
In this paper, we consider objects that share essential properties with self-similar quasiperiodic substitution tilings, such as the Penrose tiling. However, first we must replace the basic definitions used in this field with modified ones that capture the essential concepts while extending more neatly to our tilings embedded in hyperbolic space.

\subsubsection{In Euclidean Space}
Let us start by ``reviewing'' some facts about substitution tilings in Euclidean space (where ``reviewing'' is in quotes because we will modify some standard definitions). For a nice introduction to substitution tilings which introduces the key concepts, with a minimum of excess baggage and many visual examples, we recommend the Tilings Encyclopedia and its glossary \cite{TilingsEncyclopedia} (see also \cite{baake2013aperiodic, boyle2016self, boyle2016coxeter}). For an introduction to the Penrose tiling, a famous and beautiful substitution tiling of special relevance to us in this paper, we recommend Martin Gardner's original introductory articles \cite{gardner1977extraordinary, gardner1997penrose}.

We begin with the notion of a \textit{substitution tiling} (or \textit{inflation tiling}). Start with a finite set of $N$ \textit{prototiles}, $\{T_{1},\dots,T_{N}\}$, each of which has a fixed finite shape (possibly decorated) in $d$-dimensional Euclidean space. We give a rule for uniformly rescaling each prototile $T_{i}$ by a linear factor $\lambda>1$, and then for decomposing the rescaled tile $\lambda T_{i}$ into copies of the original prototiles $T_{1},\ldots,T_{N}$.  This rule is called the \textit{substitution rule} (or \textit{inflation rule}), and $\lambda$ is called the \textit{scale factor} (or \textit{inflation factor}). Repeated iteration of this substitution rule yields a substitution tiling in $\bbE^d$. 

\begin{figure*}
    \begin{minipage}{0.565\textwidth}
        \includegraphics[width=.95\linewidth]{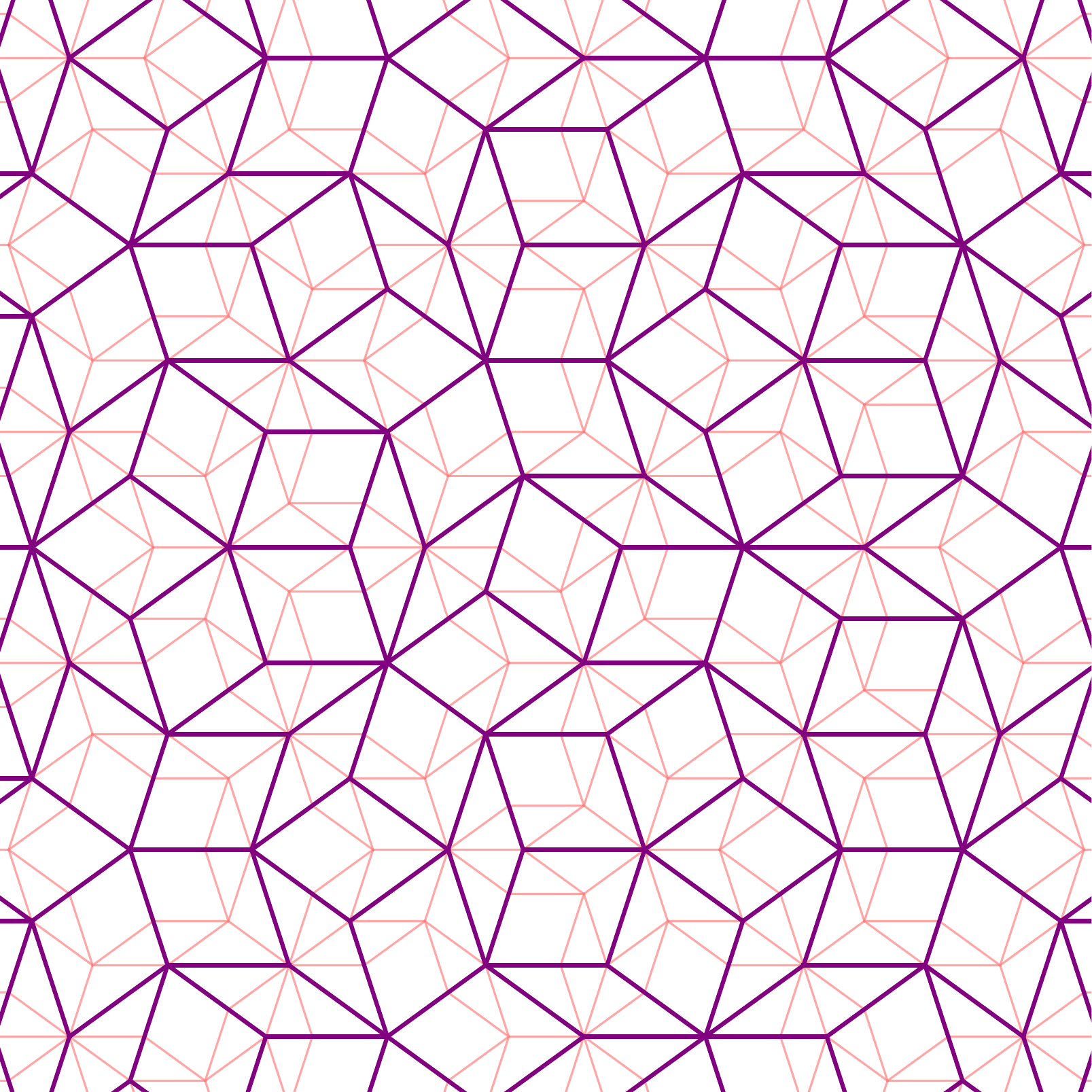}
	\end{minipage}
    \begin{minipage}{0.425\textwidth}
        \includegraphics[width=.95\linewidth]{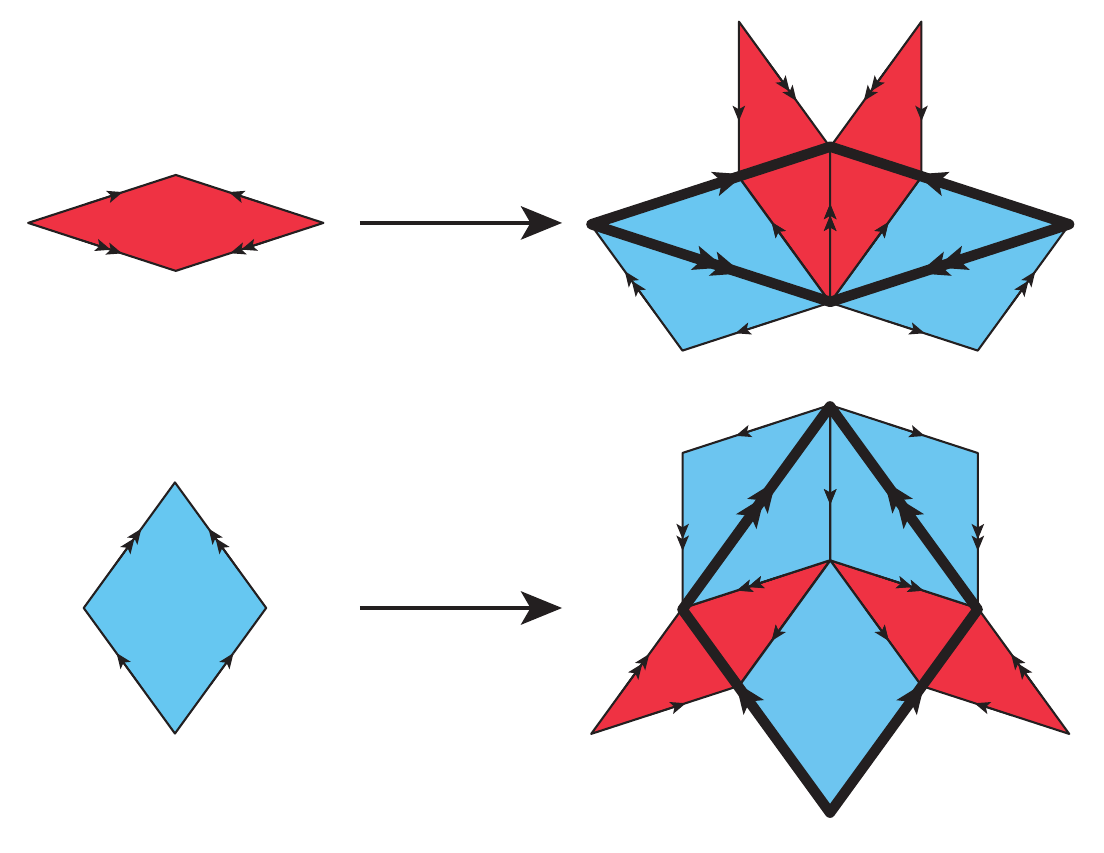}
        \includegraphics[width=.95\linewidth]{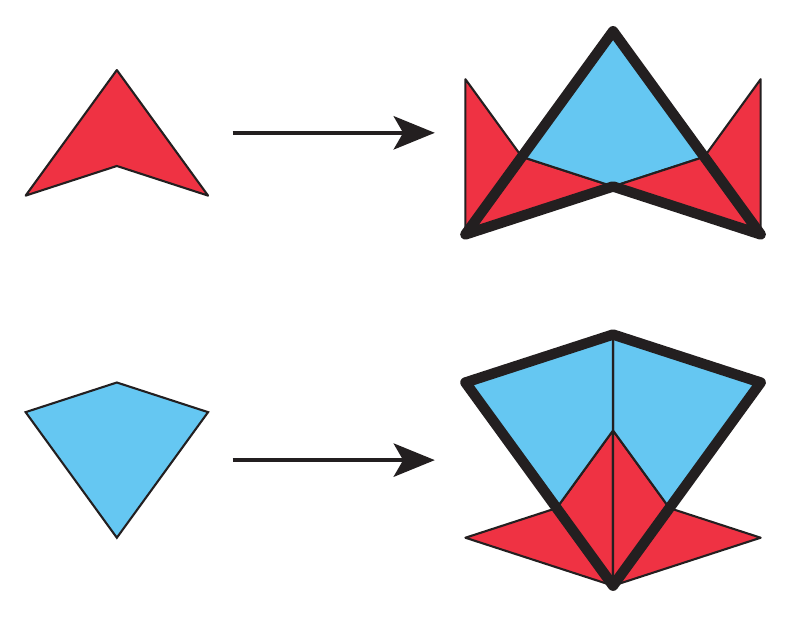}
	\end{minipage}
	\caption{\RaggedRight{Left, a patch of Penrose tiling given by rhombs in \textcolor[RGB]{255,170,170}{\textbf{pink}} and overlaid with its inflation in \textcolor[RGB]{117,0,117}{\textbf{purple}}. Right top, the ``thin'' Penrose rhomb $\tile{T}$ (\textcolor[RGB]{255,65,58}{\textbf{coral}}) and the ``thick'' or ``fat'' Penrose rhomb $\tile{F}$ (\textcolor[RGB]{73,197,241}{\textbf{light blue}}) are presented with their matching rules on edges and their inflation rules. Right bottom, the darts (\textcolor[RGB]{255,65,58}{\textbf{coral}}) and kites (\textcolor[RGB]{73,197,241}{\textbf{light blue}}) are given with their inflation rules. Unlike the rhombs, the kites and darts do not require matching rules to form an admissible Penrose tiling. The two presentations are equivalent since tilings built from kites and darts are mutually locally derivable from tilings by rhombs.}}
	\label{fig:penroseTiling}
\end{figure*}

For example, the Penrose tiling has two prototiles (a thin rhomb and a thick rhomb), and is generated by the substitution rule shown in Figure \ref{fig:penroseTiling}, with scale factor $\lambda=\varphi_+=\frac{1}{2}(1+\sqrt{5})$, the golden ratio.

Suppose that after each application of the substitution rule, the $j$'th prototile $T_{j}$ is decomposed into $M_{1j}$ prototiles of type $T_{1}$, $M_{2j}$ prototiles of type $T_{2}$, etc. The square matrix $M_{ij}$ is called the \textit{substitution matrix} (or \textit{inflation matrix}). Of special interest are substitution rules with $M_{ij}$ of unit determinant, with one eigenvalue $\lambda$ satisfying $\abs{\lambda}>1$, and the remaining eigenvalues $\lambda_{k}$ satisfying $\abs{\lambda_{k}}\leq 1$. Here $\lambda$ is the same scale factor in the corresponding substitution rule described above.

Next, let us define a notion of ``crystal'' which naturally extends to hyperbolic space. Let $\Iso(\bbE^d)$ denote the group of isometries of $d$-dimensional Euclidean space $\bbE^{d}$, and let $P$ be a ``pattern'' (tiling, honeycomb, point set, etc.) in $\bbE^{d}$. We define $P$'s symmetry (or automorphism) group $\Aut(P)$ to be the subgroup of $\Iso(\bbE^d)$ that leaves the pattern $P$ invariant, carrying it into (and onto) itself.

We say that $P$ is a $d$-dimensional \textit{crystalline} pattern (i.e. a $d$-dimensional \textit{crystal}) if its fundamental domain $\bbE^{d}/\Aut(P)$ is a $d$-dimensional polytope of finite non-zero size. Throughout this paper, a $d$-dimensional shape will be said to have finite non-zero size if it has finite non-zero $d$-dimensional volume, inradius, and circumradius. 

Finally, let us introduce four definitions which are helpful in discussing the relationship between different patterns (including tilings):
\begin{enumerate}
    \item Two patterns $P$ and $P'$ in $\bbE^{d}$ are \textit{locally indistinguishable} \cite{baake2013aperiodic} (or \textit{locally isomorphic} \cite{levine1986quasicrystals}) if for any finite portion of pattern $P$ there is an identical portion of pattern $P'$, and vice versa. Here, ``identical'' means congruent: differing only by a rigid motion of Euclidean space, i.e. an element of $\Iso(\bbE^d)$.
    
    For example, consider the Penrose Tiling shown in Figure \ref{fig:penroseTiling}, obeying the rule that adjacent tiles must have matching arrows along their shared edge. It turns out that there are an uncountably infinite number of such tilings that are globally distinct but locally indistinguishable \cite{gardner1977extraordinary, gardner1997penrose}.
    
    \item Two patterns $P$ and $P'$ are \textit{locally equivalent} (or \textit{mutually locally derivable} \cite{baake2013aperiodic}) if $P$ can be obtained from $P'$ in a unique way by local rules, and vice versa, such that any symmetry of $P'$ is also a symmetry of $P$, and vice versa.
    
    For example, Penrose tilings can be presented in terms of Penrose's rhombs or Conway's kites and darts, shown in Figure \ref{fig:penroseTiling}. The tilings are locally equivalent \cite{gardner1977extraordinary, gardner1997penrose}.
    
    \item A pattern $P$ has \textit{local scale symmetry} (or \textit{local inflation/deflation symmetry} \cite{baake2013aperiodic}), with scale factor $\lambda>1$ if it is both locally equivalent to another pattern $P'$, {\it and} locally indistinguishable from $\lambda P'$ (where $\lambda P'$ means the pattern $P'$, expanded by the factor $\lambda$).  The pattern $\lambda P'$ is the \textit{inflation} of $P$, and the local rule for obtaining $P'$ from $P$ is the corresponding inflation rule.  Or, viewed in the reverse direction, $P$ is called the \textit{deflation} of $\lambda P'$, and the local rule for obtaining $P$ from $P'$ is the corresponding \textit{deflation rule}.
    
    Note a key consequence \cite{baake2013aperiodic}: a crystalline pattern cannot have local scale symmetry with $\lambda>1$. Equivalently, the contrapositive: a pattern with local scale symmetry with $\lambda>1$ cannot be crystalline.  Proof: if the two crystalline patterns $P$ and $P'$ are locally equivalent, they have the same fundamental domain; but then $P$ and $\lambda P'$ do {\it not} have the same fundamental domain, so are not locally indistinguishable. 

    Also note: the diffraction spectrum of a self-similar substitution tiling has a non-trivial pure-point component -- like the diffraction spectrum of a crystal -- if and only if its inflation factor is a Pisot number (or PV number), i.e. a real algebraic number $\lambda>1$ whose Galois conjugates all have absolute value less than one  \cite{solomyak1997dynamics, adiceam2016open}.

    Hence, our fourth definition:

    \item A pattern with local scale symmetry, whose scale factor $\lambda$ is a Pisot number, is a {\it self-similar quasicrystal}.
    
    For example, the Penrose tiling has a local scale symmetry with scale factor $\lambda=\varphi_+$, see Figure \ref{fig:penroseTiling}; and since $\varphi_+$ is a PV number, the Penrose tiling is a self-similar quasicrystal.    
\end{enumerate}

\subsubsection{In Hyperbolic Space}
The definitions and results described above apply to $d$-dimensional patterns in $d$-dimensional Euclidean space $\bbE^{d}$. But they also extend to describe $d$-dimensional patterns that are {\it embedded} as $d$-dimensional surfaces in a higher-dimensional manifold.  

In particular, in this paper we describe patterns which naturally take the form of $d$-dimensional surfaces embedded in ($d+1$)-dimensional hyperbolic space $\bbH^{d+1}$.  
Let $\Iso(\bbH^{d+1})$ denote the group of isometries of $\bbH^{d+1}$, and let $P$ be a pattern which is also a $d$-dimensional surface embedded in $\bbH^{d+1}$. We define $P$'s symmetry (or automorphism) group $\Aut(P)$ to be the subgroup of $\Iso(\bbH^{d+1})$ that carries $P$ into itself.  We say that $P$ is a $d$-dimensional \textit{crystalline} pattern (i.e. a $d$-dimensional \textit{crystal}) if its fundamental domain $P/\Aut(P)$ is a piece of $P$ of finite non-zero size.

The remaining four Euclidean definitions now translate to hyperbolic space in a straightforward way. Definitions 2, 3, and 4 are identical, and Definition 1 is modified so that ``identical'' patterns are congruent by an element of $\Iso(\bbH^{d+1})$.
    
    

In our case, we study the quasicrystalline patterns which emerge from regular tessellations of hyperbolic space. In this paper, we discuss the patterns obtained from the $\{p,q\}$ tilings of $\bbH^2$ and the $\{p,q,r\}$ tessellations in $\bbH^3$; and we will explain how the construction extends to higher dimensions.  As a detailed example, we study the 2D self-similar quasicrystal obtained from the self-dual regular icosahedral honeycomb $\{3,5,3\}$ in $\bbH^3$. The $\{3,5,3\}$ honeycomb is built from regular icosahedra, with 12 icosahedra meeting at each vertex in the honeycomb. In particular, we consider 2D ``patterns'' obtained by choosing a subset of the icosahedral faces in this honeycomb which join together to form a 2D surface (with the topology of a 2D sphere or a 2D plane, by construction). These surfaces are stacked into natural families that discretely foliate the icosahedral honeycomb and thus hyperbolic space, and define a \textit{holographic foliation} of $\bbH^3$ (and similarly for the other regular tessellations). 

\section{1D Quasicrystals from \texorpdfstring{$\bbH^2$}{H2}}\label{sec:1DQCfrom2D} 
The essential idea that regular tessellations of $\mathbb{H}^{2}$ naturally give rise to 1D self-similar quasicrystals was first pointed out and explained in \cite{Boyle:2018uiv}. In Section \ref{sec:pqqpGeneralities}, we present a deeper and more elegant refinement of this story which clarifies a number of subtleties in the previous treatment. This refined approach also naturally prepares us for the higher-dimensional generalization presented in subsequent sections. In particular, we show that there is a natural relationship between $\{p,q\}$ and $\{q,p\}$ tilings following from their shared underlying triangulation by $[p,q] = [q,p]$ mirror planes (which we conjecture can be refined even further, see \ref{sec:OpenProblems}). In Section \ref{sec:3773Example} we briefly study the $\{7,3\}$/$\{3,7\}$ example and comment on its connection to the 1D Fibonacci quasicrystal. Finally, in Section \ref{sec:globalaspects} we further address how the growth procedure described here defines a quasicrystal and discuss the simplest examples of ``holographic foliations'' of $\bbH^2$ by quasicrystalline patterns.

\subsection{\texorpdfstring{$\{p,q\}/\{q,p\}$}{{p,q}/{q,p}} Tessellations and Half-Step Rules}\label{sec:pqqpGeneralities}
As discussed in Section \ref{sec:tessellations}, if we draw in all reflection symmetry mirrors on a regular $\{p,q\}$ tessellation, we obtain a triangulation of $\bbH^2$ by $(\pi/2, \pi/p, \pi/q)$ triangles -- i.e. by copies of the fundamental domain of the $[p,q]$ isometry group (see Figure \ref{fig:tessKaleidoscope}).  If we group together the triangles that share a common $\pi/p$ vertex, they form the $p$-gons of the $\{p,q\}$ tiling; and if we instead group together triangles that share a common $\pi/q$ vertex, they form the $q$-gons of the dual $\{q,p\}$ tiling.

\begin{figure*}
\begin{minipage}{0.32\textwidth}
        \centering
    \includegraphics[width=.95\linewidth]{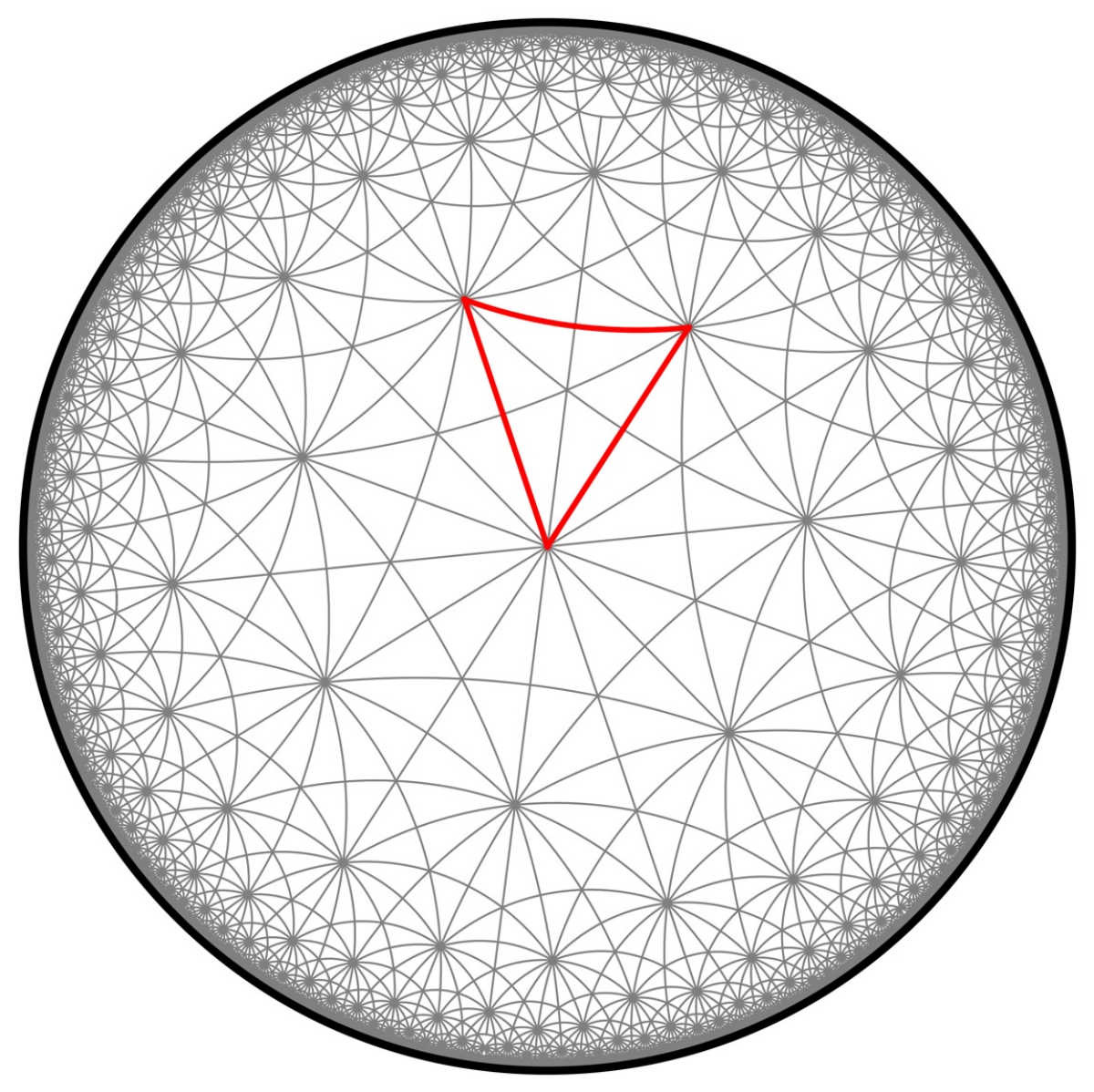}
\end{minipage}
    \begin{minipage}{0.32\textwidth}
        \centering
    \includegraphics[width=.95\linewidth]{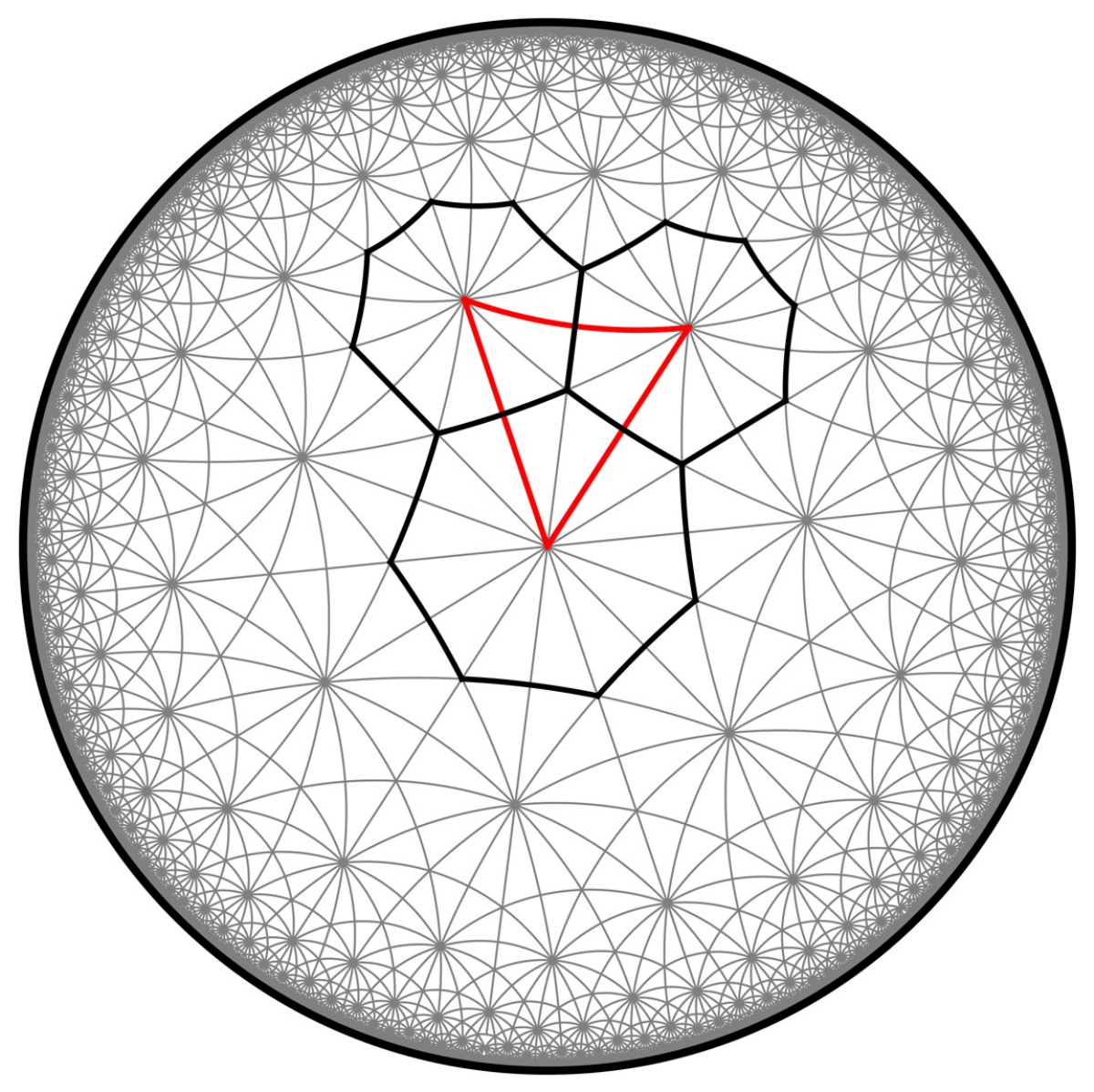}
\end{minipage}
\begin{minipage}{0.32\textwidth}
        \centering
    \includegraphics[width=.95\linewidth]{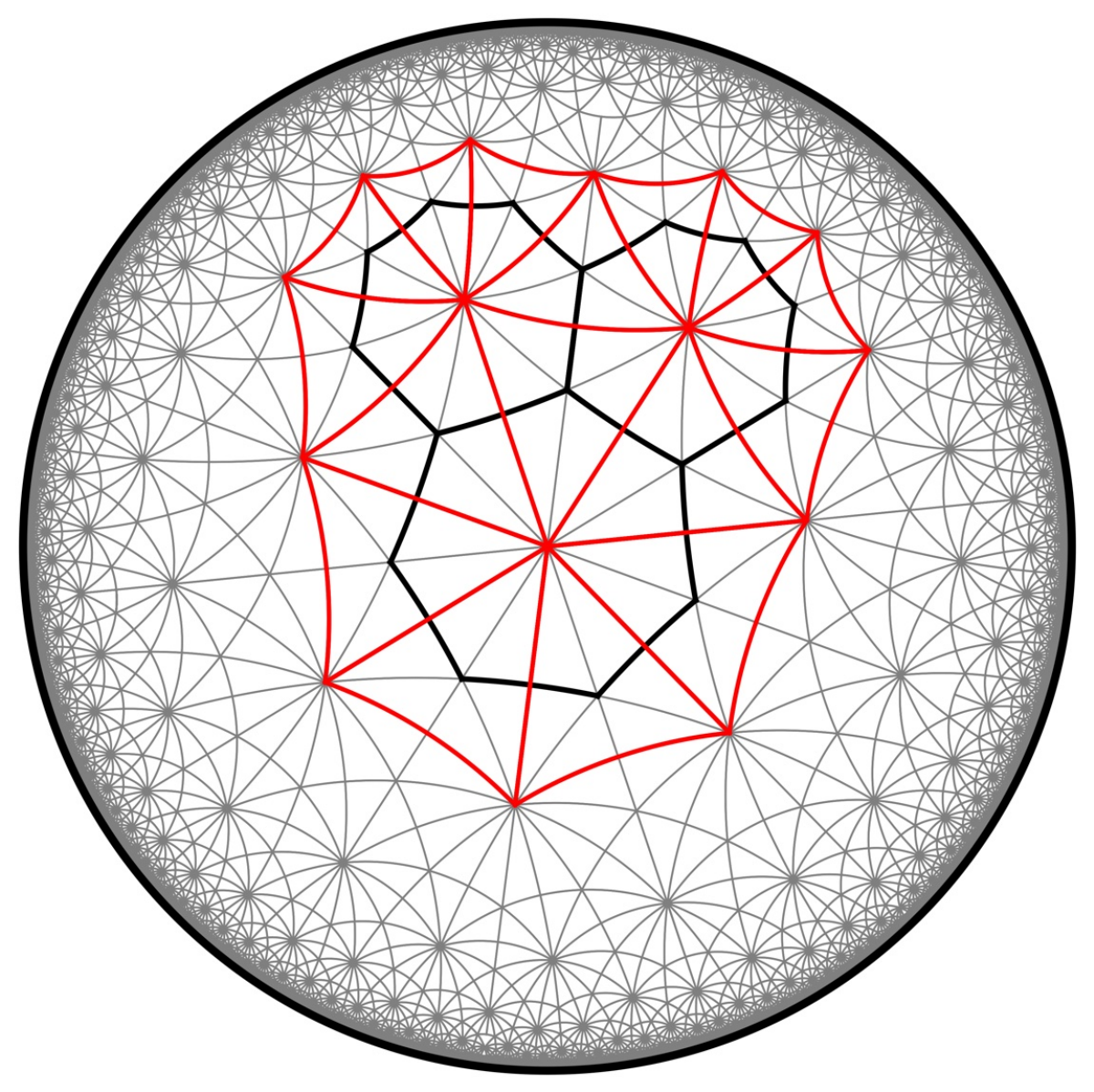}
\end{minipage}
\caption{\RaggedRight{The half-step growth procedures between the $\{3,7\}$ and $\{7,3\}$ tessellations. Left, a patch with 1 triangle (\textcolor[RGB]{232,70,52}{\textbf{red}}) in the $\{7,3\}$ tiling sits inside the triangulation of $\bbH^2$ by $(\tfrac{\pi}{2},\tfrac{\pi}{3},\tfrac{\pi}{7})$ triangles. The associated boundary string is $[\btile{1} \btile{1} \btile{1}]$. Middle, a patch of 3 heptagons (\textcolor{black}{\textbf{black}}) appear in the $\{7,3\}$ tiling, with associated boundary string $[\tile{2} \tile{1}^4 \tile{2} \tile{1}^4 \tile{2} \tile{1}^4]$. Right, another half-step inflation to the $\{3,7\}$ tiling with boundary string $[\btile{3}\btile{2}^3\btile{3}\btile{2}^3\btile{3}\btile{2}^3]$.}}
\label{fig:37GrowthRules}
\end{figure*}

Now consider some finite simply-connected patch\footnote{From here out, we will understand ``finite, simply-connected'' to be included in the word ``patch,'' unless otherwise stated.} $S_{i}$ of $p$-gons in the regular $\{p,q\}$-tiling of $\bbH^2$. If we take the union of $S_{i}$ with all of the $(\pi/2, \pi/p, \pi/q)$ triangles that touch it, we define a new, slightly larger region, denoted $\bar{S}_{i+1}$. Rather than thinking about the underlying tessellation by simplices, it is more geometrically helpful to think of this growth procedure $S_i \mapsto \bar{S}_{i+1}$ by a dualization method: in a patch $S_i$ of a $\{p,q\}$ tessellation, each  $\{p,q\}$ vertex (i.e. each vertex of a $p$-gon)
in $S_i$ may be interpreted as the center of a dual $q$-gon in the $\{q,p\}$ tiling. The union of all such $q$-gons defines the patch $\bar{S}_{i+1}$ of the $\{q,p\}$ tiling. See Figure \ref{fig:37GrowthRules}.

Having defined this growth procedure, we can proceed in a dual fashion for the $\{q,p\}$ tiling: if we take the union of $\bar{S}_{i+1}$ with all of the $(\pi/2, \pi/p, \pi/q)$ triangles that touch it -- or, equivalently, if we reinterpret each $\{q,p\}$ vertex in $\bar{S}_{i+1}$ as the center of a $p$-gon in the dual $\{p,q\}$ tiling and take the union of these $p$-gons -- we define an even larger region $S_{i+2}$ in the $\{p,q\}$ tiling. By repeating this process, we obtain a (semi-) infinite sequence of successively larger patches:
\begin{equation}
    S_i \mapsto \bar{S}_{i+1} \mapsto S_{i+2} \mapsto \bar{S}_{i+3} \mapsto \ldots\,,
\end{equation}
where each patch $S_{i+{\rm even}}$ is a subset of the $\{p,q\}$ tiling, and each patch $\bar{S}_{i+{\rm odd}}$ is a subset of the dual $\{q,p\}$ tiling. We will refer to each step in this growth process as a ``half inflation," and a sequence of two such steps as a ``full inflation."\footnote{We start our indexing at $i$ as a reminder that (based on the size of the initial $S_i$), there may be a (generically non-unique) patch $\bar{S}_{i-1}$ such that $\bar{S}_{i-1} \mapsto S_i$. i.e. deflations are possible from initial configurations.} A picture of this growth procedure is depicted in Figure \ref{fig:37GrowthRules}.

Now return to a single patch $S_i$ of $\{p,q\}$ tiling. The boundary $\partial S_i$ of $S_{i}$ is an alternating sequence of edges and vertices in the $\{p,q\}$ tiling. If a boundary vertex is shared by $n$ many $p$-gons in $S_{i}$ (note that $1\leq n\leq q-1$), we call it an ``$n$-vertex," and denote it $\tile{n}$.  Likewise, the boundary of $\bar{S}_{i+1}$ is a sequence of edges and vertices in the dual $\{q,p\}$ tiling, and if a vertex is shared by $n$ many $q$-gons in $\bar{S}_{i+1}$ (where $1\leq n\leq p-1$), we called it an $n$-vertex of $\{q,p\}$ and denote it $\btile{n}$. We use the bold-faced notation to emphasize the role of the $n$-vertices as the 1D tiles in our boundary quasicrystal.

The shape of the boundary of $S_{i}$ can be specified by giving the sequence of its vertex labels, which we write as
\begin{equation}
    \partial S_i = \cdots\bf{n}_{\ell-1}\bf{n}_{\ell}\bf{ n}_{\ell+1}\cdots\,.
\end{equation}
We explain the reconstructability of the shape of $S_i$ from the string of vertex labels 
$\partial S_i$ (and the higher dimensional analog) in more detail in Appendix \ref{sec:shootRotateShoot}. Our principle claim is that: \textit{the growth procedure $S_i \mapsto \bar{S}_{i+1}$ induces a local substitution rule $\partial S_i \mapsto \partial \bar{S}_{i+1}$ on the boundary string; and, for hyperbolic honeycombs, this local substitution rule defines substitution tilings that are self-similar quasicrystals.}

First, let's define the substitution rule. Locality of the substitution rule will be manifest since we will define it on $n$-vertices without (essential) reference to its neighbours:
\begin{enumerate}
    \item Consider an $n$-vertex ${\bf n}$ on the boundary of $S_{i}$, locally it looks like:
    \begin{equation}\label{eq:localGrowth1}
        \bf{n} \sim
\begin{tikzpicture}[every path/.style={thick}, baseline={(current bounding box.center)},scale=0.5]
\tikzstyle{every node}=[font=\normalsize]
\def\xx{10pt};
\def\yy{10pt};
\coordinate (A) at (1.25,2.5);
\coordinate (B) at (1.5,3.75);
\coordinate (C) at (2.5,4.75);
\coordinate (D) at (6.25,2.5);
\coordinate (E) at (6,3.75);
\coordinate (F) at (5,4.75);
\coordinate (G) at (1.75,1.25);
\coordinate (H) at (6,1.25);
\coordinate (I) at (2.75,0.5);
\coordinate (J) at (4.75,0.5);
\coordinate (K) at (3.75,2.5);
\coordinate (L) at (0,1.5);
\coordinate (M) at (7.5,1.5);
\coordinate (N) at (6.25,0);
\coordinate (O) at (1.25,0);
\coordinate (P) at (3.75,0);
\draw [line width=1.0pt, black] (K)--(L);
\draw [line width=1.0pt, black] (K)--(M);
\draw [line width=1.0pt, black] (K)--(N);
\draw [line width=1.0pt, black] (K)--(O);
\draw [line width=1.0pt, black] (K)--(P);
\draw (J) node[xshift = -1pt] {$\cdots$};
\draw (H) node {$n$};
\draw (I) node {$\,2$};
\draw (G) node[xshift = -5pt] {$1$};
\end{tikzpicture}
\hphantom{\sim {\bf n}}
    \end{equation}
    \item When we perform the $\{p,q\}\mapsto\{q,p\}$ half-inflation $S_{i}\mapsto\bar{S}_{i+1}$, the ${\bf n}$ becomes the center of a new $q$-gon. $n$ vertices of the $q$-gon lie interior to $S_{i}$ (\textcolor{black}{\textbf{black nodes}} below), while the remaining $n-q$ vertices are exterior to $S_{i}$. We split the exterior vertices into the two outer-most vertices (\textcolor{orange}{\textbf{left-most node}} and \textcolor{orange}{\textbf{right-most node}}) and the remaining $q-n-2$ vertices in between (\textcolor{black}{\textbf{white nodes}}):
    \begin{equation}\label{eq:localGrowth2}
        \begin{tikzpicture}[every path/.style={thick}, baseline={(current bounding box.center)}, scale=0.75]
\tikzstyle{every node}=[font=\large]
\def\xx{8pt};
\def\yy{8pt};
\coordinate (A) at (-2.5, 2.5);
\coordinate (D) at (2.5, 2.5);
\coordinate (B) at (-2.25, 3.75);
\coordinate (E) at (2.25, 3.75);
\coordinate (C) at (-1.25, 4.75);
\coordinate (F) at (1.25, 4.75);
\coordinate (G) at (-2.0, 1.25);
\coordinate (H) at (2.0, 1.25);
\coordinate (I) at (-1.0, 0.5);
\coordinate (J) at (1.0, 0.5);
\coordinate (K) at (0.0, 2.5);
\coordinate (L) at (-3.75, 1.5);
\coordinate (M) at (3.75, 1.5);
\coordinate (N) at (2.5, 0);
\coordinate (O) at (-2.5, 0);
\coordinate (P) at (0.0, 0);
\draw [line width=1.0pt, gray] (K)--(L);
\draw [line width=1.0pt, gray] (K)--(M);
\draw [line width=1.0pt, gray] (K)--(N);
\draw [line width=1.0pt, gray] (K)--(O);
\draw [line width=1.0pt, gray] (K)--(P);
\draw [black,line width=1.5pt] (A)--(B);
\draw [black,line width=1.5pt] (D)--(E);
\draw [black,line width=1.5pt] (B)--(C);
\draw [black,line width=1.5pt] (G)--(A);
\draw [black,line width=1.5pt] (H)--(D);
\draw [black,line width=1.5pt] (G)--(I);
\draw [black,line width=1.5pt] (I)--(J);
\draw [black,line width=1.5pt] (J)--(H);
\draw [black,line width=1.5pt] (E)--(F);
\draw [black, line width=1.5pt, dotted] (C) -- (F) node[midway, above] {};
\filldraw [WhiteNode] (B) circle (3pt) node[xshift={-1*\xx*0.7}, yshift={\yy*0.7}] {1};
\filldraw [WhiteNode] (E) circle (3pt) node[xshift={\xx*0.7}, yshift={\yy*0.7}] {1};
\filldraw [WhiteNode] (C) circle (3pt) node[xshift=0, yshift=\yy] {1};
\filldraw [WhiteNode] (F) circle (3pt) node[xshift=0, yshift=\yy] {1};
\filldraw [OrangeNode] (A) circle (3pt) node[xshift=-\xx, yshift=0] {2};
\filldraw [OrangeNode] (D) circle (3pt) node[xshift=+\xx, yshift=0] {2};
\filldraw [BlackNode] (J) circle (3pt) node[xshift=0, yshift=-\yy] {};
\filldraw [BlackNode] (H) circle (3pt) node[xshift=+\xx, yshift=-\yy] {};
\filldraw [BlackNode] (I) circle (3pt) node[xshift=0, yshift=-\yy] {};
\filldraw [BlackNode] (G) circle (3pt) node[xshift=-\xx, yshift=-\yy] {};
\draw [decorate,decoration={brace,amplitude=6pt,raise=9ex}]
  (B) -- (E) node[midway, yshift = 12ex]{{\small $q\!-\!n\!-\!2$ vertices}};
\end{tikzpicture}
    \end{equation}
    \item The $q-n-2$ middle vertices become $\bar{{\bf 1}}$'s on the new boundary. In contrast, the outer-most vertices become new $\bar{{\bf 2}}$'s since they also grow from the original ${\bf n}$'s left and right neighbours in the boundary of $S_{i}$.
\end{enumerate}
Since the two new $\bar{{\bf 2}}$'s in $\partial \bar{S}_{i+1}$ are only ``half contributed to'' by the ${\bf n}$ itself, with half also coming from its left and right neighbours, we opt to present the substitution rule with fractional powers with the understanding that only integer powers will actually appear in the boundary. Thus for a $\{p,q\}\mapsto\{q,p\}$ half-inflation, we have the substitution rule
\begin{equation}
    {\bf n}\mapsto \bar{{\bf 2}}^{{\frac{1}{2}}}\bar{{\bf 1}}^{q-n-2}\bar{{\bf 2}}^{{\frac{1}{2}}} \label{n_rule_v1}\,.
\end{equation}
The procedure repeats analogously for the $\{q,p\} \mapsto \{p,q\}$ half-inflation. So that, in total, we have the half-step inflation rules
\begin{empheq}[box={\mymath[colback=gray!10, sharp corners]}]{align}
\begin{split}
    {\bf n} &\mapsto\bar{{\bf 2}}^{{\frac{1}{2}}}\bar{{\bf 1}}^{q-n-2}\bar{{\bf 2}}^{{\frac{1}{2}}} \\
    \bar{{\bf n}}&\mapsto{\bf 2}^{{\frac{1}{2}}}{\bf 1}^{p-n-2}{\bf 2}^{{\frac{1}{2}}}
\end{split}\label{eq:substitutionRules2D}
\end{empheq}
Using this, we can also obtain the substitution rule affiliated with the full-step inflation $\{p,q\} \mapsto \{p,q\}$ (presented in its most symmetric form):
\begin{equation}\label{eq:fullStep2D}
    \mathbf{n} \mapsto {\mathbf{1}}^{{\frac{p-4}{2}}} {\mathbf{2}}^{{\frac{1}{2}}} ({\mathbf{2}}^{\frac{1}{2}}{\mathbf{1}}^{p-3} {\mathbf{2}}^{\frac{1}{2}})^{q-n-2} {\mathbf{2}}^{{\frac{1}{2}}} {\mathbf{1}}^{\frac{p-4}{2}}\,.
\end{equation}

So far, we have implicitly assumed that $n$ is less than its maximal value $q-1$, so that the number of middle vertices is non-negative, $q-n-2\geq0$.  When $n$ is its maximal value, $n = q-1$, the number of middle vertices is a negative integer power $q-n-2=-1$. Similar to our understanding of fractional powers like $\btile{2}^{{\frac{1}{2}}}$, we can make sense of negative powers algebraically, and in Appendix \ref{sec:shootRotateShoot} we give it a clean geometric interpretation. Actually, considering negative powers of vertices introduces a deeper formulation of the half-inflation rule.

To state this rule in its most elegant and general form, we first formally regard $n$-vertices with $n>2$ as ``composite'' strings, formed from a sequence of 
adjacent $2$-vertices, glued together by ``inverse'' $1$-vertices, in a way that preserves the angle subtended by the vertex (again, the geometric interpretation is provided in Appendix \ref{sec:shootRotateShoot}). Thus:
\begin{align}
    {\bf 3} &\cong{\bf 2}{\bf 1}^{-1}{\bf 2} \nonumber\\
    {\bf 4} &\cong{\bf 2}{\bf 1}^{-1}{\bf 2}{\bf 1}^{-1}{\bf 2} \\
    {\bf 5} &\cong{\bf 2}{\bf 1}^{-1}{\bf 2}{\bf 1}^{-1}{\bf 2} {\bf 1}^{-1}{\bf 2} \nonumber\\
    & \qquad \quad \vdots \nonumber
\end{align}
and so on, and likewise for the $\bar{\mathbf{n}}$ vertices. More generally, we can write
\begin{empheq}[box={\mymath[colback=gray!10, sharp corners]}]{align}
\begin{split}
    {\bf n} &\cong {\bf 2}({\bf 1}^{-1}{\bf 2})^{n-2}
    =({\bf 2}{\bf 1}^{-1})^{n-2}{\bf 2} \\
    \bar{{\bf n}}&\cong \bar{{\bf 2}}(\bar{{\bf 1}}^{-1}\bar{{\bf 2}})^{n-2}
    =(\bar{{\bf 2}}\bar{{\bf 1}}^{-1})^{n-2}\bar{{\bf 2}}
\end{split}\label{n_vert_equiv}
\end{empheq}
which we note are valid even for $n=1$ and $n=2$. 

With this identification, \textit{we can now regard the $1$ and $2$-vertices as fundamental building blocks for all other vertices.} As we will see below, this is also generically a good choice for describing boundary behaviour after a large number of inflations. In particular, under a $\{p,q\}\mapsto\{q,p\}$ half inflation the $1$ and $2$ vertices become:
\begin{align}
\begin{split}
    {\bf 1}&\mapsto\bar{{\bf 2}}^{{\frac{1}{2}}}\bar{{\bf 1}}^{q-3}\bar{{\bf 2}}^{{\frac{1}{2}}}\,,\\
    {\bf 2}&\mapsto\bar{{\bf 2}}^{{\frac{1}{2}}}\bar{{\bf 1}}^{q-4}\bar{{\bf 2}}^{{\frac{1}{2}}}\,, \label{2D_subs}
\end{split}
\end{align}
and likewise for the dual tiles.
For algebraic consistency, we note that we must have the half-step substitution rules on inverses:
\begin{align}
\begin{split}
    {\bf 1}^{-1}&\mapsto\bar{{\bf 2}}^{-{\frac{1}{2}}}\bar{{\bf 1}}^{3-q}
    \bar{{\bf 2}}^{-{\frac{1}{2}}} \\
    \bar{{\bf 1}}^{-1}&\mapsto{\bf 2}^{-{\frac{1}{2}}}{\bf 1}^{3-p}
    {\bf 2}^{-{\frac{1}{2}}}\,, \label{eq:inverseSubs}
\end{split}
\end{align}
which is also consistent with our rules in \eqref{eq:substitutionRules2D}. The appearance of a $\btile{1}^{-1}$ inside the substitution rule for a $(q-1)$-vertex can also be easily understood pictorially: a $(q-1)$-vertex sits adjoined to $q-1$ many $p$-gons, and thus at the center of a dual $q$-gon with one missing $p$-gonal slice. Under half-step inflation, the $\btile{1}^{-1}$ appearing captures the ``gap closing'' in this dual $q$-gon (this will be very apparent in the $\{7,3\}$ and $\{3,7\}$ example described in Section \ref{sec:3773Example}).

Working with the ``fundamental'' $1$ and $2$ vertices, our half-step substitution rules \eqref{2D_subs}
correspond to the substitution matrix\footnote{From here out, we will suppress the results for the $\{q,p\}$ tilings (i.e. the barred tiles), as they are identical to the $\{p,q\}$ tilings with $p$ and $q$ interchanged.}
\begin{equation}
    M_{pq\mapsto qp}
        =\left(\begin{array}{cc} q-3 & q-4 \\ 1 & 1\end{array}\right)\,.
\end{equation}

The full-step substitution matrix $\{p,q\} \mapsto \{p,q\}$ can be obtained by composing the two half-step substitution matrices $M_{pq\mapsto pq}=M_{qp\mapsto pq} M_{pq\mapsto qp}$:
\begin{equation}
    M_{pq\mapsto pq}
        =\left(\begin{array}{cc} \!(p\!-\!3)(q\!-\!2)\!-\!1\! & \!(p\!-\!3)(q\!-\!3)\!-\!1\! \\ q-2 & q-3 \end{array}\right)\,.
\end{equation}
This composition is consistent with the rule in \eqref{eq:fullStep2D}. The eigenvalues $\lambda_{\pm}$ are:
\begin{equation}
    \lambda_{\pm}^{}=\gamma\pm(\gamma^{2}\!-\!1)^{\frac{1}{2}}\,,\quad
    \gamma:=\frac{(p\!-\!2)(q\!-\!2)}{2}-1\,.\label{eq:2DEigenvalues}
\end{equation}
If we calculate the eigenvector $v_+$ associated with the dominant eigenvalue $\lambda_+$, its components correspond to the asymptotic frequencies of the ${\bf 1}$ and ${\bf 2}$ tiles after many rounds of inflation.

We also note that the eigenvalues of the full-step substitution matrix $M_{pq \mapsto pq}$ in \eqref{eq:2DEigenvalues} match with those for the substitution matrix in \cite{Boyle:2018uiv}. We identify a different set of basic tiles for the boundary quasicrystal than in \cite{Boyle:2018uiv}, which explains the different substitution rules and substitution matrix. However, the boundary patterns derived above are locally equivalent to the patterns defined there, and such a local equivalence induces a change of basis on the substitution matrix, hence the exact same eigenvalues. These eigenvalues are also the growth ratios of hyperbolic honeycombs, as described in \cite{nemeth2017growing} (see also Appendix \ref{sec:353Projections}).\footnote{The growth procedure also looks similar to the definition of extremal animals, see e.g. \cite{malen2023extremal}, we leave details and interplay of the connection to future works.}

If one constructs the full $q\times q$ matrix on all vertices in the full-step inflation $\{p,q\} \mapsto \{p,q\}$, one finds that it has the same eigenvalues and eigenvectors as in the $2\times 2$ matrix case, and $q-2$ additional 0-eigenvalues. This supports our earlier claim that the substitution rules can be encoded in terms of only two fundamental tiles, which we chose to identify with $1$ and $2$ vertices. The $0$-eigenvalues reflect the fact that other species of tiles go extinct after a few inflations.

In addition to being simpler overall, and generalizing better to higher dimensions, the half-step substitution rules that we identify here also explain the $p \leftrightarrow q$ symmetries in full-step formulas (e.g.\ in \eqref{eq:2DEigenvalues}), since both $\{p,q\}$ and $\{q,p\}$ tilings come from a shared underlying triangulation by fundamental domains. Indeed, while the eigenvalues/eigenvectors of the half-step matrices do not immediately make sense for tilings which are not self-dual (since the matrices do not map a space to itself), the full-step $p \leftrightarrow q$ symmetries suggest that there exist change of basis/tile matrices $\Lambda_{qp}$ and $\Lambda_{pq}$ satisfying:
\begin{align}
    \begin{split}
        \Lambda_{qp}^{-1} M_{pq \mapsto qp} M_{qp \mapsto pq} \Lambda_{qp} &= M^2\,,\\
        \Lambda_{pq}^{-1} M_{qp \mapsto pq} M_{pq \mapsto qp} \Lambda_{pq} &= M^2\,,
    \end{split}\label{eq:nonCOB}
\end{align}
where $M^2$ is some substitution matrix that is similar to both $M_{pq \mapsto pq}$ and $M_{qp \mapsto qp}$. The half-step rules then suggest the possibility of factorizing this equation (the first one, say) to the form
\begin{equation}
    (\Lambda_{qp}^{-1} M_{pq \mapsto qp} \Lambda_{pq})(\Lambda_{pq}^{-1} M_{qp \mapsto pq} \Lambda_{qp}) = M^2\,,
\end{equation}
where each parenthetical expression equals $M$. Thus, under a suitable redefinition of tiles given by $\Lambda_{qp}$ and $\Lambda_{pq}$, we have a complete unification of the $\{p,q\}$ and $\{q,p\}$ boundary quasicrystals. 

Our findings in this section can be summarized as follows:
\begin{itemize}
    \item Given a regular tessellation $\{p,q\}$, there is a natural way to ``grow'' a patch $S_i$ of the tessellation, by ``half a step,'' to a patch $\bar{S}_{i+1}$ of the dual tessellation $\{q,p\}$. This is called ``half-step inflation.''
    \item There is a natural procedure assigning a string (aka 1D tiling) $\partial S_i$ to the boundary of any patch. In our case, letters in the string are identified with vertices on the boundary. This assignment is unique up to local equivalence of tile sets.
    \item A boundary pattern $\partial S_i$ can be used to reconstruct a bulk patch $S_i$, as described geometrically in Appendix \ref{sec:shootRotateShoot}.
    \item The bulk half-step growth rule induces a \textit{local} substitution rule operation on the boundary tilings. Computing the eigenvalues of $M_{pq\mapsto pq}$, one finds that (only) for hyperbolic tessellations is the dominant eigenvalue irrational, and more specifically a Pisot number. Thus the boundary substitution rules are \textit{invertible} or, more accurately, they are invertible after sufficiently many inflations have ``washed out'' the $0$-eigenvalue directions of the full $q \times q$ substitution matrix. This reflects the underlying (eventual) self-similar quasiperiodic structure of the induced tiling.\footnote{In other words, as a patch $S_i$ grows, it becomes more convex/uniform and eventually ``stabilizes'' so that the boundary only admits two species of prototiles. Less phenomenologically, it may be cleaner to \textit{define} the growth procedure in the bulk as the process exactly dual to the substitution rule on the boundary. Then the growth/degrowth procedures in the bulk would proceed along some nested maximally convex patches.\label{footnote:Asymptotic}}
\end{itemize}
We can describe the process pictorially in the following commuting diagram:
\begin{equation*}
\begin{tikzcd}
	{\begin{matrix}\text{Bulk patch } S_i \\ \text{of $\{p,q\}$ tiling}\end{matrix}} && \begin{matrix} \text{Boundary pattern } \partial{S}_{i} \\ \cdots{\bf{n}}_{k-1}{\bf{n}}_{k}{\bf{ n}}_{k+1}\cdots\end{matrix} \\
	\\
	{\begin{matrix}\text{Bulk patch } \bar{S}_{i+1} \\ \text{of $\{q,p\}$ tiling}\end{matrix}} && \begin{matrix} \text{Boundary pattern } \partial\bar{S}_{i+1} \\ \cdots\bar{\bf{n}}_{\ell-1}\bar{\bf{n}}_{\ell}\bar{\bf{ n}}_{\ell+1}\cdots\end{matrix}
	\arrow["{\begin{matrix}\text{Pass to}\\\text{boundary}\end{matrix}}", from=1-1, to=1-3]
	\arrow["\begin{matrix}\text{Half-step}\\\text{inflation}\end{matrix}"', from=1-1, to=3-1]
	\arrow["{\begin{matrix}\text{Substitution}\\\text{Rule}\end{matrix}}", from=1-3, to=3-3]
	\arrow["{\begin{matrix}\text{Pass to}\\\text{boundary}\end{matrix}}"', from=3-1, to=3-3]
\end{tikzcd}
\end{equation*}

\subsection{Example: The \texorpdfstring{$\{7,3\}/\{3,7\}$}{{7,3}/{3,7}} Case}\label{sec:3773Example}

Let us illustrate the above formalism in the special case of a regular
$\{p,q\}=\{7,3\}$ tessellation. In this case, \eqref{2D_subs} says the $\{7,3\} \mapsto \{3,7\}$ half-step substitution rules are:
\begin{align}
\begin{split}
  \label{7_3_subs}
    {\bf 1}&\mapsto\bar{{\bf 2}}^{{\frac{1}{2}}}\bar{{\bf 1}}^{0}\bar{{\bf 2}}^{{\frac{1}{2}}}\,, \\
    {\bf 2}&\mapsto\bar{{\bf 2}}^{{\frac{1}{2}}}\bar{{\bf 1}}^{-1}\bar{{\bf 2}}^{{\frac{1}{2}}}\,,
\end{split}
\end{align}
while the $\{3,7\} \mapsto \{7,3\}$ substitution rules are:
\begin{align}
\begin{split}
  \label{3_7_subs}
    \bar{{\bf 1}}&\mapsto{\bf 2}^{{\frac{1}{2}}}{\bf 1}^{4}{\bf 2}^{{\frac{1}{2}}}\,, \\
    \bar{{\bf 2}}&\mapsto {\bf 2}^{{\frac{1}{2}}}{\bf 1}^{3}{\bf 2}^{{\frac{1}{2}}}\,.
\end{split}
\end{align}

As can be seen by example (see Figure \ref{fig:37GrowthRules}), the boundary of a simple patch of $\{3,7\}$ tiling is typically made of $2$ and $3$-vertices, not $1$ and $2$-vertices. This is not in contradiction with the results of the previous section, since the results are completely algebraically and geometrically consistent. Indeed, if we consider the rightmost picture in Figure \ref{fig:37GrowthRules} and recall the equivalence $\bar{\bf{3}} \cong \bar{\bf{2}} \bar{\bf{1}}^{-1} \bar{\bf{2}}$, we can clearly see how the substitution rule ${\bf{2}} \mapsto \bar{\bf{2}}^{{\frac{1}{2}}} \bar{\bf{1}}^{-1} \bar{\bf{2}}^{\frac{1}{2}}$ accurately captures the idea that a $\bar{\bf{3}}$ in the $\{3,7\}$ layer is produced ``above'' a $\bf{2}$ in the $\{7,3\}$ layer (with assistance from the nearest neighbouring vertices).

We note that the full-step substitution matrix for $\{7,3\}\mapsto \{7,3\}$ or $\{3,7\} \mapsto \{3,7\}$ has eigenvalues $\varphi_{\pm}^2$, where $\varphi_{\pm} = (1\pm\sqrt{5})/2$. This suggests a potential relation to the Fibonacci quasicrystal, the prototypical 1D quasicrystal \cite{bombieri1986distributions, bombieri1987quasicrystals}. As we will see, such a relationship exists: the Fibonacci quasicrystal is the boundary quasicrystal that is induced and grown by half-step inflations between the $\{3,7\}$ and $\{7,3\}$ tilings. Moreover, this demonstrates how the half-step substitution rules refine the quasicrystals that could appear as boundaries of hyperbolic honeycombs from earlier works.

The Fibonacci quasicrystal has two tiles, one ``short'' tile (\textcolor[RGB]{232,70,52}{\textbf{red}}) and one ``long'' tile (\textcolor{blue!90!black}{\textbf{blue}}) which is $\varphi_+$-times larger than the short tile. The short tile inflates to a long tile, and the long tile inflates to a short tile with two halves of a long tile on either side, as follows:
\begin{align}
    \begin{split}
        \begin{tikzpicture}[baseline={(current bounding box.center)}, scale = 1]
        	\def\r{1};
        	\def\p{1.618};
        	\node[] (ghost1) at (-1*\p/2,0) {};
        	\node[] (ghost2) at (+1*\p/2,0) {};
        	\node[] (T1) at (-1*\r/2,\r*0) {};
        	\node[] (T2) at (+1*\r/2,\r*0) {};
            \draw[-,red,very thick] (T1) -- (T2) node[midway, below] {};
        \end{tikzpicture}
            \mapsto&
        \begin{tikzpicture}[baseline={(current bounding box.center)}, scale = 1]
        	\def\r{1};
        	\def\p{1.618};
        	\node[] (T1) at (-1*\p/2,\r*0) {};
        	\node[] (T2) at (+1*\p/2,\r*0) {};
            \draw[-,blue,very thick] (T1) -- (T2) node[midway, below] {};
        \end{tikzpicture}
        \,, \\
        \begin{tikzpicture}[baseline={(current bounding box.center)}, scale = 1]
        	\def\r{1};
        	\def\p{1.618};
        	\node[] (T1) at (-1*\p/2,\r*0) {};
        	\node[] (T2) at (+1*\p/2,\r*0) {};
            \draw[-,blue,very thick] (T1) -- (T2) node[midway, below] {};
        \end{tikzpicture}
            \mapsto&
        \begin{tikzpicture}[baseline={(current bounding box.center)}, scale = 1]
        	\def\r{1};
        	\def\p{1.618};
        	\node[] (T0) at (-1*\r/2-\p/2+0.125,\r*0) {};
        	\node[] (T1e) at (-1*\r/2+0.25,\r*0) {};
        	\node[] (T1) at (-1*\r/2,\r*0) {};
        	\node[] (T2) at (+1*\r/2,\r*0) {};
        	\node[] (T2e) at (1*\r/2-0.25,\r*0) {};
        	\node[] (T3) at (+1*\r/2+\p/2-0.125,\r*0) {};
            \draw[-,blue,very thick] (T0) -- (T1e) node[midway, below] {};
            \draw[-,red,very thick] (T1) -- (T2) node[midway, below] {};
            \draw[-,blue,very thick] (T2e) -- (T3) node[midway, below] {};
        \end{tikzpicture} \,.
    \end{split}
\end{align}
The accompanying substitution matrix $M_{\mathrm{fib}}$ and eigenvalues (as promised) are:
\begin{equation}
    M_{\mathrm{fib}}=\left(\begin{array}{rr} 0 & 1 \\ 1 & 1 \end{array}\right)\,
    \!\quad\stackrel{\mathrm{Eigenvalues}}{\rightsquigarrow}\!\quad \varphi_{\pm} = \frac{1}{2}(1\pm\sqrt{5})\,.
\end{equation}

To see how the hyperbolic honeycombs give the Fibonacci tiling, we must switch from the boundary vertices to a different, locally equivalent, tile basis. Define the new tiles $\tile{S}$ and $\tile{L}$ in the $\{7,3\}$ boundary by:
\begin{align}
\begin{split}\label{eq:SLxform}
    \tile{S}&=\tile{2}^{\frac12}\tile{1}^{2}\tile{2}^{\frac12}\,,\\
    \tile{L}&=\tile{2}^{\frac12}\tile{1}^{3}\tile{2}^{\frac12}\,.
\end{split}
\end{align}
Similarly, if $\btile{2}$ and $\btile{3}$ denote the 2-vertices and 3-vertices in the $\{3,7\}$ boundary, we can define the new tiles $\btile{S}$ and $\btile{L}$ as the following sequences:
\begin{align}
\begin{split}\label{eq:SLbxform}
    \btile{S}&=\btile{2}^{\frac12}\btile{3}^{0}\btile{2}^{\frac12}\,, \\
        \btile{L}&=\btile{2}^{\frac12}\btile{3}^{1}\btile{2}^{\frac12}\,.
\end{split}
\end{align}
With this redefinition, the half-step inflation rule from the $\{7,3\}$ to the $\{3,7\}$ boundary is 
\begin{align}
\begin{split}
    \tile{S}&\mapsto\btile{L}^{\frac12}\btile{S}^{0}\btile{L}^{\frac12}\,,\\
    \tile{L}&\mapsto\btile{L}^{\frac12}\btile{S}^{1}\btile{L}^{\frac12}\,.
\end{split}
\end{align}
and the half-step inflation rule from the $\{3,7\}$ to the $\{7,3\}$ boundary is
\begin{align}
\begin{split}
    \btile{S}&\mapsto\tile{L}^{\frac12}\tile{S}^{0}\tile{L}^{\frac12}\,,\\
    \btile{L}&\mapsto\tile{L}^{\frac12}\tile{S}^{1}\tile{L}^{\frac12}\,.
\end{split}
\end{align}
These are exactly the inflation rules for the Fibonacci quasicrystal!

This demonstrates how the half-step substitution rules refine the quasicrystals that could appear as boundaries of hyperbolic honeycombs. The $\Lambda_{pp}$ and $\Lambda_{qq}$ transforms in \eqref{eq:nonCOB} may be read off from Eqs.~\eqref{eq:SLxform} and \eqref{eq:SLbxform}:
\begin{equation}
    \Lambda_{73} = 
    \begin{pmatrix}
        2 & 3\\
        1 & 1
    \end{pmatrix}\,,\quad
    \Lambda_{37} = 
    \begin{pmatrix}
        0 & -1\\
        1 & 3 
    \end{pmatrix}\,,
\end{equation}
and we can confirm that
\begin{equation}
    \Lambda_{37}^{-1}M_{73 \mapsto 37} \Lambda_{73} = 
    M_{\mathrm{fib}} = 
    \Lambda_{73}^{-1}M_{37 \mapsto 73} \Lambda_{37}\,.
\end{equation}

However, all of the 1D quasicrystals obtained from $\bbH^2$ are also obtainable in flat space, essentially because 1D tilings have no intrinsic curvature, and thus do not depend on the embedded spacetime in any interesting way. However, as we shall see in Section \ref{sec:H3Tiling}, the 2D quasicrystals obtained in $\bbH^3$ appear to be new.

\subsection{Global Aspects: 1D Quasicrystals and Holographic Foliations}\label{sec:globalaspects}

Having discussed the local inflation/deflation rules governing the evolution of our tiling from layer to layer, we now turn to some more global aspects of these tilings.

\subsubsection{Perspective 1: Growth from a Seed}
So far, we have taken the perspective that we start with a finite patch of tiles (the initial seed), and this patch then grows -- layer by layer -- through the iterative application of our half-step inflation algorithm.  Let us start with several remarks based on this perspective.

We can imagine various pathological scenarios that can arise as an initial seed grows, for example:
\begin{enumerate}
    \item[i)] The two ends of a horseshoe-shaped region can merge to form an ring-shaped region; passing from simply to non-simply connected patches.
    \item[ii)] The hole in a ring-shaped region can be filled in; passing from non-simply to simply connected patches.
    \item[iii)] A finite number of disconnected regions can merge; passing from disconnected to connected.
\end{enumerate}
In any case, after a finite (but possibly large) number of half-step inflations, any finite initial seed will eventually grow to form a single simply-connected blob. Moreover, as can be seen from examples and the analysis of our substitution matrices, given a finite initial patch $S_i$ of $\{p,q\}$ tiling, the types of tiles appearing in the boundary eventually relax to just two types (with all other tile types ``dying out'').

In this perspective, notice that some essential information about the initial configuration can be erased by the inflation process -- e.g. it is not hard to find two patches $S_i \neq S_{i}'$ such that $S_{i+k} = S_{i+k}'$ for all sufficiently large $k$ (horseshoe shapes will often do the trick).\footnote{Obviously an initial seed with a hole would also work, but it is not clear if one should assign a separate string to the ``interior boundary.'' In any case, it is not necessary, as one can find examples that stay completely within the space of simply connected configurations.}  This suggests associating an entropy ${\cal S}(S_i)$ with each finite patch $S_i$, that counts the number of distinct initial seed configurations that grow into $S_i$; we will not comment on this idea any further here.

Finally, where does our 1D quasicrystal live? From this first perspective, in which we start with a finite seed and grow it by iterating the half-step inflation process, the answer is that it lives at the {\it boundary} of $\bbH^2$.  After all, for any finite area patch $S_i$, the boundary $\partial S_i$ is a string of finite length $L$, and so is obviously not a quasicrystal since it is periodic with period $\leq L$ (with possibly shorter periods based on the symmetry of the initial seed, see e.g. Figure \ref{fig:37GrowthRules}).  But with each half-step inflation, the (minimal) period of the boundary string increases, so in the limit that we reach the boundary, we obtain a genuine quasicrystal with no finite period
\begin{equation}\label{eq:bdQuasicrystal}
    \lim_{k \to \infty} S_{i+k} \longrightarrow \text{1D Quasicrystal}\,.
\end{equation}

Now let us turn to a different perspective on our tilings, what they are, and where they live.

\subsubsection{Perspective 2: Holographic Foliations}
As a warm-up, let us return to the Penrose tiling. It is often introduced colloquially as follows: one presents the inflation rule shown in Figure \ref{fig:penroseTiling}; then, by starting from a single tile and iteratively applying this rule, one generates larger and larger (finite) patches of Penrose tiling (containing more and more tiles); in the infinite limit, one obtains an infinite Penrose tiling.  Although this {\it inflation}-based definition is helpful as an introduction, and adequately defines a {\it single} Penrose tiling, it is not really the right way to define the full {\it class} of Penrose tilings.  For one thing, it gives the incorrect impression that there is a {\it unique} Penrose tiling, when in fact there are {\it uncountably many} different Penrose tilings which are all locally indistinguishable but globally distinct \cite{gardner1977extraordinary,gardner1997penrose}. 
\begin{figure*}[ht]
    \centering
    \includegraphics[width=\linewidth]{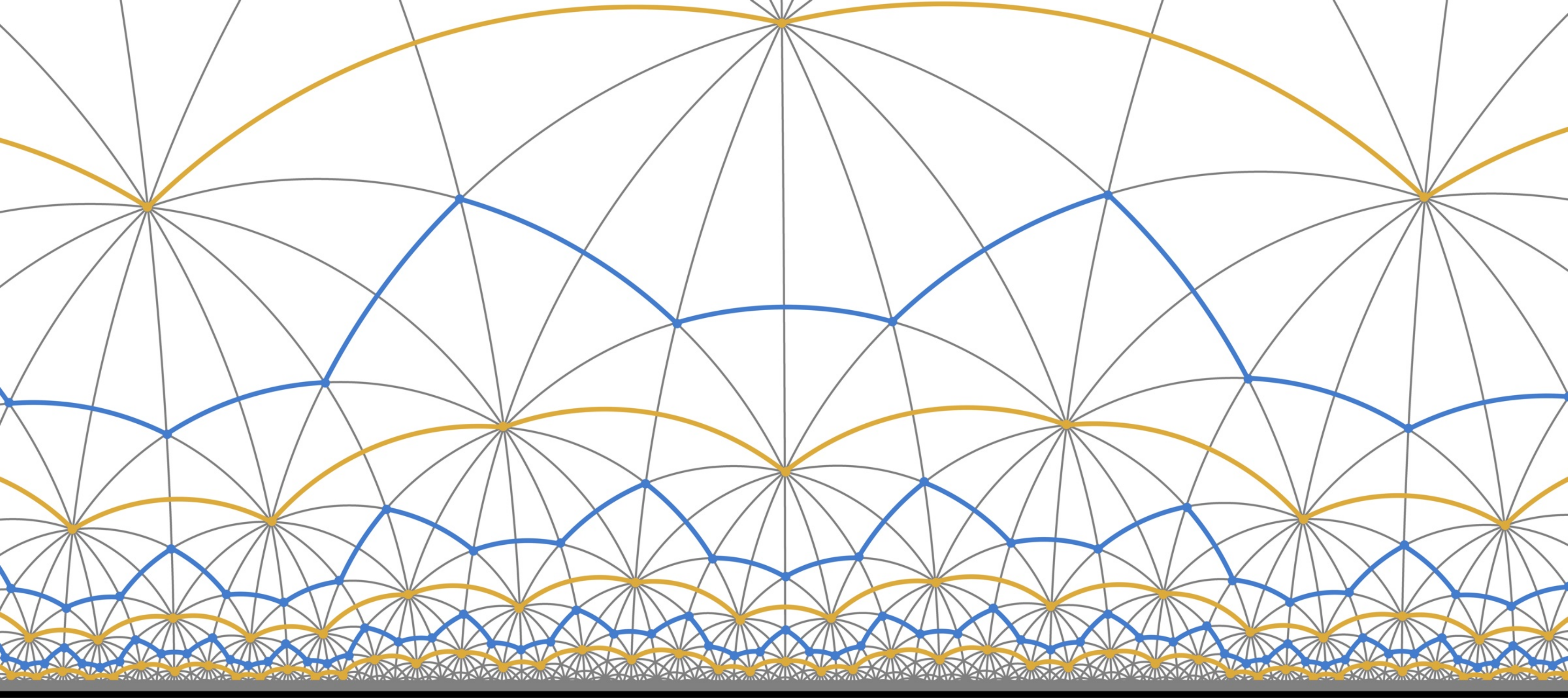}
    \caption{\RaggedRight{A holographic foliation of $\bbH^2$ by $\{7,3\}$/$\{3,7\}$ quasicrystals. \textcolor[RGB]{218,160,58}{\textbf{Orange layers}} mark the $\{7,3\}$ tiling and \textcolor[RGB]{56,114,201}{\textbf{blue layers}} mark the $\{3,7\}$ tiling. The same local substitution rules defined in Section \ref{sec:3773Example} describe the evolution from layer to layer. This holographic foliation in particular can be thought of as given by (approximately) concentric horospheres, i.e. slices of constant height.}}
    \label{fig:holographicFoliation}
\end{figure*}

Instead, as understood by de Bruijn \cite{de1981sequences, de1981algebraic}, the {\it class} of Penrose tilings is properly defined in the reverse fashion, i.e. using {\it deflation}. In inflation, we regard each tile as a ``supertile,'' to be split into special arrangements of several tiles and half-tiles. In the reverse process of deflation, we instead consider gluing arrangements of tiles and half-files to form supertiles -- see Figure \ref{fig:penroseTiling}. If we are given an infinite tiling of the plane by the Penrose kites and darts (but possibly not arranged in the right way to form a Penrose tiling), we ask: is it possible to \textit{unambiguously} group the tiles in this arrangement into supertiles, converting the original tiling uniquely into a new tiling by superkites and superdarts?  Can this process be iterated infinitely many times? If it {\it can} be repeated infinitely many times, we say that the tiling is a Penrose tiling.  Defined in this way, each Penrose tiling must be infinite in extent, and there are uncountably many such tilings (which are all locally-indistinguishable but globally-distinct).\footnote{This is the same infinite set of tilings obtained from the cut-and-project method or from the Penrose-tile matching rules.}

When it comes to defining our quasicrystals from hyperbolic tessellations, the situation is exactly parallel. Our previous inflation-based perspective started from a finite initial seed, and then constructed the corresponding quasicrystal by iterating the half-step inflation rule infinitely many times. The inflation-based perspective is easy to visualize and allows us to generate particular quasicrystals. However, now that we have gleaned the invertible inflation/deflation rule from this approach, the proper definition of the {\it class} of such quasicrystals is based on the deflation rule: i.e. \textit{we define a $\{p,q\}$/$\{q,p\}$ quasicrystal as any 1D surface in the shared $\{p,q\}$/$\{q,p\}$ tessellation to which we can successfully and unambiguously apply our half-step deflation rule infinitely many times in succession.}

This remarkable thing about this new definition is that it gives a dramatically richer picture of where the quasicrystals live and how they relate to the geometry of hyperbolic space. For starters, they no longer live at the boundary of hyperbolic space, but are simply infinite codimension-1 surfaces living within hyperbolic space. Indeed, the new definition corresponds to a picture in which a regular $\{p,q\}$ tiling of hyperbolic space naturally decomposes into an infinite sequence/stack of self-similar codimension-1 quasicrystalline slices, where each slice is uniquely related to the slice immediately above (or below) it by our invertible local inflation (or deflation) rule.  Our previous quasicrystal \eqref{eq:bdQuasicrystal} at the boundary of hyperbolic space corresponds to a special case of this construction -- the ``top" slice in the stack. See Figure \ref{fig:holographicFoliation}.

Notice that the whole tiling may be reconstructed from any single slice, using only local information. Each infinite stack of such slices is, on the one hand, a stack of all the $\{p,q\}$ and $\{q,p\}$ quasicrystals that are related to one another by inflation or deflation; and, on the other hand, a special type of discrete foliation of hyperbolic space (or of the $\{p,q\}$/$\{q,p\}$ tessellation) by such quasicrystals.  We call this special structure a {\it holographic foliation} because of its self-similar character, and the way that each leaf in the foliation encodes all the geometric information about the entire foliation.  

We also point out that, just as in the Penrose-tiling construction, we obtain the complete class of such $\{p,q\}$ quasicrystals from this deflation-based definition. This is an uncountable infinity of locally-indistinguishable but globally-distinct tilings. This corresponds to the infinitely many locally-indistinguishable, but globally-distinct, ways in which we can decompose the same regular $\{p,q\}$ tessellation into such a holographic foliation.

These holographic foliations seem potentially very rich. They seem to suggest that a weak notion of holography exists in hyperbolic space at the level of its geometry, i.e. before any fields or dynamics enter the picture. In particular, they suggest that the picture of holography as encoding information at the boundary of hyperbolic space may just be a special case of a more general holographic phenomenon, as embodied by the relationship between the entire $\{p,q\}$ honeycomb and any of its infinitely many different quasicrystaline slices. Note that the freedom to consider different bulk slicings is familiar from ordinary AdS/CFT, but in our construction, the infinite quasicrystal -- {\it i.e.} our analogue of the holographic screen -- need not live at the boundary, but can equally be embedded anywhere in the bulk.

In the subsequent section, we will see how all of these ideas extend to higher dimensional hyperbolic space.

\section{
2D Quasicrystals from \texorpdfstring{$\{3,5,3\}$}{{3,5,3}} Tiling of \texorpdfstring{$\bbH^3$}{H3}}\label{sec:H3Tiling}
To derive the logic behind the half-step inflation rule in 3D we will focus on a particular example: the regular $\{3,5,3\}$ honeycomb in $\bbH^3$. We will describe the general rule and remaining 3D honeycombs in Section \ref{sec:GeneralStory}. 

In the complete $\{3,5,3\}$ honeycomb, each cell is an icosahedron $\{3,5\}$, with 2 icosahedra meeting at each triangular face, 3 icosahedra surrounding each edge, and 12 icosahedra meeting at each vertex. This honeycomb is self-dual i.e. the dual honeycomb is also a $\{3,5,3\}$, which we will denote $\overline{\{3,5,3\}}$ to distinguish them. The $\overline{\{3,5,3\}}$ honeycomb has a vertex at the center of each $\{3,5,3\}$ icosahedron, and an icosahedral cell surrounding each vertex in the original $\{3,5,3\}$ honeycomb. Our half-step growth rules will grow:
\begin{equation}
    \{3,5,3\} \mapsto \overline{\{3,5,3\}} \mapsto \{3,5,3\} \mapsto \dots \,.    
\end{equation}
Before proceeding with the growth rules, let us get a better handle on what the icosahedral honeycomb looks like.\footnote{We also recommend the figures in \cite{nelson2017visualizing, RocchiniFigures} for additional visual aid.}

As mentioned above, in this honeycomb 12 icosahedra meet at each vertex. To see this, consider a vertex of a single icosahedron, there are 5 (hyperbolic) equilateral triangle faces which meet at that vertex:
\begin{equation}\label{eq:icosahead}
    \begin{tikzpicture}
    [   x={(\xx cm,\xy cm)},
        y={(\yx cm,\yy cm)},
        z={(\zx cm,\zy cm)},
        scale=\scaleOne,
        every node/.style={scale=\scaleOne},
        line join=round,
        every path/.style={ultra thick},
        rotate around x=0,
        rotate around y=0,
        rotate around z=0,
        baseline={(current bounding box.center)}
    ]  
        
        \coordinate (pi1) at (0,\phi,\a);
        \coordinate (pi2) at (0,\phi,-\a);
        \coordinate (pi3) at (0,-\phi,\a);
        \coordinate (pi4) at (0,-\phi,-\a);
        \coordinate (pi5) at (\a,0,\phi);
        \coordinate (pi6) at (\a,0,-\phi);
        \coordinate (pi7) at (-\a,0,\phi);
        \coordinate (pi8) at (-\a,0,-\phi);
        \coordinate (pi9) at (\phi,\a,0);
        \coordinate (pi10) at (\phi,-\a,0);
        \coordinate (pi11) at (-\phi,\a,0);
        \coordinate (pi12) at (-\phi,-\a,0);
        
        \fill[gray,fill opacity=0.15] (pi7) -- (pi1) -- (pi5) -- cycle;
        \fill[gray,fill opacity=0.15] (pi1) -- (pi9) -- (pi5) -- cycle;
        \fill[gray,fill opacity=0.15] (pi7) -- (pi11) -- (pi1) -- cycle;
        \fill[gray,fill opacity=0.15] (pi1) -- (pi11) -- (pi2) -- cycle;
        \fill[gray,fill opacity=0.15] (pi1) -- (pi9) -- (pi2) -- cycle;
        
        \draw[black, dashed] (pi2) -- (pi9);
        \draw[black, dashed] (pi2) -- (pi11);
        \draw[black, dashed] (pi2) -- (pi1);
        
        \draw[black] (pi1) -- (pi9);
        \draw[black] (pi1) -- (pi11);
        \draw[black] (pi1) -- (pi5);
        \draw[black] (pi1) -- (pi7);
        \draw[black] (pi11) -- (pi7);
        \draw[black] (pi7) -- (pi5);
        \draw[black] (pi5) -- (pi9);
        
    \end{tikzpicture}
\end{equation}
In the $\{3,5,3\}$ tessellation, there will be a ``ring'' of 5 additional icosahedra stacked on top, which share one of their 5 faces with the initial icosahedron, and all with the distinguished vertex in common. Then there will be a second ring of 5 more icosahedra stacked on top of those in the same way. Finally, there will be 1 antipodal icosahedron to the original which completes the figure around the vertex. Note how the ``head'' of each icosahedron, depicted in \eqref{eq:icosahead}, meeting at the shared vertex, contains its own 5 equilateral triangle faces all sharing the vertex in question. The union of the 12 icosahedron heads form a regular dodecahedron $\{5,3\}$, which is the ``vertex figure'' of the $\{3,5,3\}$ honeycomb.

Each 2D surface in the $\{3,5,3\}$ honeycomb is built from a collection of congruent (hyperbolic) 2D equilateral triangles (the faces of the icosahedra). In the cases of interest (growing patches), this will typically be a triangulation of a topological sphere or plane. Each edge in the full honeycomb is a junction where three faces meet at an angle of $2\pi/3$:
\begin{equation}
{
    \centering
    \tdplotsetmaincoords{40}{10}
    \begin{tikzpicture}[tdplot_main_coords, baseline={(current bounding box.center)}, scale = 0.7]
    \draw[very thick, black] (0,0,0) -- (0,0,3); 
    \def\height{3 * sqrt(3)/2}
    \foreach \angle in {0,120,240} {
        \draw[thick, fill=gray, opacity=0.25] 
            (0,0,0) 
                -- ({3*cos(\angle)},{3*sin(\angle)},1.5) 
                -- (0,0,3) 
                -- cycle;
    }
    \foreach \angle in {0,120,240} {
        \draw[thick, dashed]
            (0,0,1.5)
                -- ({3*cos(\angle)},{3*sin(\angle)},1.5);
    }

\end{tikzpicture}
}
\end{equation}
Each exposed edge in the 2D boundary surface can therefore either bow ``outward'' or ``inward'' relative to the surface, corresponding to a $1$-edge or $2$-edge respectively, see also Appendix \ref{sec:higherWRW}. 

As we will see below, the entire 2D surface can be decomposed into a countable set of pieces, each of which is congruent to one of a finite set of fundamental pieces -- our prototiles -- identified with neighbourhoods of different types of vertices in the boundary. Thus the geometry of the wrinkled surface is locally described by its relative configuration of boundary vertices. 

In the following Section \ref{sec:Growing353} we will discuss the growth of patches $S_i$ of $\{3,5,3\}$ to patches $\bar{S}_{i+1}$ of $\overline{\{3,5,3\}}$. By carefully studying the bulk 3D growth procedure around a vertex, analogous to \eqref{eq:localGrowth1} and \eqref{eq:localGrowth2}, we will find a set of 2D prototiles which describe the boundary tilings. In 2D/3D, the prototiles are once again identified with neighbourhoods of $n$ vertices in $\partial S_i$. Essentially by construction of the prototiles we will see that the surface $\partial \bar{S}_{i+1}$ is obtained by a local substitution rule acting on $\partial S_i$. In Section \ref{sec:353Combinatorics} we will pass from geometry to combinatorics and study the substitution matrix of these tiles, confirming that these adjacent surfaces are self-similar quasicrystalline patterns related by local inflation/deflation rules, analogous to the Penrose tiling. We will also confirm the relationship to the hyperbolic growth rates of surfaces described in \cite{nemeth2017growing} (see also \ref{sec:353Projections}). In Section \ref{sec:ThurstonAndPenrose} we will compare our tessellation to the Penrose tiling and discuss a conjecture of William Thurston that claims horospheres slicing the $\{3,5,3\}$ tessellation carry a Penrose tiling.


\subsection{Growing \texorpdfstring{$\{3,5,3\}$}{\{3,5,3\}} Honeycombs}\label{sec:Growing353}
The bulk half-step inflation rule in 3D works exactly analogously to the bulk 2D case: by taking a patch $S_i$ of the $\{3,5,3\}$ honeycomb and defining $\bar{S}_{i+1}$ as the union of all icosahedra in the dual $\overline{\{3,5,3\}}$ tiling which intersect the original $S_i$ tiling. To understand the boundary tiles and local substitution rule induced from this growth procedure, let us study this procedure in more detail.

Consider a patch $S_i$ of the $\{3,5,3\}$ honeycomb in $\bbH^3$. This patch has a closed 2D boundary surface $\partial S_i$: the interior of this closed surface consists of cells in the $\{3,5,3\}$ honeycomb which are in the patch $S_i$, and the exterior consists of all cells of the $\{3,5,3\}$ honeycomb which are not in $S_i$. During the bulk growth procedure $S_i \mapsto \bar{S}_{i+1}$, each vertex $v$ that is interior to the surface $\partial S_i$ or (more importantly) on $\partial S_i$ is wrapped by a dual $\overline{\{3,5\}}$ icosahedron, centered on the vertex $v$, in the $\overline{\{3,5,3\}}$ honeycomb. In particular, the boundary $\partial \bar{S}_{i+1}$ of $\bar{S}_{i+1}$ consists of pieces of icosahedra centered around vertices on the original boundary $\partial S_i$. This allows us to relate the surface geometry of $\partial \bar{S}_{i+1}$ to the vertices/geometry of $\partial S_i$.

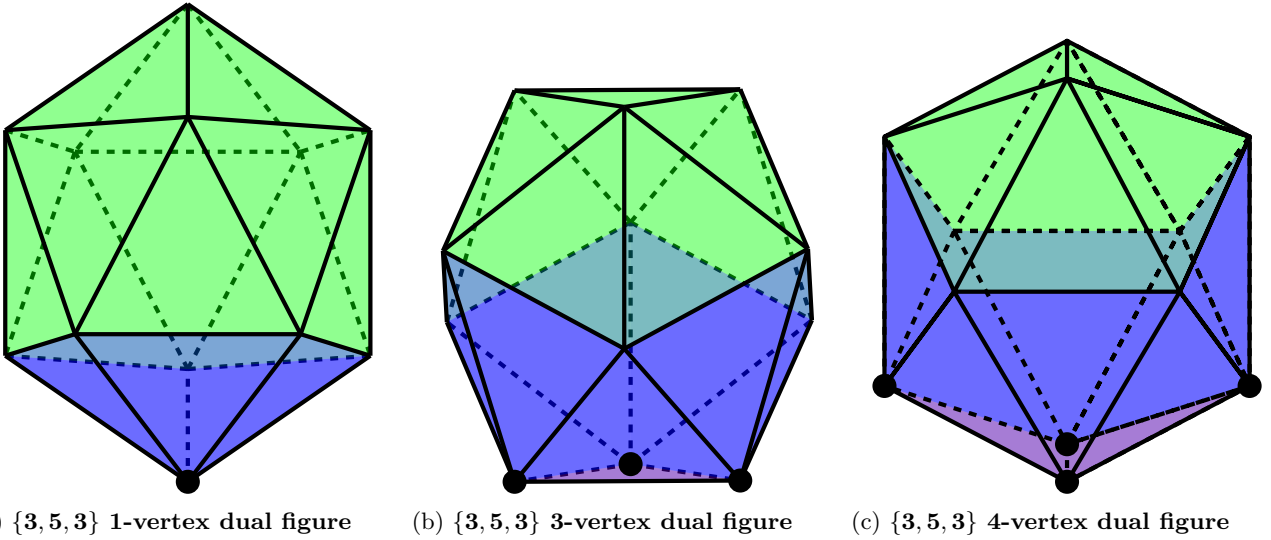
\begin{figure*}
    \begin{subfigure}[t]{0.32\textwidth}
    \centering
	\begin{tikzpicture}
    [   x={(\xx cm,\xy cm)},
        y={(\yx cm,\yy cm)},
        z={(\zx cm,\zy cm)},
        scale=\scaleOne,
        every node/.style={scale=\scaleOne},
        line join=round,
        every path/.style={ultra thick},
        rotate around x=0,
        rotate around y=0,
        rotate around z=0
    ]  
        
        \coordinate (pi1) at (0,\phi,\a);
        \coordinate (pi2) at (0,\phi,-\a);
        \coordinate (pi3) at (0,-\phi,\a);
        \coordinate (pi4) at (0,-\phi,-\a);
        \coordinate (pi5) at (\a,0,\phi);
        \coordinate (pi6) at (\a,0,-\phi);
        \coordinate (pi7) at (-\a,0,\phi);
        \coordinate (pi8) at (-\a,0,-\phi);
        \coordinate (pi9) at (\phi,\a,0);
        \coordinate (pi10) at (\phi,-\a,0);
        \coordinate (pi11) at (-\phi,\a,0);
        \coordinate (pi12) at (-\phi,-\a,0);
        
        \draw[black, dashed] (pi4) -- (pi3);
        \draw[black, dashed] (pi3) -- (pi5);
        \draw[black, dashed] (pi3) -- (pi7);
        \draw[black, dashed] (pi5) -- (pi10);
        \draw[black, dashed] (pi7) -- (pi12);
        \draw[black, dashed] (pi3) -- (pi10);
        \draw[black, dashed] (pi3) -- (pi12);
        \draw[black, dashed] (pi1) -- (pi5);
        \draw[black, dashed] (pi1) -- (pi7);
        \draw[black, dashed] (pi11) -- (pi7);
        \draw[black, dashed] (pi7) -- (pi5);
        \draw[black, dashed] (pi5) -- (pi9);
        
        \fill[blue,fill opacity=0.35] (pi4) -- (pi3) -- (pi10) -- cycle;
        \fill[blue,fill opacity=0.35] (pi4) -- (pi3) -- (pi12) -- cycle;
        
        \fill[green,fill opacity=0.25] (pi9) -- (pi10) -- (pi5) -- cycle;
        \fill[green,fill opacity=0.25] (pi10) -- (pi3) -- (pi5) -- cycle;
        \fill[green,fill opacity=0.25] (pi7) -- (pi3) -- (pi5) -- cycle;
        \fill[green,fill opacity=0.25] (pi3) -- (pi7) -- (pi12) -- cycle;
        \fill[green,fill opacity=0.25] (pi7) -- (pi11) -- (pi12) -- cycle;
        
        \fill[green,fill opacity=0.25] (pi7) -- (pi1) -- (pi5) -- cycle;
        \fill[green,fill opacity=0.25] (pi1) -- (pi9) -- (pi5) -- cycle;
        \fill[green,fill opacity=0.25] (pi7) -- (pi11) -- (pi1) -- cycle;
        \fill[blue,fill opacity=0.35] (pi12) -- (pi8) -- (pi4) -- cycle;
        \fill[blue,fill opacity=0.35] (pi8) -- (pi6) -- (pi4) -- cycle;
        \fill[blue,fill opacity=0.35] (pi10) -- (pi6) -- (pi4) -- cycle;
        \fill[green,fill opacity=0.25] (pi10) -- (pi9) -- (pi6) -- cycle;
        \fill[green,fill opacity=0.25] (pi8) -- (pi11) -- (pi12) -- cycle;
        
        \fill[green,fill opacity=0.25] (pi1) -- (pi11) -- (pi2) -- cycle;
        \fill[green,fill opacity=0.25] (pi1) -- (pi9) -- (pi2) -- cycle;
        \fill[green,fill opacity=0.25] (pi8) -- (pi11) -- (pi2) -- cycle;
        \fill[green,fill opacity=0.25] (pi8) -- (pi6) -- (pi2) -- cycle;
        \fill[green,fill opacity=0.25] (pi6) -- (pi9) -- (pi2) -- cycle;
        
        \draw[black] (pi4) -- (pi10);
        \draw[black] (pi4) -- (pi6);
        \draw[black] (pi4) -- (pi8);
        \draw[black] (pi4) -- (pi12);
        \draw[black] (pi10) -- (pi9);
        \draw[black] (pi9) -- (pi6);
        \draw[black] (pi6) -- (pi2);
        \draw[black] (pi2) -- (pi8);
        \draw[black] (pi8) -- (pi11);
        \draw[black] (pi11) -- (pi12);
        \draw[black] (pi2) -- (pi9);
        \draw[black] (pi2) -- (pi11);
        \draw[black] (pi2) -- (pi1);
        \draw[black] (pi1) -- (pi9);
        \draw[black] (pi1) -- (pi11);
        \draw[black] (pi12) -- (pi8);
        \draw[black] (pi8) -- (pi6);
        \draw[black] (pi6) -- (pi10);
        
        \node[BlackNode] at (pi4) {};
        
    \end{tikzpicture}
	\caption{\RaggedRight{\textbf{$\mathbf{\{3,5,3\}}$ 1-vertex dual figure}
    }}
    \label{fig:111}
	\end{subfigure}
    \begin{subfigure}[t]{0.32\textwidth}
        \centering
	\begin{tikzpicture}
    [   x={(\xx cm,\xy cm)},
        y={(\yx cm,\yy cm)},
        z={(\zx cm,\zy cm)},
        scale=\scaleOne,
        every node/.style={scale=\scaleOne},
        line join=round,
        every path/.style={ultra thick},
        rotate around x=-47,
        rotate around y=0,
        rotate around z=1
    ]  
        
        \coordinate (pi1) at (0,\phi,\a);
        \coordinate (pi2) at (0,\phi,-\a);
        \coordinate (pi3) at (0,-\phi,\a);
        \coordinate (pi4) at (0,-\phi,-\a);
        \coordinate (pi5) at (\a,0,\phi);
        \coordinate (pi6) at (\a,0,-\phi);
        \coordinate (pi7) at (-\a,0,\phi);
        \coordinate (pi8) at (-\a,0,-\phi);
        \coordinate (pi9) at (\phi,\a,0);
        \coordinate (pi10) at (\phi,-\a,0);
        \coordinate (pi11) at (-\phi,\a,0);
        \coordinate (pi12) at (-\phi,-\a,0);
        
        \draw[black, dashed] (pi4) -- (pi6);
        \draw[black, dashed] (pi4) -- (pi8);
        \draw[black, dashed] (pi4) -- (pi10);
        \draw[black, dashed] (pi4) -- (pi12);
        \draw[black, dashed] (pi4) -- (pi3);
        \draw[black, dashed] (pi3) -- (pi5);
        \draw[black, dashed] (pi3) -- (pi7);
        \draw[black, dashed] (pi3) -- (pi10);
        \draw[black, dashed] (pi3) -- (pi12);
        \draw[black, dashed] (pi10) -- (pi5);
        \draw[black, dashed] (pi12) -- (pi7);
        
        \fill[red,fill opacity=0.25] (pi6) -- (pi4) -- (pi8) -- cycle;
        \fill[blue,fill opacity=0.35] (pi6) -- (pi10) -- (pi4) -- cycle;
        \fill[blue,fill opacity=0.35] (pi8) -- (pi4) -- (pi12) -- cycle;
        \fill[blue,fill opacity=0.35] (pi10) -- (pi4) -- (pi3) -- cycle;
        \fill[blue,fill opacity=0.35] (pi12) -- (pi4) -- (pi3) -- cycle;
        
        \fill[green,fill opacity=0.25] (pi3) -- (pi5) -- (pi7) -- cycle;
        \fill[green,fill opacity=0.25] (pi3) -- (pi5) -- (pi10) -- cycle;
        \fill[green,fill opacity=0.25] (pi5) -- (pi9) -- (pi10) -- cycle;
        \fill[green,fill opacity=0.25] (pi3) -- (pi7) -- (pi12) -- cycle;
        \fill[green,fill opacity=0.25] (pi7) -- (pi11) -- (pi12) -- cycle;
        \fill[blue,fill opacity=0.35] (pi8) -- (pi6) -- (pi2) -- cycle;
        \fill[blue,fill opacity=0.35] (pi8) -- (pi2) -- (pi11) -- cycle;
        \fill[blue,fill opacity=0.35] (pi8) -- (pi11) -- (pi12) -- cycle;
        \fill[blue,fill opacity=0.35] (pi6) -- (pi2) -- (pi9) -- cycle;
        \fill[blue,fill opacity=0.35] (pi6) -- (pi9) -- (pi10) -- cycle;
        
        \fill[green,fill opacity=0.25] (pi1) -- (pi11) -- (pi2) -- cycle;
        \fill[green,fill opacity=0.25] (pi1) -- (pi9) -- (pi2) -- cycle;
        \fill[green,fill opacity=0.25] (pi1) -- (pi7) -- (pi5) -- cycle;
        \fill[green,fill opacity=0.25] (pi1) -- (pi7) -- (pi11) -- cycle;
        \fill[green,fill opacity=0.25] (pi1) -- (pi5) -- (pi9) -- cycle;
        
        \node[BlackNode] at (pi4) {};
        
        \draw[black] (pi1) -- (pi7);
        \draw[black] (pi1) -- (pi5);
        \draw[black] (pi1) -- (pi9);
        \draw[black] (pi1) -- (pi11);
        \draw[black] (pi1) -- (pi2);
        \draw[black] (pi8) -- (pi6);
        \draw[black] (pi6) -- (pi10);
        \draw[black] (pi10) -- (pi9);
        \draw[black] (pi9) -- (pi5);
        \draw[black] (pi5) -- (pi7);
        \draw[black] (pi7) -- (pi11);
        \draw[black] (pi11) -- (pi12);
        \draw[black] (pi12) -- (pi8); 
        \draw[black] (pi2) -- (pi9);
        \draw[black] (pi2) -- (pi11);
        \draw[black] (pi2) -- (pi8);
        \draw[black] (pi2) -- (pi6);
        \draw[black] (pi9) -- (pi6);
        \draw[black] (pi11) -- (pi8);
        
        
        \node[BlackNode] at (pi6) {};
        \node[BlackNode] at (pi8) {};
        
    \end{tikzpicture}
    \caption{\RaggedRight{\textbf{$\mathbf{\{3,5,3\}}$ 3-vertex dual figure}
    }}
    \label{fig:333}
	\end{subfigure}
	\begin{subfigure}[t]{0.32\textwidth}
        \centering
	\begin{tikzpicture}
    [   x={(\xx cm,\xy cm)},
        y={(\yx cm,\yy cm)},
        z={(\zx cm,\zy cm)},
        scale=\scaleOne,
        every node/.style={scale=\scaleOne},
        line join=round,
        every path/.style={ultra thick},
        rotate around x=18,
        rotate around y=0,
        rotate around z=0
    ]  
        
        \coordinate (pi1) at (0,\phi,\a);
        \coordinate (pi2) at (0,\phi,-\a);
        \coordinate (pi3) at (0,-\phi,\a);
        \coordinate (pi4) at (0,-\phi,-\a);
        \coordinate (pi5) at (\a,0,\phi);
        \coordinate (pi6) at (\a,0,-\phi);
        \coordinate (pi7) at (-\a,0,\phi);
        \coordinate (pi8) at (-\a,0,-\phi);
        \coordinate (pi9) at (\phi,\a,0);
        \coordinate (pi10) at (\phi,-\a,0);
        \coordinate (pi11) at (-\phi,\a,0);
        \coordinate (pi12) at (-\phi,-\a,0);
        
        \fill[red,fill opacity=0.25] (pi12) -- (pi3) -- (pi4) -- cycle;
        \fill[red,fill opacity=0.25] (pi10) -- (pi3) -- (pi4) -- cycle;
        
        \fill[blue,fill opacity=0.35] (pi12) -- (pi3) -- (pi7) -- cycle;
        \fill[blue,fill opacity=0.35] (pi10) -- (pi3) -- (pi5) -- cycle;
        \fill[blue,fill opacity=0.35] (pi7) -- (pi3) -- (pi5) -- cycle;
        \fill[blue,fill opacity=0.35] (pi10) -- (pi9) -- (pi5) -- cycle;
        \fill[blue,fill opacity=0.35] (pi7) -- (pi11) -- (pi12) -- cycle;
        
        \fill[green,fill opacity=0.25] (pi7) -- (pi1) -- (pi5) -- cycle;
        \fill[green,fill opacity=0.25] (pi1) -- (pi9) -- (pi5) -- cycle;
        \fill[green,fill opacity=0.25] (pi7) -- (pi11) -- (pi1) -- cycle;
        \fill[blue,fill opacity=0.35] (pi12) -- (pi8) -- (pi4) -- cycle;
        \fill[blue,fill opacity=0.35] (pi8) -- (pi6) -- (pi4) -- cycle;
        \fill[blue,fill opacity=0.35] (pi10) -- (pi6) -- (pi4) -- cycle;
        \fill[blue,fill opacity=0.35] (pi10) -- (pi9) -- (pi6) -- cycle;
        \fill[blue,fill opacity=0.35] (pi8) -- (pi11) -- (pi12) -- cycle;
        
        \fill[green,fill opacity=0.25] (pi1) -- (pi11) -- (pi2) -- cycle;
        \fill[green,fill opacity=0.25] (pi1) -- (pi9) -- (pi2) -- cycle;
        \fill[green,fill opacity=0.25] (pi8) -- (pi11) -- (pi2) -- cycle;
        \fill[green,fill opacity=0.25] (pi8) -- (pi6) -- (pi2) -- cycle;
        \fill[green,fill opacity=0.25] (pi6) -- (pi9) -- (pi2) -- cycle;
        
        \draw[ultra thick]    
        (pi2) -- (pi8) -- (pi6) 
        (pi4) -- (pi12) -- (pi8) --cycle 
        (pi11) -- (pi12) -- (pi8) --cycle
        (pi2) -- (pi11) 
        (pi1) -- (pi11)
        (pi1) -- (pi9) -- (pi2) --cycle 
        (pi6) -- (pi9) -- (pi2) --cycle 
        (pi6) -- (pi9) -- (pi10) --cycle
        (pi6) -- (pi4) -- (pi10) --cycle
        ;
        
        \draw[ultra thick, dashed]
        (pi11) -- (pi12) -- (pi7) --cycle
        (pi3) -- (pi5) -- (pi10) --cycle
        (pi9) -- (pi5) -- (pi10) --cycle
        (pi9) -- (pi5) -- (pi1) --cycle
        (pi7) -- (pi5) -- (pi1) --cycle
        (pi7) -- (pi5) -- (pi3) --cycle
        (pi3) -- (pi4) -- (pi10) --cycle
        (pi3) -- (pi12)
        ;
        
        \node[BlackNode] at (pi3) {};
        \node[BlackNode] at (pi10) {};
        \node[BlackNode] at (pi4) {};
        \node[BlackNode] at (pi12) {};
        
    \end{tikzpicture}
    \caption{\RaggedRight{\textbf{$\mathbf{\{3,5,3\}}$ 4-vertex dual figure}
    }}
    \label{fig:444}
	\end{subfigure}
    \caption{\RaggedRight{The dual vertex diagrams in the shared $\{3,5,3\}$ and $\overline{\{3,5,3\}}$ tessellation, from which we derive the corresponding 1, 3 and 4-vertex tiles in Figure \ref{fig:353VertexTiles}. \textcolor{black}{\textbf{Black nodes}} are vertices that are centers of cells in the previous layer. \textcolor{capRed}{\textbf{Red faces}} are dual to completely submerged edges in the previous layer and are themselves completely submerged in the previous layer. \textcolor{capBlue}{\textbf{Blue faces}} are dual to boundary edges in the previous layer and are hidden in the new layer. \textcolor{capGreen}{\textbf{Green faces}} mark every new exposed boundary face.}}
	\label{fig:vvvv}
\end{figure*}

To derive the local rules relating the shape of the old $\{3,5,3\}$ boundary $\partial S_i$ to the new $\overline{\{3,5,3\}}$ boundary $\partial\bar{S}_{i+1}$ lying a half-step above it, let's track the local growth around a particular vertex $v \in \partial S_i$ in the original $\{3,5,3\}$ boundary. Here we imagine that the surface is oriented so that the local neighbourhood of $v$ exterior to the surface is ``upwards.'' We analogize the growth procedure to a rising water level: the interior pieces of $S_i$ are ``submerged'' under the water surface $\partial S_i$. The following procedure is analogous to the 1D/2D procedure described around \eqref{eq:localGrowth1} and \eqref{eq:localGrowth2}:
\begin{enumerate}
    \item At the vertex $v \in \partial S_i$, a particular configuration of $n$ many submerged $\{3,5\}$ icosahedra in $S_i$ meet at $v$.
    \item When we inflate by a half-step, we surround the vertex $v$ by a dual $\overline{\{3,5\}}$ icosahedron. The dual $\overline{\{3,5\}}$ icosahedron has some submerged vertices interior to $S_i$ and some vertices exterior to $S_i$. The interior vertices of the $\overline{\{3,5\}}$ are precisely those $n$ at the center of the aforementioned $\{3,5\}$ icosahedra in $S_i$ meeting at $v$. We colour those submerged vertices \textcolor{black}{\textbf{black}} in Figure \ref{fig:vvvv}.
    \item Now we colour the rest of the $\overline{\{3,5\}}$ surrounding the vertex $v$ as follows:
    \begin{enumerate}
        \item Any face of the $\overline{\{3,5\}}$ whose vertices are {\it all} black is coloured \textcolor{capRed}{\textbf{red}}.\footnote{In our figures, red faces often appear purple because they are behind blue faces.} Each red face is dual to a completely submerged edge in $S_i$ that is hanging down from the vertex $v$, and completely surrounded by a full complement of three ``underwater'' $\{3,5\}$ icosahedra. From Figure \ref{fig:vvvv}, we see that at a $\{3,5,3\}$ boundary, the $1$, $3$ and $4$ vertices have $0$, $1$ and $2$ completely submerged edges respectively.
        \item Any face of the $\overline{\{3,5\}}$ with {\it some} (but not all) of its vertices coloured black is coloured \textcolor{capBlue}{\textbf{blue}}. Each blue face is dual to an edge that is attached to $v$, but is {\it not} completely surrounded by $\{3,5\}$'s, and hence is {\it exposed} and lying on the original boundary $\partial S_i$.
        \item All remaining faces are coloured \textcolor{capGreen}{\textbf{green}}. These are the exposed faces in the new $\overline{\{3,5,3\}}$ surface $\partial \bar{S}_{i+1}$.
    \end{enumerate}
\end{enumerate}

\begin{figure*}
    \begin{subfigure}[t]{0.32\textwidth}
        \includegraphics[width=.95\linewidth]{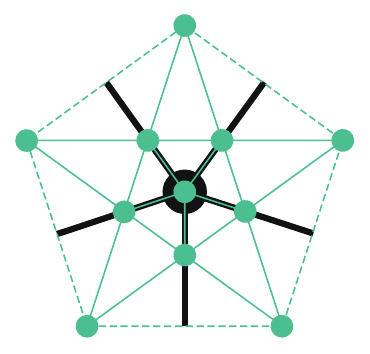}
        \caption{\RaggedRight{\textbf{$\mathbf{\{3,5,3\}}$ 1-vertex tile}
    }}
    \end{subfigure}
    \begin{subfigure}[t]{0.32\textwidth}
        \includegraphics[width=.95\linewidth]{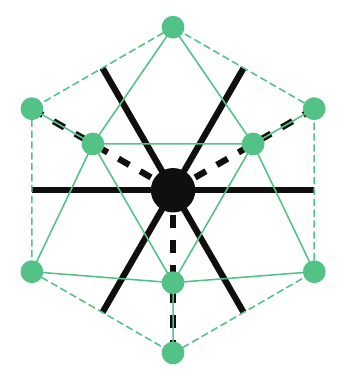}
        \caption{\RaggedRight{\textbf{$\mathbf{\{3,5,3\}}$ 3-vertex tile}
    }}
    \end{subfigure}
	\begin{subfigure}[t]{0.32\textwidth}
        \includegraphics[width=.95\linewidth]{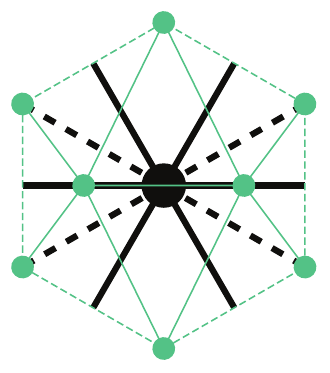}
        \caption{\RaggedRight{\textbf{$\mathbf{\{3,5,3\}}$ 4-vertex tile}
    }}
    \end{subfigure}
    \caption{\RaggedRight{The vertex tiles showing the half-step inflation from a $\{3,5,3\}$ patch to its dual $\overline{\{3,5,3\}}$ patch, derived from the dual vertex diagrams in Figure \ref{fig:vvvv}. The initial $\{3,5,3\}$ boundary is shown in \textcolor{black}{\textbf{black}} and the subsequent $\overline{\{3,5,3\}}$ boundary is shown in \textcolor[RGB]{45,183,112}{\textbf{green}}. $1$-edges are denoted by solid lines and $2$-edges are denoted by dashed lines.}}
	\label{fig:353VertexTiles}
\end{figure*}

Now we can identify the boundary tiles from this growth procedure, the results are shown in Figure \ref{fig:353VertexTiles}. In the tiles, each edge of $\partial S_i$ is dual to a blue face in Figure \ref{fig:vvvv}; we depict the edge by a black line
in Figure \ref{fig:353VertexTiles}. More specifically, if a blue face in Figure \ref{fig:vvvv} has:
\begin{itemize}
    \item $1$ black vertex, it is dual to an exposed $1$-edge in $\partial S_i$ (bowing outwards from the surface) and is shown as a solid black line  in Figure \ref{fig:353VertexTiles}.
    \item $2$ black vertices, it is dual to an exposed $2$-edge (bowing interior to the surface) and is shown as a dashed black line in Figure \ref{fig:353VertexTiles}.
\end{itemize}
To obtain the substitution rules, we simply look at the icosahedral figures in Figure \ref{fig:vvvv} grown above the different species of boundary vertex $v$, then ``project them down'' onto the vertex neighbourhood underneath:
\begin{itemize}
    \item Edges of $\overline{\{3,5\}}$ completely in the green regions in Figure \ref{fig:vvvv} correspond to exposed $1$-edges in $\partial \bar{S}_{i+1}$, and are shown as 
    solid green lines in Figure \ref{fig:353VertexTiles}.
    \item Vertices of $\overline{\{3,5\}}$ completely in the green regions are $1$-vertices in $\partial \bar{S}_{i+1}$.
    \item Edges of $\overline{\{3,5\}}$ at the boundary between blue and green regions in Figure \ref{fig:vvvv} correspond to exposed $2$-edges, and are shown as dashed green lines in Figure \ref{fig:353VertexTiles}.
    \item Vertices of $\overline{\{3,5\}}$ on the boundary between blue and green regions can be $3$ or $4$ vertices in $\partial \bar{S}_{i+1}$ (more details on that in the next section).
\end{itemize}

Thus we obtain the three vertex tiles shown in Figure \ref{fig:353VertexTiles} with their black and green decorations. Each vertex tile summarizes the shape of the original $\{3,5,3\}$ boundary in the vicinity of $v$ in black, and the shape of the $\overline{\{3,5,3\}}$ surface lying one half-step ``higher'' in green. As with the 1D/2D case, in the $\{3,5,3\}$ example we find that even if one starts with a surface with many different vertex configurations, it appears that, after sufficiently many iterations of the half-step inflation rule, only $1$, $3$ and $4$-vertices remain on the boundary. We do not prove this explicitly, as we did in the 1D/2D case, but it can be proven (analogous to the 1D/2D case) by following the general methods described in Section \ref{sec:GeneralStory} by studying every possible dual figure configuration and writing the general substitution matrix. In Appendix \ref{sec:353Projections} we depict the first few inflation layers starting from a single icosahedron in the $\{3,5,3\}$ tiling; this also supports the previous assertion that only $1$, $3$, and $4$-vertices appear.

\subsection{Combinatorial Checks}\label{sec:353Combinatorics}
Now that we have our prescription for boundary tiles, labelled (generically) by neighbourhoods of $1$, $3$, and $4$-vertices in the boundary $\partial S_i$, as well as their substitution rules, we can now track the numbers $V_1$, $V_3$, and $V_4$ of tiles in a layer of $\{3,5,3\}$ boundary and relate it to the numbers $\overline{V}_1$, $\overline{V}_3$, and $\overline{V}_4$ in the subsequent $\overline{\{3,5,3\}}$ boundary (and vice-versa by self-duality). To do this, it's helpful to view the 3 and 4-vertices as being composed of smaller ``wedges'' sitting between (or bounded by) adjacent dashed lines in the tiles, i.e. we define:
\begin{eqnarray}
    \begin{array}{c}
            A\;{\rm Wedge}\,\sim \\
            \\\\\\\\\\
            B\;{\rm Wedge}\,\sim
        \end{array}  
        &
        \begin{array}{c}
            \includegraphics[width=.45\linewidth] {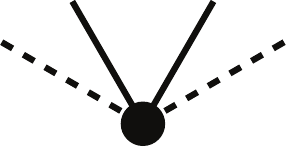} \\
            \\
            \includegraphics[width=.35\linewidth]{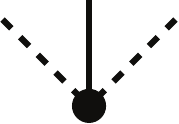} \\
        \end{array}
        &
\end{eqnarray}
So that the 3-vertex is made by gluing three ``A'' wedges, and the 4-vertex by gluing two ``A'' and two ``B'' wedges. This also makes clear an ambiguity from the previous section: the boundaries between blue and green regions in Figure \ref{fig:vvvv} supported both $3$ and $4$ vertices. The wedges show us that the green vertices on the tiles in Figure \ref{fig:353VertexTiles} that touch a black dashed line are necessarily the $4$-vertices. Using $A$ and $B$ to also denote the number of $A$ and $B$ wedges, we can turn the geometric identifications into combinatorial formulas. That is, we have:
\begin{equation}
  \begin{array}{rcl}
    V_3&=&3A\,, \\
    V_{4}&=&2A+2B\,, 
  \end{array}
  \quad\Rightarrow\quad
  \begin{array}{rcl}
    A&=&+\tfrac{1}{3}V_3\,, \\
    B&=&-\tfrac{1}{3}V_3+\tfrac{1}{2}V_{4}\,.
  \end{array}
\end{equation}

Now we can easily read off the substitution rules from Figure \ref{fig:353VertexTiles}:
\begin{align}\label{eq:134Subs}
\begin{split}
V_{1}
    &\mapsto 6\overline{V}_1+5\overline{A} \\
    &= 6\overline{V}_1+\frac{5}{3}\overline{V}_3\,, \\
  V_{3}
    &\mapsto 3\overline{V}_{1}+3\overline{A}+3\overline{B} \\
    &= 3\overline{V}_{1}+\frac{3}{2}\overline{V}_{4}\,, \\
  V_{4}
    &\mapsto 2\overline{V}_{1}+2\overline{A}+4\overline{B}\\
    &= 2\overline{V}_1-\frac{2}{3}\overline{V}_3+2\overline{V}_4 \,,
  \end{split}
\end{align}
or, in terms of a substitution matrix $M_{353}$ 
\begin{equation}\label{eq:353SubMat}
  \begin{pmatrix} \overline{V}_1 \\ \overline{V}_3 \\ \overline{V}_4 \end{pmatrix}
        = \begin{pmatrix}
          6  & 3 & 2 \\ 
          \frac{5}{3} & 0 & -\frac{2}{3} \\
          0 & \frac{3}{2} & 2
      \end{pmatrix}
        \begin{pmatrix} V_1 \\ V_3 \\ V_4 \end{pmatrix}\,.
\end{equation}
This matrix has the eigenvalues and (right) eigenvectors\footnote{Note: since the tiling is self-dual, the half-step eigenvalues/eigenvectors make sense without having to do further re-definitions of tiles as in \eqref{eq:nonCOB}. In other words, $\Lambda_{pqr} = \mathds{1} = \Lambda_{rqp}$ automatically.}
\begin{alignat}{3}
  \lambda_{\pm}
    &=\varphi_{\pm}^4\,,\qquad
  &v_{\pm}^T
    & =\left(\varphi_{\pm}^4,\varphi_{\pm},\tfrac{1}{2}\right)\,,\\
  \lambda_{0}
    &=1\,,\qquad 
  &v_{0}^T
    &=\tfrac{1}{6}\left(0,-2,3\right)\,.
\end{alignat}

As in the previous sections, the dominant $\lambda_{+}$ eigenvalue, when squared, agrees with the ``full-step'' growth factor found by N\'emeth for the $\{3,5,3\}$ honeycomb in \cite{nemeth2017growing}. The components of the corresponding dominant (right) eigenvector $v_{+}$ give the relative frequencies of the three types of vertices, $V_1$, $V_3$ and $V_4$, in the corresponding quasicrystal. As with all substitution matrices, the left eigenvector connected to the dominant eigenvalue $\lambda_+$ gives the relative length/areas/volumes of prototiles.

As a simple check, if we start with 1 single icosahedron, then it has $V_1 = 12$ vertices. Under successive inflations
\begin{equation}
    \begin{pmatrix}
        12\\ 0 \\ 0
    \end{pmatrix}
    \mapsto
    \begin{pmatrix}
        72\\ 20 \\ 0
    \end{pmatrix}
    \mapsto
    \begin{pmatrix}
        492\\ 120 \\ 30
    \end{pmatrix}
    \mapsto \dots\,.
\end{equation}
This agrees with the total number of boundary vertices, $12$, $92$, $642$, $\dots$, described by the $\mathbf{M}$ matrix in \cite{nemeth2017growing}. Note how $4$-vertices do not appear until the third inflation in this case. We draw (projections of) these inflations in Appendix \ref{sec:353Projections} and, by inspection, one can confirm that only $1$ and $3$ vertices are present until the third layer is added.

Until now, we have considered a matrix with a dominant eigenvalue which is a quadratic Pisot number, and one eigenvalue is $1$. By the logic of Section \ref{sec:SubTilingsAndQuasi}, this defines a quasicrystal either by inflating our seed infinitely many times, or by using the deflation logic of Section \ref{sec:globalaspects}. Numerically, the $1$ eigenvalue suggests (by general substitution matrix lore) that there should be a way to reduce our prototile set to just a set of two. We will return to this question after first discussing some generalities about the curvature of surfaces and tiles.

\subsubsection{Topological Constraints and Tile Curvature}\label{sec:topConstraints}
Let us start by investigating the topological properties of a surface, as captured by the Euler characteristic. The total number of vertices in a boundary $\partial S_i$ is just $V=V_1+V_3+V_4$. To obtain the total number of edges, note that any edge in the tiling stretches between two vertices, and we know how many edges each of the $1$, $3$, and $4$ vertices are connected to, hence $2E=5V_1+9V_3+10V_4$. Likewise, we can count the number of faces on the surface as $3 F = 2 E$. Putting this together, we find that
\begin{align} \label{eq:Euler_constraint}
    \chi 
        &:= V - E + F\\
        &= \frac{1}{6}(V_1 - 3 V_3 - 4V_4)\,.\label{eq:Euler_constraint2}
\end{align}

Applying the substitution rules \eqref{eq:134Subs} to the Euler formula \eqref{eq:Euler_constraint2}, we see that: \textit{if a surface has only $1$, $3$, and $4$ vertices, then its Euler characteristic is preserved under inflation.} As a result, if we have a surface with only this set of vertices and the topology of a sphere (or plane), then we expect it to continue to be a topological sphere (or plane) under successive inflations, e.g. a sphere cannot degenerate into a topological surface of genus $g\neq 0$. 

There is a neat way to repackage the above fact. The Euler characteristic of any 2D surface $\Sigma$ is just the integral of the Gaussian curvature. Famously, the formula works even when the surface has boundary and is only piecewise smooth:
\begin{equation}\label{eq:GBGen}
    2\pi \chi(\Sigma) = \int_\Sigma K \, dA + \sum_{i} \int_{C_i} k_g \, ds + \sum_i \theta_{i}\,.
\end{equation}
Here the computation involves an integral of Gaussian curvature over surfaces, an integral of geodesic curvature along the boundary curves, and a discrete contribution from the ``sharp'' external angles (see e.g. \cite{do2016differential}). Perhaps the most important fact about the Euler characteristic is that it is a topological invariant; this means that under homeomorphisms, the local curvature contributions in \eqref{eq:GBGen} often ``shift around'' between terms.

Given a discrete surface/polytope in Euclidean space, made from straight Euclidean lines and flat Euclidean faces, the integral curvature contributions along faces and edges vanish and all of the curvature becomes localized at vertices. Indeed, it is a straightforward exercise to show that the Gauss-Bonnet formula becomes:
\begin{equation} \label{eq:DiscreteGB}
    2\pi \chi(\Sigma) = \sum_{v \in \Sigma} K_v + \sum_{v \in \partial \Sigma} \theta_v\,.
\end{equation}
Here $K_v = 2\pi - \sum_{i \in \mathrm{Ang}(v)} \theta_i$ is the ``curvature at $v$'', and includes a sum over all interior angles $\theta_i$ of faces meeting at $v$.

Let us specify to a discrete surface made only of $p$-gons, assumed without boundary for simplicity. Then we can use \eqref{eq:Euler_constraint} to give another purely combinatorial version of the Euler characteristic. Specifically, we have $p F = 2E$, and if we let $n(v) = \{\text{\# Edges (or Faces) at $v$}\}$, then
\begin{equation}\label{eq:FlatTriangleGB}
    \chi(\Sigma) = \frac{1}{2p}\sum_{v\in \Sigma} \left(2p-(p-2)n(v)\right)\,.
\end{equation}
It is combinatorial in the sense that it is independent of whether or not the discrete surface is defined as a polytope in flat, spherical, or hyperbolic space.

If we compare \eqref{eq:DiscreteGB} to \eqref{eq:FlatTriangleGB}, we see that $\sum_v \sum_{i\in\mathrm{Ang}(v)} \theta_i = \frac{(p-2)\pi}{p}\sum_v n(v)$, but it is crucial that the two sides be summed over all $v$. That is, the quantities are only equal upon averaging over all $v$, otherwise the ``combinatorial curvature'' at a vertex and geometric curvature $K_v$ are very obviously different. However, one special case when the two summands are actually equal is when all the $p$-gons are regular and have the same internal angle. In this case, the curvature stored at a vertex is exactly given by its combinatorial curvature:
\begin{equation}
    K_v^{\text{reg,flat}} = 2\pi - \tfrac{(p-2)\pi}{p} n(v)\,.
\end{equation}

How does this change in curved space? Fix a symmetric space of constant curvature $\lambda = +1,0,-1$. By discrete surface/polytope, we will mean a surface made analogously to the Euclidean case, e.g. with edges given by geodesics, etc. For example, any patch of hyperbolic honeycomb, or any 3D regular polytope ``projected outwards'' onto the surface of the sphere (i.e. a spherical honeycomb). Given such a surface $\Sigma$, and assuming again no boundary, the formula \eqref{eq:DiscreteGB} becomes \cite{thurstonNotes, chow2003combinatorial, ge2017discrete}:
\begin{equation} \label{eq:DiscreteGB2}
    2\pi \chi(\Sigma) = \sum_{v \in \Sigma} K_v + \lambda \mathrm{Area}(\Sigma)\,.
\end{equation}
In curved space, the curvature at a vertex is still the same: $K_v = 2\pi - \sum_{i \in \mathrm{Ang}(v)} \theta_i$, intuitively this follows since on infinitesimal scales around the point there is no ability to distinguish the background space. Clearly \eqref{eq:DiscreteGB2} is the flat space formula \eqref{eq:DiscreteGB} with curvature contributions coming from the faces of $\Sigma$. On the other hand, \eqref{eq:FlatTriangleGB} is still equally as valid in curved space since it was purely combinatorial. We recall that the area of a general $p$-gon in hyperbolic space satisfies
\begin{equation}
    -\lambda A = (p-2)\pi - \sum_i \alpha_i\,,
\end{equation}
where $\alpha_i$ are the internal angles of the $p$-gon. 

As before, one interesting special case is when the $p$-gons in the discretized surface $\Sigma$ are regular, then we can link the vertex curvature to the ``combinatorial curvature.'' In curved space, the internal angles of $p$-gons can be less (hyperbolic) or greater (spherical) than flat space, but there is no constraint on precisely what the angle is; so we will denote it $\alpha$. The curvature at any vertex is:
\begin{equation}
    K_v^{\text{reg,gen}} = 2\pi - \alpha n(v)\,.
\end{equation}
Clearly, the curvature ``stored'' at any vertex in curved space is less/more than the curvature stored at the vertex in flat space. This is completely consistent with the fact that the Euler characteristic is preserved under homeomorphism and compensates for the curvature contributions from the faces/area term in \eqref{eq:DiscreteGB}.

Now we can ask about the curvature of a prototile. We identify our prototiles with neighbourhoods of vertices $v$ in the tiling. If the tiling is by regular $p$-gons with internal angle $\alpha$, then the curvature contribution from the single vertex in the prototile is $K_v^{\text{reg,gen}}$. However, the faces making the tile will also be curved. There are $n(v)$ faces that meet at a vertex, but each face is a part of $p$ equal sized neighbourhoods. Hence the total curvature carried by a prototile built around an $n$-vertex $v$ is:
\begin{align}
    K^{\text{tile}}_{\tile{n}} 
        &= K_{v}^{\text{reg,gen}} + \frac{1}{p} n(v) \lambda A^{\mathrm{reg.}}\\
        &= 2\pi - \tfrac{(p-2)\pi}{p} n(v)\,.
\end{align}
In other words, the total curvature carried by a prototile around $v$ in curved space is the same as the as the vertex curvature in flat space.\footnote{It is interesting to compare this formula with the one usually used in Regge calculus \cite{Regge:1961px}, where the faces are taken to be flat, and the curvature is entirely supported at the vertices.}  For $1$, $3$, and $4$ tiles in the $\{3,5,3\}$ tiling we have
\begin{equation}\label{eq:353Curvature}
    K^{\text{tile}}_{\tile{1}} = \frac{\pi}{3}\,,\quad
    K^{\text{tile}}_{\tile{3}} = -3\frac{\pi}{3}\,,\quad
    K^{\text{tile}}_{\tile{4}} = -4\frac{\pi}{3}\,.
\end{equation}
We recognize the numbers in \eqref{eq:353Curvature} as the (up to scaling) coefficients in \eqref{eq:Euler_constraint2}.

We can give additional insight to this non-coincidence in terms of our substitution matrix. We can map our 2D surface $\Sigma = \partial S_i$ to a ``vector''
$\ket{\Sigma}$. Our choice of prototiles as neighbourhoods of $1$, $3$, and $4$ vertices in $\Sigma$ gives a basis where $\ket{\Sigma} = (V_1, V_3, V_4)^T$. We can view \eqref{eq:Euler_constraint2} as an inner product
\begin{equation}
    \chi(\Sigma) = \bra{K}\ket{\Sigma}\,,
\end{equation}
where $\bra{K} = \frac{1}{6}\begin{pmatrix}1 & -3 & -4 \end{pmatrix}$ is the dual curvature vector. As it turns out, $\bra{K}$ is a left eigenvector of $M_{353}$ corresponding to the eigenvalue $1$, i.e.
\begin{equation}
    \mel*{K}{M_{353}}{\Sigma} = \braket{K}{\Sigma}\,.
\end{equation}
The fact that the eigenvalue is $1$ means that the Euler characteristic is an invariant under the substitution rule. In conclusion, while the left eigenvector associated to the dominant eigenvalue gives the relative volumes of the prototiles, the left eigenvector associated to the eigenvalue $1$ gives the relative curvatures of the prototiles. 

\subsubsection{Removing the Unit Eigenvalue}
The fact that the dominant eigenvalue is a quadratic Pisot number, and that one eigenvalue is just an integer, $1$, suggests there is a way to reduce our prototile set to just two prototiles. As explained above, the $1$-eigenvalue of $M_{353}$ is linked directly to the fact that the Euler characteristic is preserved under inflation. 

In the previous section, we derived the fact that $\chi$ was invariant under substitution on $1$, $3$, and $4$ vertices from our set of prototiles and their inflation combinatorics. Here instead, we would like to view this as coming from a constraint. That is, we forget our previous prescriptions for $1$, $3$ and $4$ vertex tiles and \textit{declare} that whatever $1$-vertex tiles are, that they satisfy:
\begin{align}
    V_1 
        &= 6\chi + 3V_3 + 4 V_4\,,\\
    \overline{V}_1
        &= 6\bar{\chi} + 3\overline{V}_3 + 4 \overline{V}_4\,,
\end{align}
and that they're all related by the substitution rule \eqref{eq:353SubMat}. In this case, we find that $\chi = \bar{\chi}$ and
\begin{align}
    \overline{V}_3 
        &= 5 V_3 + 6 V_4 + 10\chi\,,\\
    \overline{V}_4 
        &= \frac{3}{2} V_3 + 2 V_4\,.
\end{align}
Thus we have effectively used the Euler characteristic to remove our $V_1$ tiles.

As we know, we only have a genuine quasicrystal in the limit of many inflations and/or when we are dealing with infinite patches, i.e. when the numbers $V_3$ and $V_4$ are formally infinite. Since we want to study the behaviour as $V_3$ and $V_4$ become large numbers, we can safely discard the constant $+10\chi$ term to get the substitution matrix:
\begin{equation}\label{eq:353SubMat_2}
  \begin{pmatrix} \overline{V}_3 \\ \overline{V}_4 \end{pmatrix}
        = \begin{pmatrix}
          5  & 6 \\ 
          \frac{3}{2} & 2
      \end{pmatrix}
        \begin{pmatrix} V_3 \\ V_4 \end{pmatrix}\,.
\end{equation}
For the subsequent discussion, it is convenient to re-express the above substitution matrix to track $2V_4$ (pairs of 4-vertices), in which case it becomes
\begin{equation}\label{eq:353SubMat_3}
  \begin{pmatrix} \overline{V}_3 \\ 2\overline{V}_4 \end{pmatrix}
        = 
        M_{353}'
        \begin{pmatrix} V_3 \\ 2V_4 \end{pmatrix}\quad
        {\rm with}\quad
        M_{353}'=\begin{pmatrix}
          5  & 3 \\ 
          3 & 2
      \end{pmatrix}\,.
\end{equation}
This substitution matrix describes the quasicrystal at the boundary of the $\{3,5,3\}$ tiling in hyperbolic space,
and has eigenvalues and eigenvectors:
\begin{alignat}{3}
  \lambda_{\pm}
    &=\varphi_{\pm}^4\,,\qquad
  &v_{\pm}^T
    & =\left(\varphi_{\pm},1\right)\,.
\end{alignat}

\subsection{Comparison with Penrose Rhombs and Thurston's Conjecture}\label{sec:ThurstonAndPenrose}
In the 1D quasicrystal $\{7,3\}/\{3,7\}$ obtained from $\bbH^2$ in Section \ref{sec:3773Example}, we noted that the boundary substitution matrix had the same eigenvalues as the Fibonacci quasicrystal. We used this to conjecture, and then find, a local equivalence of our 1D boundary data to the Fibonacci quasicrystal. A similar geometric and numerical argument suggests a relationship between our substitution rules derived for the 2D quasicrystal at the boundary of the $\{3,5,3\}$ honeycomb and the Penrose tiling.

Penrose tilings are constructed from two tiles, one ``fat'' rhomb $\tile{F}$ and one ``thin'' rhomb $\tile{T}$ which is $\varphi_+$-times smaller (in area) than $\tile{F}$, see Figure \ref{fig:penroseTiling} for inflation rules. The accompanying substitution matrix and eigenvalues are:
\begin{equation}
    M_{\mathrm{Pen}} = \begin{pmatrix} 2 & 1 \\ 1 & 1 \end{pmatrix}\,
    \quad\stackrel{\mathrm{Eigenvalues}}{\rightsquigarrow}\quad \varphi_{\pm}^2 = \frac{1}{2}(3\pm\sqrt{5})\,.
\end{equation}
Famously, Penrose tilings admit points of (approximate) 5-fold rotational symmetry, and at most one point of exact 5-fold rotational symmetry. Moreover, it is known that (in flat 2D space) all quasicrystalline patterns with points of (approximate) $5$-fold rotational symmetry are essentially a Penrose tiling or a closely related cousin (see e.g. \cite{senechal1996quasicrystals, baake2013aperiodic, boyle2016self, boyle2016coxeter, gardner1977extraordinary, rokhsar1988two}). 

Similarly, the $\{3,5,3\}$ honeycomb is constructed of icosahedra, which have various axes of $5$-fold rotational symmetry. For example, if we start with an initial patch $S_i$ consisting of 1 single $\{3,5\}$ icosahedron, then its boundary has 12 point of exact $5$-fold symmetry; and when we grow this initial seed outwards to the boundary, we obtain a 2D tiling that inherits these exact 5-fold points (in addition to many other approximate 5-fold points).   Moreover, the inflation factor from one full-step of growth of the $\{3,5,3\}$ honeycomb was known to be \cite{nemeth2017growing} $\varphi_{+}^{8}$ (corresponding to four successive inflations of the Penrose tiling).  Based on this evidence, it was conjectured in \cite{Boyle:2018uiv} that the boundary quasicrystal obtained from $\{3,5,3\}$ would be closely related (and perhaps even locally equivalent) to the Penrose tiling.

A similar conjecture was apparently also made by William Thurston \cite{PenroseCommunication}: 
\begin{conjecture}[Thurston, rough]
    Take the $\{3,5,3\}$ honeycomb in $\bbH^3$ and cut it along a horosphere. Then, by a procedure analogous to the ``cut and project'' method for constructing quasicrystals in flat space, one can construct the Penrose tiling on this (flat) horosphere.
\end{conjecture}
To the best of our knowledge, the exact conjecture was not recorded and the exact procedure was not made precise\footnote{So we leave potential alternative interpretations to the reader.}.  But we {\it suspect} the procedure would be based on the following ingredients: In the upper half-space model of $\bbH^3$, a horosphere may be regarded as a copy of the flat Euclidean plane, embedded ``horizontally" (parallel to the boundary).  Given two parallel horospheres (two parallel horizontal planes), and following the normal (vertical) geodesics connecting them, we can naturally identify each point in the upper plane with a corresponding point (the point vertically below it) in the lower plane.  In this way, we can interpret the lower plane as a rescaling of the upper plane (with the rescaling factor depending on the separation between the two planes).  Now imagine taking an infinite stack of such parallel horizontal planes, separated such that each plane is rescaled relative to its neighbor by the Penrose tiling inflation factor $\varphi_{+}^{2}$.  It is natural to expect that, if one copy of the Penrose tiling lives on one such plane, its inflations and deflations should live on the higher and lower planes in the stack; and we suspect the hypothetical construction of the Penrose tiling would involve projecting onto a given plane the portion of the $\{3,5,3\}$ honeycomb sufficient close to that plane (where the precise definitions of ``projecting" and ``sufficiently close" are analogous to the choices of a projection scheme and a ``window" in the traditional cut-and-project/model-set construction of quasicrystals in flat space \cite{baake2013aperiodic}).
See also \cite{bulatov2} for computer generated images of horospheres slicing hyperbolic honeycombs.

Let us slightly modify this conjecture to our framework:\footnote{It is unclear if this is merely a rephrasing of the previous conjecture or an essential change.} 
\begin{conjecture}[Thurston, modified]
    Take the $\{3,5,3\}$ honeycomb in $\bbH^3$ and cut it along a horosphere. Consider the union of all icosahedra interior to, or intersecting, the horosphere, this defines a patch $S_i$ of $\{3,5,3\}$ honeycomb in $\bbH^3$ with infinite boundary $\partial S_i$. This boundary $\partial S_i$ is equivalent to (seeds the growth of) a Penrose tiling.
\end{conjecture}
Since the substitution tilings we study are local, we argue that the conjecture in \cite{Boyle:2018uiv} and this modified conjecture of Thurston are equivalent. In the remainder of this section we will argue that while there is combinatorial evidence that the conjecture is true, i.e. that Penrose tilings live at the boundary of the $\{3,5,3\}$ honeycomb, a more careful analysis reveals that this is not the case.

The eigenvalues of the inflation matrix given by the half-step rule \eqref{eq:353SubMat_3} are $\varphi^4_{\pm}$, while the Penrose tiling has eigenvalues $\varphi^2_{\pm}$.  This implies that, if the two tilings are related, it must be that a single half-step inflation of the $\{3,5,3\}$ quasicrystal must correspond to two successive inflations of the Penrose tiling.  And, indeed, we see that the substitution matrix $M_{Pen}^{2}$ describing two successive inflations of the Penrose tiling precisely corresponds to the substitution matrix $M_{353}'$
describing a half-step inflation of the $\{3,5,3\}$ tiling
\begin{equation}
    M_{\mathrm{Pen}}^2 = \begin{pmatrix} 5 & 3 \\ 3 & 2 \end{pmatrix}=M_{353}'\,.
\end{equation}


Our construction produces a self-similar quasicrystal with points of $5$-fold symmetry and the same substitution matrix as two consecutive inflations of the Penrose tiling, but our class of $\{3,5,3\}$ quasicrystals cannot be mutually locally derivable to the class of Penrose tilings. The argument follows from a fact about the uncountable class of all Penrose tilings. While each of the (uncountably many) Penrose tilings has (countably) infinite numbers of (arbitrarily large) neighbourhoods of $5$-fold symmetry, there can be at most $1$-point of global $5$-fold rotational symmetry. Moreover, it is known that only $4$ Penrose tilings have this single point of global $5$-fold symmetry and that they inflate to each other in a cycle of period $4$ under inflation. That is, if we call them $\{A, B, A^R, B^R\}$ then:\footnote{The superscript $R$ emphasizes the fact that two of the tilings in this cycle are just reflections of the other two.}
\begin{equation}
    A \mapsto B \mapsto A^R \mapsto B^R \mapsto A \mapsto \dots\,.
\end{equation}
However, our local substitution rules show that a $\{3,5,3\}$ quasicrystal with global $5$-fold symmetry would inflate to itself after a single inflation. Thus the substitution matrices want us to identify the $\{3,5,3\}$ quasicrystal with two successive Penrose inflations, but the global structure would like us to identify the $\{3,5,3\}$ quasicrystal with four successive inflations, a contradiction.

As a result, we determine that the $\{3,5,3\}$ quasicrystal is a distinctly new type of 2D quasicrystalline pattern. This is in contradistinction the 1D case, where the lack of curvature/geometry was less constraining, and the substitution rules followed essentially entirely from their combinatorics.

\section{The General Case and Remaining 2D Quasicrystals}\label{sec:GeneralStory}
In the previous section, we focused on the particular example of a hyperbolic quasicrystal arising from the self-dual $\{3,5,3\}$ tessellation of hyperbolic space $\bbH^3$. Building on this analysis, in this section, we recast this construction so that it applies in higher dimensions with arbitrary honeycombs. As an illustration, we apply this general formalism to construct the remaining 2D hyperbolic quasicrystals arising from the remaining (relevant) regular tessellations of 3D hyperbolic space, i.e.~the $\{5,3,5\}$, $\{4,3,5\}$ and $\{5,3,4\}$ honeycombs \cite{coxeterHyperbolic}.

\subsection{The General Case}
First, let us recall how the half-step inflation works in 2D. We start with a $\{p,q\}$ boundary, which defines our ``waterline,'' and consider an $n$-vertex in the boundary; by definition, $n$ many ``submerged'' $p$-gons meet at the $n$ vertex. To grow by a half-step, we surround this vertex with a $q$-gon, so that $n$-many of the new $q$'s vertices are submerged (they are located at the centers of the aforementioned submerged $p$-gons). As a result, the remaining $q-n$ vertices lie above the waterline; the inner $q-n-2$ of them are $\btile{1}$'s in the new $\{q,p\}$ boundary, while the remaining outer two are $\btile{2}^{\frac{1}{2}}$'s in the new  boundary.

As we saw in the $\{3,5,3\}$ example, the half-step inflation works analogously in 3D. For a general $\{p,q,r\}$ tiling we can understand the local effect of a half-step inflation and then use it to colour our dual vertex diagrams (similar to Figure \ref{fig:vvvv}) and define our vertex tiles (similar to Figure \ref{fig:353VertexTiles}). 

We start with a $\{p,q,r\}$ boundary, which now defines our ``waterline,'' and consider an $n$-vertex $v$ in the boundary. $n$ many submerged $\{p,q\}$'s meet at the $n$-vertex. To grow by a half-step, we surround the vertex $v$ with an $\{r,q\}$ in the dual $\{r,q,p\}$ tiling, so that $n$ many of the new $\{r,q\}$'s vertices are submerged (at the center of the previous $n$ many $\{p,q\}$'s). As in Figure \ref{fig:vvvv}, we imagine colouring these submerged vertices in \textcolor{black}{\textbf{black}}. Now we colour the rest of the dual $\{r,q\}$ as follows: 
\begin{enumerate}
    \item Any face of the $\{r,q\}$ whose vertices are {\it all} black is coloured \textcolor{capRed}{\textbf{red}}. Each red face is dual to a completely submerged edge that is hanging down from the vertex $v$ in the $\{p,q,r\}$ boundary, with only its upper end touching the waterline.
    \item Any face of the $\{r,q\}$ with {\it some} (but not all) of its vertices coloured black, is coloured \textcolor{capBlue}{\textbf{blue}}. Each blue face is dual to an exposed edge lying in the original $\{p,q,r\}$ boundary. If this blue face has one black vertex, it is dual to an exposed 1-edge, and the $1$-edge is drawn with a solid black line in the corresponding vertex tile. If the face has two black vertices, it is dual to an exposed 2-edge, and the $2$-edge is drawn with a dashed black line in the corresponding vertex tile.
    \item Any remaining faces of the $\{r,q\}$ are coloured \textcolor{capGreen}{\textbf{green}}. The edges of the $\{r,q\}$ lying in the green region (or its boundary) are exposed edges in the new $\{r,q,p\}$ surface and are shown as green edges in the corresponding vertex tile. Specifically, the edges forming the boundary between the blue and green regions in the $\{r,q\}$, correspond to the exposed 2-edges, and are drawn with dashed green lines in the corresponding vertex tile; whereas the remaining interior edges correspond to exposed 1-edges, and are drawn with solid green lines in the corresponding vertex tile.
\end{enumerate}
Just as we could express all of the 2D substitution rules in terms of the fundamental vertices ${\bf 1}$ and ${\bf 2}$, we can express all of the 3D substitution rules in terms of tiles involving just $1$-edges and $2$-edges (see also Appendix \ref{sec:higherWRW}). Generically, we find that there are three basic types of boundary tiles, each containing only $1$ and $2$-edges, and the three can be related by Gauss-Bonnet constraints as in the $\{3,5,3\}$ example above.

This strategy now continues to 4D and higher dimensions, although we explain it in 4D for simplicity. Note that a 4D polytope has 0D vertices, 1D edges, 2D faces, and 3D {\it facets}. 

We start with a $\{p,q,r,s\}$ boundary, which defines our waterline. At a given vertex $v$ in this boundary, a certain configuration of $\{p,q,r\}$'s meet below the waterline. The vertex $v$ will become the black dot at the center of the corresponding vertex tile. When we grow by a half step, we surround the vertex $v$ by its dual $\{s,r,q\}$, with a certain number of these $\{s,r,q\}$ vertices submerged at the centers of the aforementioned $\{p,q,r\}$'s; we colour these vertices in \textcolor{black}{\textbf{black}}. Now we colour the rest of the dual $\{s,r,q\}$ as follows: 
\begin{enumerate}
    \item Any facet (or face) of the $\{s,r,q\}$ with {\it all} of its vertices coloured black, is coloured \textcolor{capRed}{\textbf{red}}. Each such red facet (or face) is dual to a completely submerged edge (or completely submerged face) hanging down from the original $\{p,q,r,s\}$ boundary.  
    \item Any facet (or face) of the $\{s,r,q\}$ with {\it some} (but not all) of its vertices coloured black, is coloured \textcolor{capBlue}{\textbf{blue}}. Each blue facet (or face) is dual to an exposed edge (or exposed face) lying in the original $\{p,q,r,s\}$ boundary. This corresponds to a black edge (or face) in the corresponding vertex tile. If this blue facet (or face) has one black vertex, it is dual to an exposed 1-edge (or exposed 1-face) -- i.e. a solid black line (or face) in the corresponding vertex tile; and if it has two black vertices, it is dual to an exposed 2-edge (or exposed 2-face) -- i.e. a dashed black line (or face) in the corresponding vertex tile; and so on.
    \item Any remaining facets (or faces) of the $\{s,r,q\}$ are coloured \textcolor{capGreen}{\textbf{green}}. The edges (or faces) of the $\{s,r,q\}$ lying in the green region (or its boundary) are exposed edges (or faces) in the new $\{s,r,q,p\}$ boundary, and are shown as green edges (or faces) in the corresponding vertex tile.  In particular, the edges (or faces) forming the boundary between the blue and green regions in the $\{s,r,q\}$, correspond to exposed 2-edges (or 2-faces) -- shown as dashed green lines (or faces) in the corresponding vertex tiles; the remaining edges (or faces) in the interior of the green region correspond to the exposed 1-edges (or 1-faces) -- shown as solid green lines (or faces) in the corresponding vertex tile; and so on.
\end{enumerate}
Analogously to the 2D case (where we could express all substitution rules in terms of 1-vertices and 2-vertices) and the 3D case (where we could express all substitution rules in terms of tiles involving just 1-edges and 2-edges), in the 4D case, we can express all of the 4D substitution rules in terms of tiles involving just 1-faces and 2-faces. This set of tiles could then be further reduced using the higher dimensional version of the Gauss-Bonnet formula \cite{coxeter1973regular}.

In the 2D case these 1-vertices and 2-vertices turned out to be all of the tiles that could appear on the boundary, in the 3D case, we will see that only 3 tiles appear on the boundary (which reduces to 2 independent tiles when Gauss constraints are included), in the 4D case there will be some unknown number of independent vertex tiles (supporting only 1 and 2 faces), which will be further constrained by Gauss constraints.

In the following subsections, we will use the process described above to obtain the vertex tiles describing the 2D quasicrystals (and their inflation rules) produced from the remaining 3D hyperbolic honeycombs. This process may also be used to write down the vertex tiles encoding the 3D quasicrystals (and their inflation rules) produced from 4D hyperbolic honeycombs, but we leave this for future work.

\subsection{Example: The \texorpdfstring{$\{5,3,5\}$}{{5,3,5}} Honeycomb}
In the $\{5,3,5\}$ honeycomb, each cell is a dodecahedron $\{5,3\}$, with $5$ dodecahedra meeting around each edge and $20$ dodecahera meeting at each vertex. Like the $\{3,5,3\}$ tessellation, this honeycomb is self-dual. 

The boundary of a patch of dodecahedral tiling is a 2D surface consisting of pentagonal faces, stitched together at various types of edges and vertices. In principle, these can include $1,\dots, 4$ edges and $1,\dots, 19$ vertices.  However, after sufficiently many half-step inflations the boundary will only include only $1$ and $2$ edges (as described in the previous section), and $1$, $2$, and $5$ vertices.

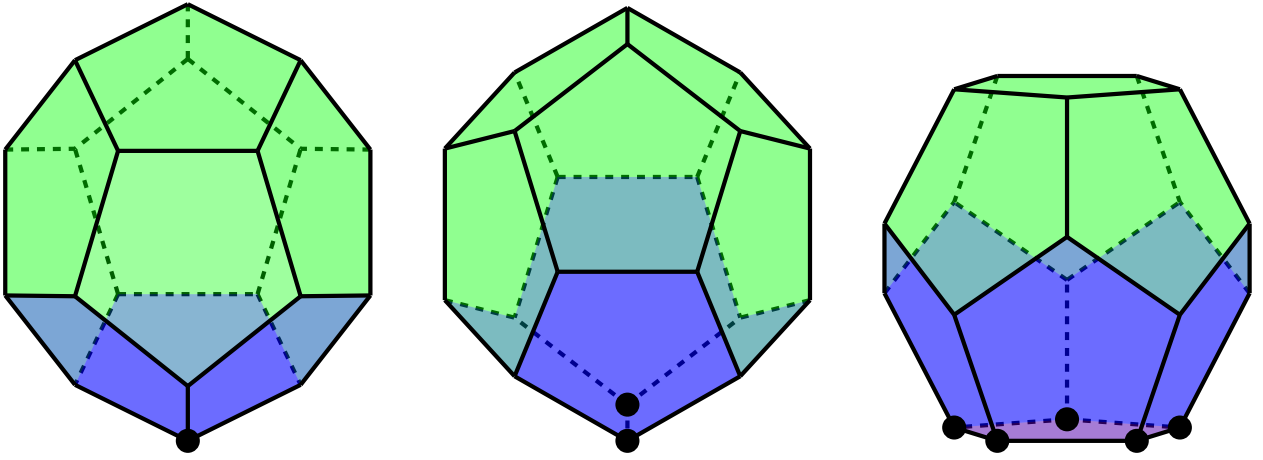
\begin{figure*}[t]
    \begin{subfigure}[t]{0.32\textwidth}
        \centering
	\begin{tikzpicture}
    [   x={(\xx cm,\xy cm)},
        y={(\yx cm,\yy cm)},
        z={(\zx cm,\zy cm)},
        scale=\scaleOne,
        every node/.style={scale=\scaleOne},
        line join=round,
        every path/.style={ultra thick},
        rotate around x=96,
        rotate around z=0,
        rotate around y=0
    ]
    
        \coordinate (pd1) at (-\a,-\a,-\a);
        \coordinate (pd2) at (-\a,-\a,\a);
        \coordinate (pd3) at (-\a,\a,-\a);
        \coordinate (pd4) at (-\a,\a,\a);
        \coordinate (pd5) at (\a,-\a,-\a);
        \coordinate (pd6) at (\a,-\a,\a);
        \coordinate (pd7) at (\a,\a,-\a);
        \coordinate (pd8) at (\a,\a,\a);
        \coordinate (pd9) at (0,-\igr,-\gr);
        \coordinate (pd10) at (0,-\igr,\gr);
        \coordinate (pd11) at (0,\igr,-\gr);
        \coordinate (pd12) at (0,\igr,\gr);
        \coordinate (pd13) at (-\igr,-\gr,0);
        \coordinate (pd14) at (-\igr,\gr,0);
        \coordinate (pd15) at (\igr,-\gr,0);
        \coordinate (pd16) at (\igr,\gr,0);
        \coordinate (pd17) at (-\gr,0,-\igr);
        \coordinate (pd18) at (-\gr,0,\igr);
        \coordinate (pd19) at (\gr,0,-\igr);
        \coordinate (pd20) at (\gr,0,\igr);
        
        \draw[black, dashed] (pd9) -- (pd11);
        \draw[black, dashed] (pd9) -- (pd1);
        \draw[black, dashed] (pd9) -- (pd5);
        \draw[black, dashed] (pd13) -- (pd15);
        \draw[black, dashed] (pd13) -- (pd1);
        \draw[black, dashed] (pd13) -- (pd2);
        \draw[black, dashed] (pd15) -- (pd5);
        \draw[black, dashed] (pd15) -- (pd6);
        \draw[black, dashed] (pd19) -- (pd5);
        \draw[black, dashed] (pd17) -- (pd1);

        \fill[blue,fill opacity=0.35] (pd10) -- (pd6) -- (pd15) -- (pd13) -- (pd2) -- cycle;
        \fill[green,fill opacity=0.25] (pd1) -- (pd13) -- (pd2) -- (pd18) -- (pd17) -- cycle;
        \fill[green,fill opacity=0.25] (pd1) -- (pd9) -- (pd5) -- (pd15) -- (pd13) -- cycle;
        \fill[green,fill opacity=0.25] (pd20) -- (pd19) -- (pd5) -- (pd15) -- (pd6) -- cycle;
        \fill[green,fill opacity=0.25] (pd11) -- (pd9) -- (pd5) -- (pd19) -- (pd7) -- cycle;
        \fill[green,fill opacity=0.25] (pd11) -- (pd9) -- (pd1) -- (pd17) -- (pd3) -- cycle;
        \fill[blue,fill opacity=0.35] (pd12) -- (pd8) -- (pd20) -- (pd6) -- (pd10) -- cycle;
        \fill[blue,fill opacity=0.35] (pd12) -- (pd10) -- (pd2) -- (pd18) -- (pd4) -- cycle;
        \fill[green,fill opacity=0.18] (pd14) -- (pd16) -- (pd8) -- (pd12) -- (pd4) -- cycle;
        \fill[green,fill opacity=0.25] (pd3) -- (pd14) -- (pd4) -- (pd18) -- (pd17) -- cycle;
        \fill[green,fill opacity=0.25] (pd8) -- (pd16) -- (pd7) -- (pd19) -- (pd20) -- cycle;
        \fill[green,fill opacity=0.25] (pd11) -- (pd7) -- (pd16) -- (pd14) -- (pd3) -- cycle;

        \draw[black] (pd11) -- (pd3);
        \draw[black] (pd11) -- (pd7);
        \draw[black] (pd14) -- (pd16);
        \draw[black] (pd16) -- (pd8);
        \draw[black] (pd16) -- (pd7);
        \draw[black] (pd14) -- (pd3);
        \draw[black] (pd14) -- (pd4);
        \draw[black] (pd10) -- (pd12);
        \draw[black] (pd12) -- (pd4);
        \draw[black] (pd12) -- (pd8);
        \draw[black] (pd10) -- (pd2);
        \draw[black] (pd10) -- (pd6);
        \draw[black] (pd20) -- (pd19); 
        \draw[black] (pd19) -- (pd7);
        \draw[black] (pd20) -- (pd8);
        \draw[black] (pd20) -- (pd6);
        \draw[black] (pd17) -- (pd18);
        \draw[black] (pd17) -- (pd3);
        \draw[black] (pd18) -- (pd2);
        \draw[black] (pd18) -- (pd4);
        
        \node[BlackNode] at (pd10) {};
     
    \end{tikzpicture}
    \caption{\RaggedRight{\textbf{$\mathbf{\{p,3,5\}}$ 1-vertex dual figure}
    }}
    \label{fig:535_1vertexcap}
	\end{subfigure}
 \begin{subfigure}[t]{0.32\textwidth}
        \centering
	\begin{tikzpicture}
    [   x={(\xx cm,\xy cm)},
        y={(\yx cm,\yy cm)},
        z={(\zx cm,\zy cm)},
        scale=\scaleOne,
        every node/.style={scale=\scaleOne},
        line join=round,
        every path/.style={ultra thick},
        rotate around x=130,
        rotate around z=0,
        rotate around y=0
    ]
    
        \coordinate (pd1) at (-\a,-\a,-\a);
        \coordinate (pd2) at (-\a,-\a,\a);
        \coordinate (pd3) at (-\a,\a,-\a);
        \coordinate (pd4) at (-\a,\a,\a);
        \coordinate (pd5) at (\a,-\a,-\a);
        \coordinate (pd6) at (\a,-\a,\a);
        \coordinate (pd7) at (\a,\a,-\a);
        \coordinate (pd8) at (\a,\a,\a);
        \coordinate (pd9) at (0,-\igr,-\gr);
        \coordinate (pd10) at (0,-\igr,\gr);
        \coordinate (pd11) at (0,\igr,-\gr);
        \coordinate (pd12) at (0,\igr,\gr);
        \coordinate (pd13) at (-\igr,-\gr,0);
        \coordinate (pd14) at (-\igr,\gr,0);
        \coordinate (pd15) at (\igr,-\gr,0);
        \coordinate (pd16) at (\igr,\gr,0);
        \coordinate (pd17) at (-\gr,0,-\igr);
        \coordinate (pd18) at (-\gr,0,\igr);
        \coordinate (pd19) at (\gr,0,-\igr);
        \coordinate (pd20) at (\gr,0,\igr);

        \draw[black, dashed] (pd13) -- (pd15);
        \draw[black, dashed] (pd13) -- (pd1);
        \draw[black, dashed] (pd13) -- (pd2);
        \draw[black, dashed] (pd15) -- (pd5);
        \draw[black, dashed] (pd15) -- (pd6);
        \draw[black, dashed] (pd10) -- (pd12);
        \draw[black, dashed] (pd10) -- (pd2);
        \draw[black, dashed] (pd10) -- (pd6);
        \draw[black, dashed] (pd20) -- (pd6);
        \draw[black, dashed] (pd18) -- (pd2);

        \fill[green,fill opacity=0.25] (pd1) -- (pd13) -- (pd2) -- (pd18) -- (pd17) -- cycle;
        \fill[green,fill opacity=0.25] (pd1) -- (pd9) -- (pd5) -- (pd15) -- (pd13) -- cycle;
        \fill[green,fill opacity=0.25] (pd20) -- (pd19) -- (pd5) -- (pd15) -- (pd6) -- cycle;
        \fill[blue,fill opacity=0.35] (pd10) -- (pd6) -- (pd15) -- (pd13) -- (pd2) -- cycle;
        \fill[blue,fill opacity=0.35] (pd12) -- (pd10) -- (pd2) -- (pd18) -- (pd4) -- cycle;
        \fill[blue,fill opacity=0.35] (pd12) -- (pd8) -- (pd20) -- (pd6) -- (pd10) -- cycle;
        \fill[blue,fill opacity=0.35] (pd14) -- (pd16) -- (pd8) -- (pd12) -- (pd4) -- cycle;
        \fill[green,fill opacity=0.25] (pd11) -- (pd7) -- (pd16) -- (pd14) -- (pd3) -- cycle;
        \fill[green,fill opacity=0.25] (pd3) -- (pd14) -- (pd4) -- (pd18) -- (pd17) -- cycle;
        \fill[green,fill opacity=0.25] (pd8) -- (pd16) -- (pd7) -- (pd19) -- (pd20) -- cycle;
        \fill[green,fill opacity=0.25] (pd11) -- (pd9) -- (pd5) -- (pd19) -- (pd7) -- cycle;
        \fill[green,fill opacity=0.25] (pd11) -- (pd9) -- (pd1) -- (pd17) -- (pd3) -- cycle;
    
        \draw[black] (pd9) -- (pd11);
        \draw[black] (pd11) -- (pd3);
        \draw[black] (pd11) -- (pd7);
        \draw[black] (pd9) -- (pd1);
        \draw[black] (pd9) -- (pd5);
        \draw[black] (pd14) -- (pd16);
        \draw[black] (pd16) -- (pd8);
        \draw[black] (pd16) -- (pd7);
        \draw[black] (pd14) -- (pd3);
        \draw[black] (pd14) -- (pd4);
        \draw[black] (pd12) -- (pd4);
        \draw[black] (pd12) -- (pd8);
        \draw[black] (pd20) -- (pd19); 
        \draw[black] (pd19) -- (pd7);
        \draw[black] (pd19) -- (pd5);
        \draw[black] (pd20) -- (pd8);
        \draw[black] (pd17) -- (pd18);
        \draw[black] (pd17) -- (pd3);
        \draw[black] (pd17) -- (pd1);
        \draw[black] (pd18) -- (pd4);
        
        \node[BlackNode] at (pd10) {};
        \node[BlackNode] at (pd12) {};
    \end{tikzpicture}
    \caption{\RaggedRight{\textbf{$\mathbf{\{p,3,5\}}$ 2-vertex dual figure}
    }}
    \label{fig:535_4vertexcap}
	\end{subfigure}
	\begin{subfigure}[t]{0.32\textwidth}
        \centering
	\begin{tikzpicture}
    [   x={(\xx cm,\xy cm)},
        y={(\yx cm,\yy cm)},
        z={(\zx cm,\zy cm)},
        scale=\scaleOne,
        every node/.style={scale=\scaleOne},
        line join=round,
        every path/.style={ultra thick},
        rotate around x=0,
        rotate around z=0,
        rotate around y=0
    ]
    
        \coordinate (pd1) at (-\a,-\a,-\a);
        \coordinate (pd2) at (-\a,-\a,\a);
        \coordinate (pd3) at (-\a,\a,-\a);
        \coordinate (pd4) at (-\a,\a,\a);
        \coordinate (pd5) at (\a,-\a,-\a);
        \coordinate (pd6) at (\a,-\a,\a);
        \coordinate (pd7) at (\a,\a,-\a);
        \coordinate (pd8) at (\a,\a,\a);
        \coordinate (pd9) at (0,-\igr,-\gr);
        \coordinate (pd10) at (0,-\igr,\gr);
        \coordinate (pd11) at (0,\igr,-\gr);
        \coordinate (pd12) at (0,\igr,\gr);
        \coordinate (pd13) at (-\igr,-\gr,0);
        \coordinate (pd14) at (-\igr,\gr,0);
        \coordinate (pd15) at (\igr,-\gr,0);
        \coordinate (pd16) at (\igr,\gr,0);
        \coordinate (pd17) at (-\gr,0,-\igr);
        \coordinate (pd18) at (-\gr,0,\igr);
        \coordinate (pd19) at (\gr,0,-\igr);
        \coordinate (pd20) at (\gr,0,\igr);

        \draw[black, dashed] (pd9) -- (pd11);
        \draw[black, dashed] (pd11) -- (pd3);
        \draw[black, dashed] (pd11) -- (pd7);
        \draw[black, dashed] (pd9) -- (pd1);
        \draw[black, dashed] (pd9) -- (pd5);
        \draw[black, dashed] (pd16) -- (pd7);
        \draw[black, dashed] (pd14) -- (pd3);
        \draw[black, dashed] (pd19) -- (pd7);
        \draw[black, dashed] (pd17) -- (pd3);

        \fill[red,fill opacity=0.25] (pd1) -- (pd9) -- (pd5) -- (pd15) -- (pd13) -- cycle;
        \fill[blue,fill opacity=0.35] (pd11) -- (pd9) -- (pd5) -- (pd19) -- (pd7) -- cycle;
        \fill[blue,fill opacity=0.35] (pd11) -- (pd9) -- (pd1) -- (pd17) -- (pd3) -- cycle;
        \fill[green,fill opacity=0.25] (pd3) -- (pd14) -- (pd4) -- (pd18) -- (pd17) -- cycle;
        \fill[green,fill opacity=0.25] (pd11) -- (pd7) -- (pd16) -- (pd14) -- (pd3) -- cycle;
        \fill[green,fill opacity=0.25] (pd8) -- (pd16) -- (pd7) -- (pd19) -- (pd20) -- cycle;
        \fill[blue,fill opacity=0.35] (pd1) -- (pd13) -- (pd2) -- (pd18) -- (pd17) -- cycle;
        \fill[blue,fill opacity=0.35] (pd20) -- (pd19) -- (pd5) -- (pd15) -- (pd6) -- cycle;
        \fill[blue,fill opacity=0.35] (pd10) -- (pd6) -- (pd15) -- (pd13) -- (pd2) -- cycle;
        \fill[green,fill opacity=0.25] (pd12) -- (pd8) -- (pd20) -- (pd6) -- (pd10) -- cycle;
        \fill[green,fill opacity=0.25] (pd12) -- (pd10) -- (pd2) -- (pd18) -- (pd4) -- cycle;
        \fill[green,fill opacity=0.25] (pd14) -- (pd16) -- (pd8) -- (pd12) -- (pd4) -- cycle;

        \draw[black] (pd14) -- (pd16);
        \draw[black] (pd16) -- (pd8);
        \draw[black] (pd14) -- (pd4);
        \draw[black] (pd13) -- (pd15);
        \draw[black] (pd13) -- (pd1);
        \draw[black] (pd13) -- (pd2);
        \draw[black] (pd15) -- (pd5);
        \draw[black] (pd15) -- (pd6);
        \draw[black] (pd10) -- (pd12);
        \draw[black] (pd12) -- (pd4);
        \draw[black] (pd12) -- (pd8);
        \draw[black] (pd10) -- (pd2);
        \draw[black] (pd10) -- (pd6);
        \draw[black] (pd20) -- (pd19); 
        \draw[black] (pd19) -- (pd5);
        \draw[black] (pd20) -- (pd8);
        \draw[black] (pd20) -- (pd6);
        \draw[black] (pd17) -- (pd18);
        \draw[black] (pd17) -- (pd1);
        \draw[black] (pd18) -- (pd2);
        \draw[black] (pd18) -- (pd4);
        
        \node[BlackNode] at (pd9) {};
        \node[BlackNode] at (pd1) {};
        \node[BlackNode] at (pd13) {};
        \node[BlackNode] at (pd15) {};
        \node[BlackNode] at (pd5) {};
        
    \end{tikzpicture}
    \caption{\RaggedRight{\textbf{$\mathbf{\{p,3,5\}}$ 5-vertex dual figure}
    }}
    \label{fig:535_5vertexcap}
	\end{subfigure}
    \caption{\RaggedRight{The dual vertex diagrams in a $\{5,3,5\}$ or $\{4,3,5\}$ layer of tessellation, from which we derive the corresponding 1, 2 and 5 vertex tiles in Figure \ref{fig:535VertexTiles}. Note that all $\{p,3,5\}$ tilings share the same dual figures since they both half-step inflate to a tiling by dodecahedra. \textcolor{black}{\textbf{Black nodes}} are vertices that are centers of cells in the previous layer. \textcolor{capRed}{\textbf{Red faces}} are dual to completely submerged edges in the previous layer and are themselves completely submerged in the previous layer. \textcolor{capBlue}{\textbf{Blue faces}} are dual to boundary edges in the previous layer and are hidden in the new layer. \textcolor{capGreen}{\textbf{Green faces}} mark every new boundary facet.}}
	\label{fig:vertexCapping535}
\end{figure*}

\begin{figure*}
    \begin{subfigure}[t]{0.32\textwidth}
        \includegraphics[width=.95\linewidth]{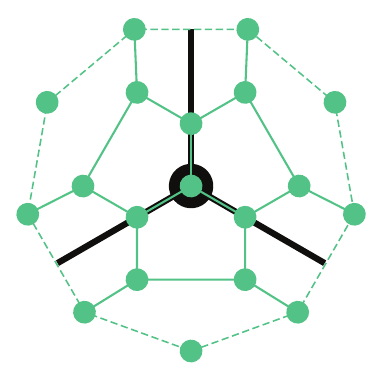}
        \caption{\RaggedRight{\textbf{$\mathbf{\{p,3,5\}}$ 1-vertex tile}
    }}
    \end{subfigure}
    \begin{subfigure}[t]{0.32\textwidth}
        \includegraphics[width=.95\linewidth]{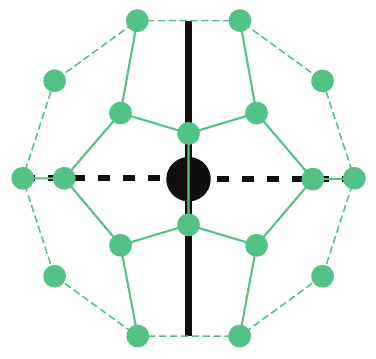}
        \caption{\RaggedRight{\textbf{$\mathbf{\{p,3,5\}}$ 2-vertex tile}
    }}
    \end{subfigure}
	\begin{subfigure}[t]{0.32\textwidth}
        \includegraphics[width=.95\linewidth]{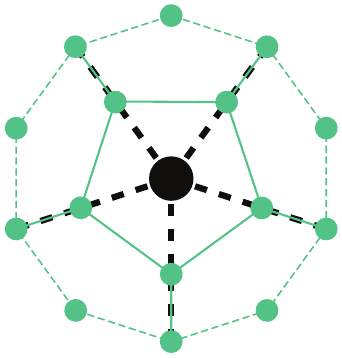}
        \caption{\RaggedRight{\textbf{$\mathbf{\{p,3,5\}}$ 5-vertex tile}
    }}
    \end{subfigure}
    \caption{\RaggedRight{The vertex tiles showing the half-step inflation from a $\{p,3,5\}$ patch to a $\{5,3,p\}$ patch, derived from the dual vertex diagrams in Figure \ref{fig:vertexCapping535}. The initial $\{p,3,5\}$ boundary is shown in \textcolor{black}{\textbf{black}} and the subsequent $\overline{\{5,3,p\}}$ boundary is shown in \textcolor[RGB]{45,183,112}{\textbf{green}}. $1$-edges are denoted by solid lines and $2$-edges are denoted by dashed lines.}}
	\label{fig:535VertexTiles}
\end{figure*}

Following the procedure described in the previous section, we can produce the vertex diagrams
drawn in Figure \ref{fig:vertexCapping535}. By looking at the blue faces in the figure, we can read off the dual edges in a patch $\partial S_i$, and denote them with black in Figure \ref{fig:535VertexTiles}. Likewise, by considering the boundary between the blue and green regions and the edges of the totally green tilings in Figure \ref{fig:vertexCapping535}, we can obtain the new edges (denoted in green in Figure \ref{fig:535VertexTiles}) in $\partial\bar{S}_{i+1}$.

Next we can perform combinatorial checks of our geometric substitution rules. Let $V_1$, $V_2$, and $V_5$ denote the number of $1$, $2$, and $5$ vertices respectively, and likewise for $\overline{V}_1$, $\overline{V}_2$, and $\overline{V}_5$. As in Section \ref{sec:353Combinatorics}, it is helpful to break up the tiles into smaller wedges bounded by the dashed black lines in Figure \ref{fig:535VertexTiles}, we identify them as:
\begin{eqnarray}\label{eq:535AB}
    \begin{array}{c}
            A\;{\rm Wedge}\,\sim \\
            \\\\\\\\\\
            B\;{\rm Wedge}\,\sim
        \end{array}  
        &
        \begin{array}{c}
            \includegraphics[width=.45\linewidth] {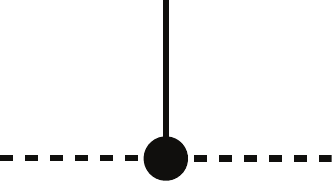} \\
            \\
            \includegraphics[width=.3\linewidth]{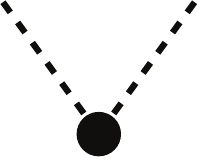}
        \end{array}
        &
\end{eqnarray}
The resulting replacement rules on wedges are
\begin{equation}
  \begin{array}{rcl}
    V_2&=&2A\,, \\
    V_{5}&=&5B\,, 
  \end{array}
  \quad\Rightarrow\quad
  \begin{array}{rcl}
    A&=&\tfrac{1}{2}V_2\,, \\
    B&=&\tfrac{1}{5}V_5\,.
  \end{array}
\end{equation}

Thus we determine the substitution rules from Figure \ref{fig:535VertexTiles} to be:
\begin{align}\label{eq:535Subs}
\begin{split}
V_{1}
    &\mapsto 10\overline{V}_1+6\overline{A}+3\overline{B} \\
    &= 10\overline{V}_1+3\overline{V}_2+\frac{3}{5}\overline{V}_5\,, \\
  V_{2}
    &\mapsto 8\overline{V}_{1}+6\overline{A}+4\overline{B} \\
    &= 8\overline{V}_{1}+3\overline{V}_2+\frac{4}{5}\overline{V}_5\,, \\
  V_{4}
    &\mapsto 5\overline{V}_{1}+5\overline{A}+5\overline{B}\\
    &= 5\overline{V}_{1}+\frac{5}{2}\overline{V}_2+\overline{V}_5 \,,
  \end{split}
\end{align}
or, in terms of a substitution matrix $M_{535}$ 
\begin{equation}\label{eq:535SubMat}
  \begin{pmatrix} \overline{V}_1 \\ \overline{V}_2 \\ \overline{V}_5 \end{pmatrix}
        = \begin{pmatrix}
          10  & 8 & 5 \\ 
          3 & 3 & \frac{5}{2} \\
          \frac{3}{5} & \frac{4}{5} & 1
      \end{pmatrix}
        \begin{pmatrix} V_1 \\ V_2 \\ V_5 \end{pmatrix}\,.
\end{equation}
This matrix has the eigenvalues:
\begin{equation}
  \lambda_{\pm}
    =\frac{1}{2}(13 \pm \sqrt{165})\,,\quad
  \lambda_{0}
    =1\,,
\end{equation}
which (when squared) matches the eigenvalues in \cite{nemeth2017growing} and also Appendix \ref{sec:353Projections}. The curvature of the tiles is
\begin{equation}\label{eq:535Curvature}
    K^{\text{tile}}_{\tile{1}} = \frac{\pi}{5}\,,\quad
    K^{\text{tile}}_{\tile{2}} = -2\frac{\pi}{5}\,,\quad
    K^{\text{tile}}_{\tile{5}} = -5\frac{\pi}{5}\,,
\end{equation}
which, as explained in Section \ref{sec:topConstraints}, defines a left eigenvector for the $1$ eigenvalue of $M_{353}$ and enforces that the Euler characteristic is preserved when working with a configuration of $1$, $2$, and $5$ vertices.

\subsection{Example: The \texorpdfstring{$\{4,3,5\}$}{{4,3,5}}/\texorpdfstring{$\{5,3,4\}$}{{5,3,4}} Honeycombs}
In the $\{4,3,5\}$ honeycomb, each cell is a cube $\{4,3\}$, with $5$ cubes meeting around each edge and $20$ cubes meeting at each vertex. Unlike the previous two tessellations, this honeycomb is NOT self-dual and is dual to the $\{5,3,4\}$ honeycomb, i.e. they both naturally live in a triangulation of $\bbH^3$ by fundamental domains of the $[4,3,5]=[5,3,4]$ Coxeter group. The $\{5,3,4\}$ honeycomb has dodecahedral cells $\{5,3\}$, with $4$ dodecahedra meeting around each edge and $8$ dodecahedra meeting at each vertex. 

Since a patch $S_i$ of $\{4,3,5\}$ honeycomb inflates to a patch $\bar{S}_{i+1}$ of $\{5,3,4\}$ dodecahedral tiling, we have the same dual tiles and vertex tiles as in the $\{5,3,5\}$ honeycomb. As a result, we only expect $1$, $2$, and $5$ vertices in a generic patch of $\{4,3,5\}$ honeycomb, and we have the same dual vertex diagrams shown in Figure \ref{fig:vertexCapping535} and vertex tiles shown in Figure \ref{fig:535VertexTiles} (see also the honeycomb in the center of Figure \ref{figure:DisperseBendWalk}).

\begin{figure*}[t]
    \begin{subfigure}[t]{0.32\textwidth}
        \centering
	\begin{tikzpicture}
    [   x={(\xx cm,\xy cm)},
        y={(\yx cm,\yy cm)},
        z={(\zx cm,\zy cm)},
        scale=\scaleOne,
        every node/.style={scale=\scaleOne},
        line join=round,
        every path/.style={ultra thick},
        rotate around x=85,
        rotate around z=0,
        rotate around y=45
    ]
        \coordinate (pd1) at (\a,\a,\a);
        \coordinate (pd2) at (-\a,\a,\a);
        \coordinate (pd3) at (\a,-\a,\a);
        \coordinate (pd4) at (\a,\a,-\a);
        \coordinate (pd5) at (\a,-\a,-\a);
        \coordinate (pd6) at (-\a,\a,-\a);
        \coordinate (pd7) at (-\a,-\a,\a);
        \coordinate (pd8) at (-\a,-\a,-\a);
        
        \draw[black, dashed] (pd3) -- (pd5);
        \draw[black, dashed] (pd4) -- (pd5);
        \draw[black, dashed] (pd5) -- (pd8);

        \fill[blue,fill opacity=0.35] (pd3) -- (pd5) -- (pd8) -- (pd7) -- cycle;
        \fill[blue,fill opacity=0.35] (pd1) -- (pd3) -- (pd7) -- (pd2) -- cycle;
        \fill[blue,fill opacity=0.35] (pd2) -- (pd7) -- (pd8) -- (pd6) -- cycle;
        \fill[green,fill opacity=0.35] (pd3) -- (pd5) -- (pd4) -- (pd1) -- cycle;
        \fill[green,fill opacity=0.35] (pd4) -- (pd5) -- (pd8) -- (pd6) -- cycle; 
        \fill[green,fill opacity=0.35] (pd4) -- (pd6) -- (pd2) -- (pd1) -- cycle;

        \node[BlackNode] at (pd7) {};

        \draw[black, solid] (pd3) -- (pd7);
        \draw[black, solid] (pd4) -- (pd6);
        \draw[black, solid] (pd6) -- (pd8);
        \draw[black, solid] (pd7) -- (pd8);
        \draw[black, solid] (pd1) -- (pd2);
        \draw[black, solid] (pd1) -- (pd3);
        \draw[black, solid] (pd1) -- (pd4);
        \draw[black, solid] (pd2) -- (pd6);
        \draw[black, solid] (pd2) -- (pd7);
     
    \end{tikzpicture}
    \caption{\RaggedRight{\textbf{$\mathbf{\{5,3,4\}}$ 1-vertex dual figure}
    }}
    \label{fig:435_1vertexcap}
	\end{subfigure}
 \begin{subfigure}[t]{0.32\textwidth}
        \centering
	\begin{tikzpicture}
    [   x={(\xx cm,\xy cm)},
        y={(\yx cm,\yy cm)},
        z={(\zx cm,\zy cm)},
        scale=\scaleOne,
        every node/.style={scale=\scaleOne},
        line join=round,
        every path/.style={ultra thick},
        rotate around x=125,
        rotate around z=0,
        rotate around y=45
    ]
    
        \coordinate (pd1) at (\a,\a,\a);
        \coordinate (pd2) at (-\a,\a,\a);
        \coordinate (pd3) at (\a,-\a,\a);
        \coordinate (pd4) at (\a,\a,-\a);
        \coordinate (pd5) at (\a,-\a,-\a);
        \coordinate (pd6) at (-\a,\a,-\a);
        \coordinate (pd7) at (-\a,-\a,\a);
        \coordinate (pd8) at (-\a,-\a,-\a);
        
        \draw[black, dashed] (pd2) -- (pd7);
        \draw[black, dashed] (pd3) -- (pd7);
        \draw[black, dashed] (pd7) -- (pd8);

        \fill[blue,fill opacity=0.35] (pd3) -- (pd5) -- (pd8) -- (pd7) -- cycle;
        \fill[blue,fill opacity=0.35] (pd1) -- (pd3) -- (pd7) -- (pd2) -- cycle;
        \fill[blue,fill opacity=0.35] (pd2) -- (pd7) -- (pd8) -- (pd6) -- cycle;
        \fill[green,fill opacity=0.35] (pd3) -- (pd5) -- (pd4) -- (pd1) -- cycle;
        \fill[green,fill opacity=0.35] (pd4) -- (pd5) -- (pd8) -- (pd6) -- cycle; 
        \fill[blue,fill opacity=0.35] (pd4) -- (pd6) -- (pd2) -- (pd1) -- cycle;

        \node[BlackNode] at (pd7) {};
        \node[BlackNode] at (pd2) {};

        \draw[black, solid] (pd4) -- (pd5);
        \draw[black, solid] (pd4) -- (pd6);
        \draw[black, solid] (pd5) -- (pd8);
        \draw[black, solid] (pd6) -- (pd8);
        \draw[black, solid] (pd3) -- (pd5);
        \draw[black, solid] (pd1) -- (pd2);
        \draw[black, solid] (pd1) -- (pd3);
        \draw[black, solid] (pd1) -- (pd4);
        \draw[black, solid] (pd2) -- (pd6);
     
    \end{tikzpicture}
    \caption{\RaggedRight{\textbf{$\mathbf{\{5,3,4\}}$ 2-vertex dual figure}
    }}
    \label{fig:435_2vertexcap}
	\end{subfigure}
	\begin{subfigure}[t]{0.32\textwidth}
        \centering
	\begin{tikzpicture}
    [   x={(\xx cm,\xy cm)},
        y={(\yx cm,\yy cm)},
        z={(\zx cm,\zy cm)},
        scale=\scaleOne,
        every node/.style={scale=\scaleOne},
        line join=round,
        every path/.style={ultra thick},
        rotate around x=35,
        rotate around z=0,
        rotate around y=80
    ]
    
        \coordinate (pd1) at (\a,\a,\a);
        \coordinate (pd2) at (-\a,\a,\a);
        \coordinate (pd3) at (\a,-\a,\a);
        \coordinate (pd4) at (\a,\a,-\a);
        \coordinate (pd5) at (\a,-\a,-\a);
        \coordinate (pd6) at (-\a,\a,-\a);
        \coordinate (pd7) at (-\a,-\a,\a);
        \coordinate (pd8) at (-\a,-\a,-\a);
        
        \draw[black, dashed] (pd3) -- (pd5);
        \draw[black, dashed] (pd4) -- (pd5);
        \draw[black, dashed] (pd5) -- (pd8);
        
        \fill[red,fill opacity=0.35] (pd3) -- (pd5) -- (pd8) -- (pd7) -- cycle;
        \fill[blue,fill opacity=0.35] (pd1) -- (pd3) -- (pd7) -- (pd2) -- cycle;
        \fill[blue,fill opacity=0.35] (pd2) -- (pd7) -- (pd8) -- (pd6) -- cycle;
        \fill[blue,fill opacity=0.35] (pd3) -- (pd5) -- (pd4) -- (pd1) -- cycle;
        \fill[blue,fill opacity=0.35] (pd4) -- (pd5) -- (pd8) -- (pd6) -- cycle; 
        \fill[green,fill opacity=0.35] (pd4) -- (pd6) -- (pd2) -- (pd1) -- cycle;

        \node[BlackNode] at (pd7) {};
        \node[BlackNode] at (pd5) {};
        \node[BlackNode] at (pd3) {};
        \node[BlackNode] at (pd8) {};

        \draw[black, solid] (pd1) -- (pd2);
        \draw[black, solid] (pd1) -- (pd3);
        \draw[black, solid] (pd1) -- (pd4);
        \draw[black, solid] (pd2) -- (pd6);
        \draw[black, solid] (pd2) -- (pd7);
        \draw[black, solid] (pd3) -- (pd7);
        \draw[black, solid] (pd4) -- (pd6);
        \draw[black, solid] (pd6) -- (pd8);
        \draw[black, solid] (pd7) -- (pd8);
     
    \end{tikzpicture}
    \caption{\RaggedRight{\textbf{$\mathbf{\{5,3,4\}}$ 4-vertex dual figure}
    }}
    \label{fig:435_4vertexcap}
	\end{subfigure}
        \caption{\RaggedRight{The dual vertex diagrams in a $\{5,3,4\}$ layer of tessellation, from which we derive the corresponding 1, 2 and 4 vertex tiles in Figure \ref{fig:435VertexTiles}. \textcolor{black}{\textbf{Black nodes}} are vertices that are centers of cells in the previous layer. The one \textcolor{capRed}{\textbf{red faces}} is dual to a completely submerged edge in the previous layer and is itself completely submerged in the previous layer. \textcolor{capBlue}{\textbf{Blue faces}} are dual to boundary edges in the previous layer and are hidden in the new layer. \textcolor{capGreen}{\textbf{Green faces}} mark every new boundary facet.}}
	\label{fig:vertexCapping435}
\end{figure*}
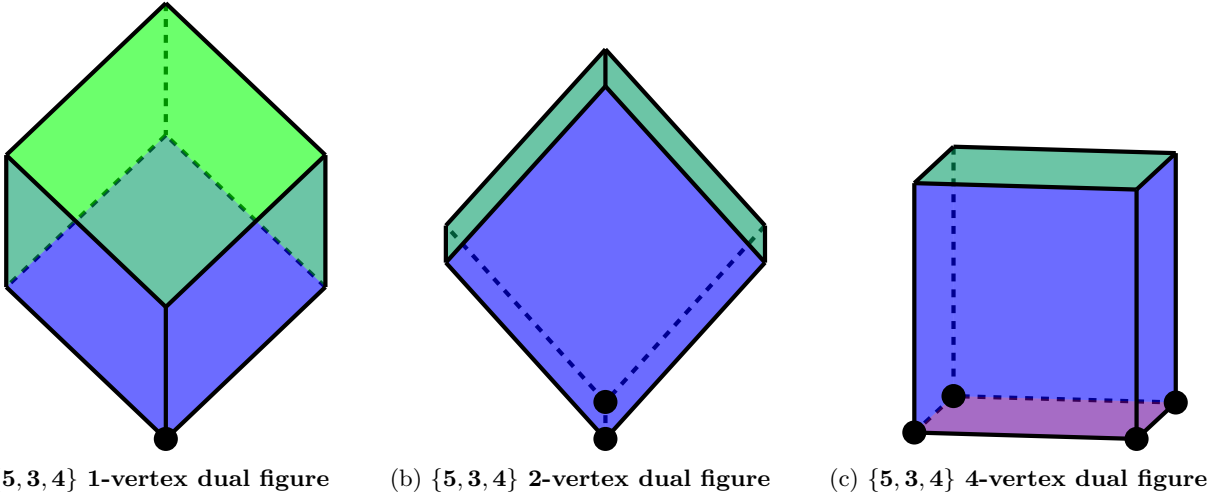

\begin{figure*}[t]
    \begin{subfigure}[t]{0.32\textwidth}
        \includegraphics[width=.95\linewidth]{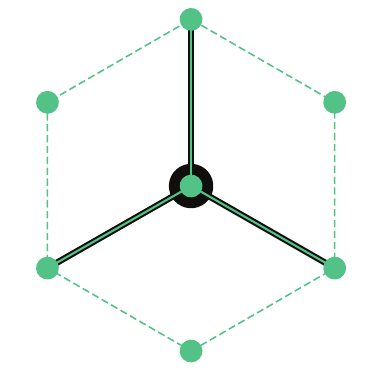}
        \caption{\RaggedRight{\textbf{$\mathbf{\{5,3,4\}}$ 1-vertex tile}
    }}
    \end{subfigure}
    \begin{subfigure}[t]{0.32\textwidth}
        \includegraphics[width=.95\linewidth]{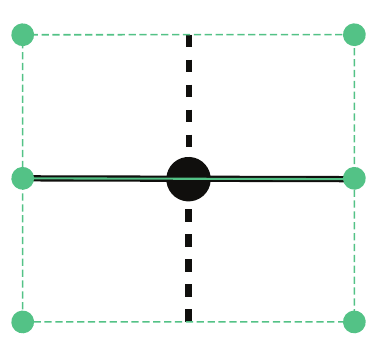}
        \caption{\RaggedRight{\textbf{$\mathbf{\{5,3,4\}}$ 2-vertex tile}
    }}
    \end{subfigure}
    \begin{subfigure}[t]{0.32\textwidth}
        \includegraphics[width=.95\linewidth]{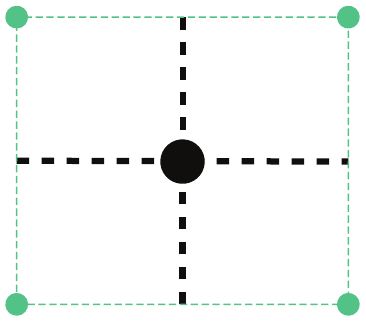}
        \caption{\RaggedRight{\textbf{$\mathbf{\{5,3,4\}}$ 4-vertex tile}
    }}
    \end{subfigure}
        \caption{\RaggedRight{The vertex tiles showing the half-step inflation from a $\{5,3,4\}$ patch to a $\{4,3,5\}$ patch, derived from the dual vertex diagrams in Figure \ref{fig:vertexCapping435}. The initial $\{5,3,4\}$ boundary is shown in \textcolor{black}{\textbf{black}} and the subsequent $\overline{\{4,3,5\}}$ boundary is shown in \textcolor[RGB]{45,183,112}{\textbf{green}}. $1$-edges are denoted by solid lines and $2$-edges are denoted by dashed lines.}}
	\label{fig:435VertexTiles}
\end{figure*}

Conversely, a patch of $\{5,3,4\}$ honeycomb inflates to a patch of $\{4,3,5\}$ honeycomb and, by the same procedure as in previous examples, we can obtain the dual vertex diagrams, depicted in Figure \ref{fig:vertexCapping435}. The vertex tiles are shown in Figure \ref{fig:435VertexTiles}. We find that the boundary generically only contains $1$, $2$, and $4$ vertices in the $\{5,3,4\}$ honeycomb.

Now we can perform combinatorial checks of the geometric substitution rules. Denote the number of $1$, $2$, and $5$ vertices of the $\{4,3,5\}$ honeycomb by $V_1$, $V_2$, and $V_5$. Conversely, denote the number of $1$, $2$, and $4$ vertices of the $\{5,3,4\}$ honeycomb by $\overline{V}_1$, $\overline{V}_2$, and $\overline{V}_4$. 

As in the previous $\{5,3,5\}$ example, we can break the $1$, $2$, and $5$ vertices appearing in the $\{4,3,5\}$ layers into $A$ and $B$ wedges as shown in \eqref{eq:535AB}. The new $1$, $2$, and $4$ tiles in the boundary of the $\{5,3,4\}$ honeycomb can similarly be broken up into $\overline{A}$-wedges and $\overline{B}$-wedges which are also identical to those shown in \eqref{eq:535AB}.

The resulting replacement rules on wedges are,
\begin{equation}
  \begin{array}{rcl}
    V_2&=&2A\,, \\
    V_{5}&=&5B\,, 
  \end{array}
  \quad\Rightarrow\quad
  \begin{array}{rcl}
    A&=&\tfrac{1}{2}V_2\,, \\
    B&=&\tfrac{1}{5}V_5\,,
  \end{array}
\end{equation}
\vspace{-0.5cm}
\begin{equation}
  \begin{array}{rcl}
    \overline{V}_2&=&2\overline{A}\,, \\
    \overline{V}_4&=&4\overline{B}\,, 
  \end{array}
  \quad\Rightarrow\quad
  \begin{array}{rcl}
    \overline{A}&=&\tfrac{1}{2}\overline{V}_2\,, \\
    \overline{B}&=&\tfrac{1}{4}\overline{V}_4\,.
  \end{array}
\end{equation}

We can now determine the $\{4,3,5\} \mapsto \{5,3,4\}$ substitution rules from Figure \ref{fig:535VertexTiles}:
\begin{align}\label{eq:435Subs}
\begin{split}
V_{1}
    &\mapsto 10\overline{V}_1+6\overline{A}+3\overline{B} \\
    &= 10\overline{V}_1+3\overline{V}_2+\frac{3}{4}\overline{V}_4\,, \\
  V_{2}
    &\mapsto 8\overline{V}_{1}+6\overline{A}+4\overline{B} \\
    &= 8\overline{V}_{1}+3\overline{V}_2+\overline{V}_4\,, \\
  V_{5}
    &\mapsto 5\overline{V}_{1}+5\overline{A}+5\overline{B}\\
    &= 5\overline{V}_{1}+\frac{5}{2}\overline{V}_2+\frac{5}{4}\overline{V}_4 \,,
  \end{split}
\end{align}
which corresponds to the substitution matrix $M_{435}$
\begin{equation}\label{eq:435SubMat}
  \begin{pmatrix} \overline{V}_1 \\ \overline{V}_2 \\ \overline{V}_4 \end{pmatrix}
        = \begin{pmatrix}
          10  & 8 & 5 \\ 
          3 & 3 & \frac{5}{2} \\
          \frac{3}{4} & 1 & \frac{5}{4}
      \end{pmatrix}
        \begin{pmatrix} V_1 \\ V_2 \\ V_5 \end{pmatrix}\,.
\end{equation}

Likewise, we can determine the $\{5,3,4\} \mapsto \{4,3,5\}$ substitution rules from Figure \ref{fig:435VertexTiles}:
\begin{align}\label{eq:534Subs}
\begin{split}
\overline{V}_{1}
    &\mapsto {V}_1+3{A}+3{B} \\
    &= V_1 + \frac{3}{2}{V}_2+\frac{3}{5}{V}_5\,, \\
\overline{V}_{2}
    &\mapsto 2{A}+4{B} \\
    &= {V}_2+\frac{4}{5}{V}_5\,, \\
\overline{V}_{4}
    &\mapsto 4{B}\\
    &= \frac{4}{5}{V}_5 \,,
  \end{split}
\end{align}
which corresponds to the substitution matrix $M_{534}$
\begin{equation}\label{eq:534SubMat}
  \begin{pmatrix} {V}_1 \\ {V}_2 \\ {V}_5 \end{pmatrix}
        = \begin{pmatrix}
          1  & 0 & 0 \\ 
          \frac{3}{2} & 1 & 0 \\
          \frac{3}{5} & \frac{4}{5} & \frac{4}{5}
      \end{pmatrix}
        \begin{pmatrix} \overline{V}_1 \\ \overline{V}_2 \\ \overline{V}_4 \end{pmatrix}\,.
\end{equation}
Since the tilings are not self-dual, the eigenvalues and eigenvectors of the half-step matrices do not a priori make sense without conjugating them by some matrices $\Lambda_{pqr}$ and $\Lambda_{rqp}$. However, the matrices can be composed to produce the full-step matrices
\begin{align}
    M_{435 \mapsto 435} = M_{534} M_{435}
    &= \begin{pmatrix}
          10 & 8 & 5 \\ 
          18 & 15 & 10 \\
          8 & 8 & 6
      \end{pmatrix}\,,\\
    M_{534 \mapsto 534} = M_{435} M_{534}
    &= \begin{pmatrix}
          25 & 12 & 4 \\ 
          9 & 5 & 2 \\
          3 & 2 & 1
      \end{pmatrix}\,,
\end{align}
both of which have eigenvalues
\begin{equation}
  \lambda_{\pm}
    =15 \pm 4 \sqrt{14}\,,\quad
  \lambda_{0}
    =1\,,
\end{equation}
which matches the values in \cite{nemeth2017growing}.
 

\section{Conclusion}\label{sec:Conclusion}
In this paper, we have shown that there naturally exists a $d$-dimensional self-similar quasicrystalline tiling (analogous to the Penrose tiling) associated with every regular tessellation of $(d+1)$-dimensional hyperbolic space -- i.e. that every bulk tessellation supports a natural boundary quasicrystal (or, more generally, a holographic foliation by such quasicrystals). As this paper shows, this is an intrinsic property of discrete geometry/kinematics in hyperbolic space, and is not a priori related in any special way to dynamics e.g. of local lattice models that may be placed in the hyperbolic space.

While an earlier version of this construction had been realized in \cite{Boyle:2018uiv} and used for various physics applications (see the numerous references in Section \ref{sec:Introduction}). The construction presented in this work vastly improves upon previous discussions in a few key ways:
\begin{enumerate}
    \item \textbf{Half-step rules}. We refine the 1D/2D inflation rules for $\{p,q\} \mapsto \{p,q\}$ tilings to ``half-step'' rules, relating any tiling $\{p,q,\dots,r,s\}$ to its dual tiling $\{s,r,\dots,q,p\}$. This explains a number of numerical coincidences in inflation rules and further unifies the stories built on the same underlying discrete hyperbolic geometry.
    \item \textbf{Higher dimensions}. We give the first extension of this procedure to higher dimensions. In the $\{3,5,3\}$ example we discovered a tiling that shares essential properties with the Penrose tiling, but is demonstrably distinct. We then use this to explicitly argue the negative to a conjecture of William Thurston, i.e. slicing the $\{3,5,3\}$ tessellation of hyperbolic space along a horosphere \textit{does not} lead to a Penrose tiling. Instead, we find a completely new type of tiling.
    \item \textbf{Global issues}. We give further commentary on global issues, such as the ability to ``reconstruct'' bulk geometries from boundary quasicrystals, and introduce the notion of a ``holographic foliation'' of hyperbolic space by quasicrystals. In particular, the latter concept allows us to define the entire class of $\{p,q\}/\{q,p\}$-quasicrystals, in analogy to the definition of the class of Penrose tilings by de Bruijn.
\end{enumerate}

\subsection{Open Questions}\label{sec:OpenProblems}
Our studies of discrete hyperbolic geometry and the discretized ``quasicrystalline'' conformal geometry are not exhaustive, and have raised a number of interesting questions for future directions. Some questions are purely mathematical in nature, while some are more physically motivated.\footnote{Although, we cannot always tell which are which. As a result, we have only organized the following list to the best of our ability.}
\begin{itemize}
    \item \textbf{$p$-Adic AdS/CFT}. What is the connection and application of the tessellation/quasicrystal correspondence described here to $p$-adic strings and the Bruhat-Tits trees and hyperbolic buildings \cite{gesteau2022holographic} appearing in the $p$-adic AdS/CFT correspondence (see e.g. \cite{Yan:2023lmj} and references within)?
    \item \textbf{Error Correcting Codes}. It has been shown that 2D quantum error correcting codes based on hyperbolic lattices have lower thresholds than their counterparts on Euclidean lattices \cite{2015arXiv150604029B, 2017QS&T....2c5007B}. Is the same true for 3D codes (e.g. the Haah code)?  
    
    And what light do our new notions of quasicrystal and holographic foliation shed on the connection pointed out in \cite{li2023penrose} between quantum error correcting codes, self-similar quasicrystals (like the Penrose tiling), and quantum geometry (see also \cite{Almheiri:2014lwa,Pastawski:2015qua,Boyle:2018uiv}).   
    \item \textbf{BCFT and Numerics}. There is a large amount of work devoted to studying lattice models in hyperbolic space and on quasicrystals separately. In particular, there exist many studies of critical phenomena on quasicrystals (see e.g. \cite{luck1993critical, luck1993classification, grimm1994nonperiodic, grimm1996aperiodic, hermisson1997aperiodic, agrawal2020universality}). Given that $(d+1)$-dimensional hyperbolic space is conformally equivalent to the upper half plane, our tessellation/quasicrystal correspondence could be used to study BCFT correlators, and also suggests a relationship between pre-existing computations.
    \item \textbf{Refining Substitution Rules}. The half-step substitution rules link together $\{p,q,\dots, r,s\}$ tessellations with $\{s,r,\dots,q,p\}$ tessellations, as described in the text. This follows from their shared discrete isometry group $[p,q,\dots,r,s]$. Is there any further refinement of the story presented, so that patches are not subsets of $\{p,q,\dots,r\}$ and/or $\{s,r,\dots,q\}$, but are instead constructed from $(\tfrac{\pi}{2}, \tfrac{\pi}{p}, \dots, \tfrac{\pi}{s})$-simplicies inside the shared triangulation of hyperbolic space by $[p,q,\dots,r,s]$ reflection mirrors?
    \item \textbf{Connecting Tiles to Bulk Surfaces}. The refined growth procedure we describe in the text involves a dualization algorithm, which connects objects of dimension $k$ in $S_i$ to objects of codimension $k$ in $\bar{S}_{i+1}$. On the other hand, the procedure described in Appendix \ref{sec:shootRotateShoot} focuses only on describing the surface by local decision procedures on objects of low codimension. It seems profitable to understand the relationship between these two pictures in more detail in higher dimensions, given the usefulness of them both in the 1D/2D case.
    \item \textbf{Higher Dimensional Space}. There are a finite number of regular tessellations of hyperbolic space in any dimension, and a maximal dimension in which regular hyperbolic tessellations can exist: $\bbH^4$, if they have finite cells and vertex figures (and $\bbH^5$, even if we allow for non-finite cells and vertex figures) \cite{coxeterHyperbolic}. Does this have any implications for physics? For our ability to create interesting (even in principle) discrete lattice models in general $\bbH^d$, it at least has strong implications for solubility, since we will be forced to break symmetry in less agreeable ways.
    \item \textbf{More Exotic Honeycombs}.
    So far we have only considered regular hyperbolic honeycombs with finite cells, and finite vertex figures.  But what about the other regular honeycombs, with infinite cells, and vertices that reach out to the boundary of hyperbolic space, or where an infinite number of cells meet at a single vertex \cite{coxeterHyperbolic}? One can also consider other regular structures, such as ``star honeycombs''  which tessellate space many times \cite{coxeterHyperbolic}. Do these and other interesting discrete structures in hyperbolic space also lead to quasicrystalline boundaries or foliations?
    \item \textbf{Pathological Patches}. Is there any way to algebraically capture the growth of pathological bulk patches, e.g.~with topologically non-trivial configurations? For example, if a 1D string describes a horseshoe shaped patch which pinches closed into a topological annulus after some number of iterations, can the boundary string encode that there are now two boundaries? Are there more general axiomatizations of the algebraic rules appearing in Appendix \ref{sec:shootRotateShoot}?
    \item \textbf{Discrete Surfaces}. The patches we grow form discrete curves/surfaces in hyperbolic honeycombs (and possibly more general spaces). The study of discrete 
    curves and surfaces is already a rich and widely studied topic e.g. in the form of random walks (which includes random walks in hyperbolic spaces) and random geometry. 
    It would be interesting to see if quasicrystalline structures are relevant in such cases, or if the mathematical formalism developed there gives insight into our quasicrystals.
    \item \textbf{Fractality of Bulk Patches}. One interesting question is whether or not the bulk surface becomes more ``fractal'' as it grows. For example, one might wonder if given $S_i$, if the area divided by the volume $A(S_i)/V(S_i)$ diverges under inflations. Considering figures see e.g. \cite{bulatov1, nelson2017visualizing}, and some elementary numerical estimates seems to indicate that the surface does become fractal, but we do not have a convincing argument either way.\footnote{In particular, when growing a $\{3,5,3\}$ tessellation, we are performing a sequence of Pachner moves and evolving from one triangulation to another (finer) triangulation of the sphere. The triangulated surface can be viewed as having a metric, with curvature given by the (discrete) Gaussian curvature. We can view the growth procedure as defining a discrete Perelman-Ricci flow, and see if the curvature proliferates or ``bunches up'' \cite{jin2008discrete}. Indeed, under substitution, a $1$-vertex leads to many $1$ vertices in the subsequent layer directly above it, and hence we guess there is a ``bunching up'' of positive curvature around initial $1$-vertex locations.}
\end{itemize}

\begin{acknowledgments}
We would like to thank Roger Penrose and Davide Gaiotto for helpful discussions that spurred on the completion of this project. LB would like to express his gratitude to Felix Flicker and the late Madeline Dickens for many interactions on this topic.  LB acknowledges support from the School of Physics and Astronomy at the University of Edinburgh and the Perimeter Institute for Theoretical Physics.  JK would like to thank helpful feedback over a number of years from Pablo Basteiro, Rathindra Nath Das, Bianca Dittrich, Johanna Erdmenger, Davide Gaiotto, Theo Johnson-Freyd, Yu Leon Liu, and Ryan Thorngren. JK acknowledges support from the NSERC of Canada and Perimeter Institute during the majority of this project. Research at Perimeter Institute is supported in part by the Government of Canada through the Department of Innovation, Science and Economic Development Canada and by the Province of Ontario through the Ministry of Colleges and Universities.
\end{acknowledgments}

\appendix
\section{Bulk Reconstruction and Boundary Geometry}\label{sec:shootRotateShoot}
In this appendix, we explain how to reconstruct a 2D bulk patch from a 1D boundary string, as discussed in Section \ref{sec:1DQCfrom2D}. Using this, we further justify the decomposition
\begin{equation}
    \tile{n} 
        \cong \tile{2}\;(\tile{1}^{-1}\; \tile{2})^{n-2} 
        =  (\tile{2}\; \tile{1}^{-1})^{n-2}\; \tile{2}\,.\label{eq:ntileDecompositionApp1}
\end{equation}
In particular, we describe to what extent one can ``uniquely'' view an $n$-vertex this way, and why other inequivalent decompositions are inadmissible. At the end, we provide some commentary on higher-dimensional generalizations of the 1D/2D bulk reconstruction procedure.

\subsection{1D/2D Reconstruction: Walk-Rotate-Walk Rule}
When dealing with tiles and substitution rules, the boundary vertices are understood to follow the basic rules of matrix algebra. However, the geometric interpretation of certain tiles, like inverses, is not immediately clear. Moreover, the understanding of the associated bulk patch is not obvious. Both of these points are clarified by the ``Walk-Rotate-Walk rule'' for reading boundary strings. 

Our goal here is to give a local geometric description of an $n$-vertex. Start by fixing an orientation (in this section, we read strings clockwise), and consider an incoming edge to the $n$-vertex.\footnote{If this is the first tile/vertex in the string, then one can choose the location of the vertex without loss of generality (by homogeneity), and choose any edge you want in your drawing and mark it with an incoming arrow towards the desired location of your $n$-vertex (by isotropy).} i.e. our local starting information is an incoming edge towards a (proposed) $n$-vertex. To construct the $n$-vertex $\tile{n}$, we:
\begin{enumerate}
    \item \textbf{Walk}. Walk along the incoming edge, in the direction of the arrow, and end at the (proposed) location of the $n$-vertex.
    \item \textbf{Rotate.} In a $\{p,q\}$ tiling, $q$ many $p$-gons will meet at this point, so we rotate \textit{clockwise} by $\pi(1-2n/q)$, so that we are pointing along an outgoing edge.
    \item \textbf{Walk.} Walk along the outgoing edge, marking our direction with an arrow as we leave.
\end{enumerate}
Given our clockwise orientation for reading strings, their should be $n$ many $p$-gons to the ``right'' of our tiling in this local neighbourhood. For example, in a $\{3,7\}$ tiling, $\tile{2}$ looks locally like:
\begin{equation}
    \tile{2} 
        \sim 
        \vcenter{\hbox{\includegraphics[width=.5\linewidth]{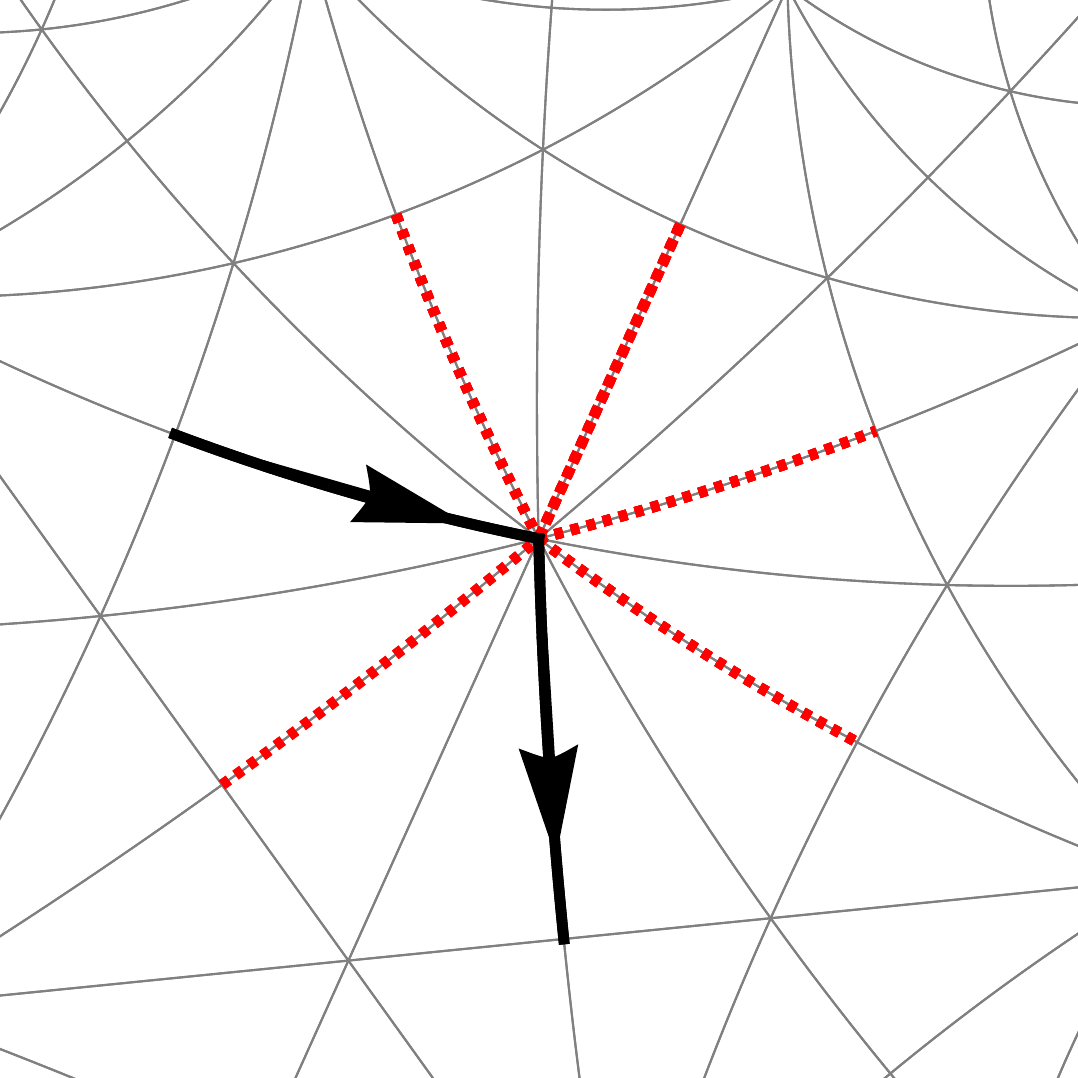}}}\,\quad.
\end{equation}

Inverse vertices $\tile{n}^{-1}$ are now easy to make consistent. Given a seed marked edge we:
\begin{enumerate}
    \item \textbf{Walk} the \textit{opposite direction} to what is marked on the initial edge.
    We are now at the location of our $\tile{n}^{-1}$-vertex.
    \item \textbf{Rotate} \textit{counter-clockwise} by $\pi(1- 2 n/q)$.
    \item \textbf{Walk} along this outgoing edge, this time marking it with an arrow in the {\it opposite} direction from our direction of travel.
\end{enumerate}
There should now be $n$-many $p$-gons to the ``left'' of our tiling. It is now easy to see geometrically how an algebraic expression like $\tile{a}\tile{n} \tile{n}^{-1}\tile{b}=\tile{a}\tile{n}^{-1}\tile{n}\tile{b} = \tile{a}\tile{b}$ is true. This also gives meaning to what a whole sequence of non-collapsing tiles to negative powers means: it is just the same patch of tiling with its orientation reversed. For example, a triangle in the $\{3,7\}$ is $[\tile{1}^3]$ or $[\tile{1}^{-3}]$, but each carries a different orientation.

Note, what really matters is the relative sign on the exponents of subsequent tiles. For example, in an expression with alternating $\pm$ exponents $\tile{a}_1^{+1} \tile{a}_2^{-1} \tile{a}_3^{+1} \tile{a}_4^{-1} \cdots \tile{a}_n^{\pm 1}$, all of the bends/vertices are ``at the same location,'' i.e. they are all effectively travelling $0$ overall distance in the tiling and simply changing the angle being rotated at the vertex.

In summary, the way that a boundary string $\partial S_i$ describes a bulk patch $S_i$ (up to overall translations and rotations) is as a series of concatenated (oriented) curves. In general, there is more information to a 1D curve segment in 2D than an angle, but since the curve is restricted to living on a hyperbolic honeycomb, it can only drive by multiples of some discrete length scale and turn by certain discrete angles. The bulk $S_i$ is the interior of the curve described in this way. Of course, not every general collection of numbers defines a valid boundary $\partial S_i$ of a patch $S_i$, as the curve may be self-intersecting or not close at all. However, the essential ingredient is just that any string of this type describes a curve with very discrete and local rules.

\subsection{Decomposing \texorpdfstring{$\tile{n} \cong \tile{2}\tile{1}^{-1}\tile{2}\cdots\tile{2}\tile{1}^{-1}\tile{2}$}{n=21**-12...21**-12}}
We now justify the equality of an $n$-vertex to an alternating combination of $2$ and inverse $1$-vertices, as in Equation \eqref{eq:ntileDecompositionApp1}. By equality, we really mean that they can be substituted without concern in any formulas, and that it is compatible with inflation rules.

To see that the rule correctly reproduces the behaviour of an $n$-vertex using the Walk-Rotate-Walk rule is straightforward. So let us briefly try to illuminate the uniqueness of such a rule. In other words, suppose our goal is to write $\tile{n}$ as a product of some set of primitive tiles/vertices, while requiring that the rules of matrix algebra and a geometric interpretation of tiles holds. Then what options do we have?

Define the index of a string to be
\begin{equation}
    Z[\prod_i \tile{a_i}^{s_i}] := \sum_{i}  s_i a_i\,.
\end{equation}
Preserving the index is clearly a coarser requirement than the Walk-Rotate-Walk rule, since the index just confirms that the same angle is subtended at a vertex. As we will see, it is a surprisingly powerful constraint.

Start by assuming that $\tile{1}$ is the only primitive tile. Just from the index alone, we will need more than just one primitive tile for any decomposition, so we proceed with both $\tile{1}$ and $\tile{k}$. Then an $n$-vertex can hypothetically be decomposed as:
\begin{equation}
    \tile{n} \stackrel{?}{\cong} \prod_i \tile{a}_i^{s_i}
\end{equation}
for some $s_i$ and $\tilde{a}_i \in \{\tile{1}, \tile{k}\}$. As explained at the end of the previous section, in order to avoid ``interacting'' with vertices at locations where the decomposition is not actually occurring: \textit{Any decomposition must consist of alternating $\pm1$ exponents.} In a sense, this keeps the decomposition local.

If we also take into consideration the relative orientation of incoming/outgoing edges in the Walk-Rotate-Walk rule, then: \textit{we require the decomposition of $\tile{n}$ to start and end with positive powers of tiles}. Thus the decomposition is essentially restricted to take the form:
\begin{equation}
    \tile{1}(\tile{k}^{-1} \tile{1})^{a} \qquad \text{or} \qquad
    \tile{k}(\tile{1}^{-1} \tile{k})^{b}
\end{equation}
for some $a$ and $b$ compatible with the index. 

Now, for compatibility with the subtended angle index, it is clear that we need to determine if $n = 1+(1-k)a$ or $n = k + (k-1)b$ for some $a$, $b$, and fixed choice of $k$. Since we are just looking at the angle turned at a vertex, it is possible that any quantities should only be considered $\Mod q$. It is unclear to us whether or not this is something that should be done for full mathematical interest, i.e. whether or not winding number could be an interesting piece of information at a vertex in some generalized mathematical context. In any case, there is always a universal solution, independent of $q$, where we choose $\tile{k}$ to be $\tile{2}$ and:
\begin{equation}
    \tile{n} 
        \cong \tile{2}\;(\tile{1}^{-1}\; \tile{2})^{n-2} 
        =  (\tile{2}\; \tile{1}^{-1})^{n-2}\; \tile{2}\,, \label{eq:ntileDecompositionApp2}
\end{equation}

There is still a slight ambiguity in the previous identification. We can define two different 0-vertex tiles, with the decompositions:
\begin{align}
    \tile{0}^+
        &:= \tile{1}\tile{2}^{-1}\tile{1}\,,\\
    \tile{0}^-
        &:= \tile{1}^{-1}\tile{2}\tile{1}^{-1}\,.
\end{align}
In particular, $\tile{0}^+$ is the limit of Equation \eqref{eq:ntileDecompositionApp2}. One can still
prepend, insert, or append strings of $\tile{0}^{\pm}$ in a way compatible with our Walk-Rotate-Walk rule, although we do not find them particularly concerning, as they neither force one to make full revolutions around a vertex nor violate the Walk-Rotate-Walk rule.

There are some interesting results that can be obtained by studying the compatibility of decomposition with inflation rules. For example, if we know that
\begin{equation}
    \tile{n} \mapsto \btile{2}^{\tfrac{1}{2}}\btile{1}^{q-n-2}\btile{2}^{\tfrac{1}{2}}\,,
\end{equation}
then we can compute the index before and after inflation:
\begin{equation}
\begin{tikzcd}
    {\tile{n}=\prod_i \tile{a_i}^{s_i}} \arrow[d, mapsto, "\text{Inflate}"] \arrow[r,mapsto, "Z"]  & n = \sum_i s_i a_i \arrow[d, mapsto, "\text{Inflate}"]\\
    {\begin{tabular}{c} $\btile{2}^{\tfrac12}\btile{1}^{q-n-2}\btile{2}^{\tfrac12}=$\\$\prod_i(\btile{2}^{\tfrac12}\btile{1}^{q-a_i-2}\btile{2}^{\tfrac12})^{s_i}$ \end{tabular}} \arrow[r,mapsto, "Z"]  & {\begin{tabular}{c} $q-n = $\\$\sum_i s_i (q-a_i)$ \end{tabular}}
\end{tikzcd}\label{eq:indexCompatibleInflation}\,.
\end{equation}
Now consider the RHS of \eqref{eq:indexCompatibleInflation}, we have both
\begin{equation}
    n = \sum_i s_i a_i\qquad \text{and} \qquad q-n = \sum_{i} s_i (q-a_i)\,.
\end{equation}
This means that $1 = \sum_i s_i$ for the decomposition of an $n$-vertex. Hence we arrive at the same conclusion as before, that the signs in the decomposition must alternate $\pm 1$ with $+1$ first and last. It is interesting that just algebraic properties of strings plus inflation rules gives constraints that otherwise required a careful geometric argument. 

It would be exciting to pursue the various algebra involved with these 2D tiles further in other works. For example, studying the algebraic properties of discrete walks through hyperbolic honeycombs.

\subsection{Generalization to Higher Dimensions}\label{sec:higherWRW}
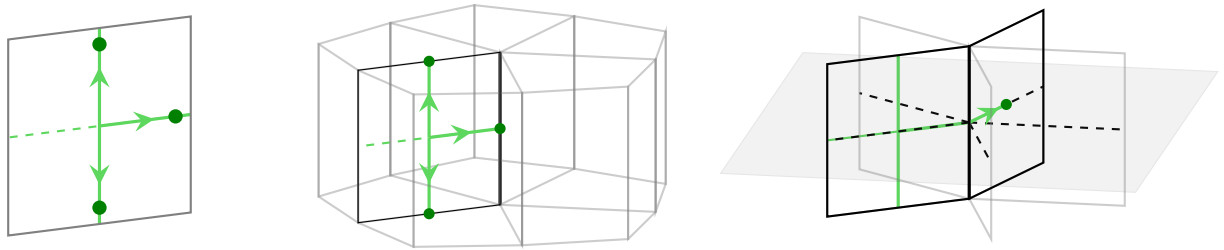
\begin{figure*}
	\begin{minipage}{0.25\textwidth}
        {
        \centering
\tdplotsetmaincoords{75}{10}
\begin{tikzpicture}[tdplot_main_coords,baseline={(current bounding box.center)}, scale = 0.9]
    \draw[thick, dashed, capGreen]
            (0,0,1.5)
                -- ({3*cos(216)},{3*sin(216)},1.5);
    \draw[very thick, capGreen, ->- = .4 rotate 180]
            (0,0,1.5)
                -- ({3/2*cos(216)},{3/2*sin(216)},1.5);
    \draw[very thick, capGreen, ->- = .4 rotate 180]
            ({3/2*cos(216)},{3/2*sin(216)},0)
                -- ({3/2*cos(216)},{3/2*sin(216)},1.5);
    \draw[very thick, capGreen, ->- = .4 rotate 180]
            ({3/2*cos(216)},{3/2*sin(216)},3)
                -- ({3/2*cos(216)},{3/2*sin(216)},1.5)
            ;

    \fill[green!50!black] ({0.25*cos(216)},{0.25*sin(216)},1.5) circle (3pt);
    \fill[green!50!black] ({3/2*cos(216)},{3/2*sin(216)},0.25) circle (3pt);
    \fill[green!50!black] ({3/2*cos(216)},{3/2*sin(216)},2.75) circle (3pt);
    
    \draw[thick, gray, opacity = 1] 
            (0,0,0) 
                -- ({3*cos(216)},{3*sin(216)},0) 
                -- ({3*cos(216)},{3*sin(216)},3) 
                -- (0,0,3) 
                -- cycle;
\end{tikzpicture}
}
	\end{minipage}
    \begin{minipage}{0.32\textwidth}
        {
        \centering
\tdplotsetmaincoords{75}{10}
\begin{tikzpicture}[tdplot_main_coords, baseline={(current bounding box.center)}, scale = 0.7]
    \draw[very thick, black] (0,0,0) -- (0,0,3); 
    \def\ptLen{3.5};

    \foreach \angle in {0,72,...,288} {
        \draw[thick, gray, opacity=0.4] 
            ({3*cos(\angle)},{3*sin(\angle)},0)
                -- ({3*cos(\angle)},{3*sin(\angle)},3) 
                -- ({\ptLen*cos(\angle+36)},{\ptLen*sin(\angle+36)},3)
                -- ({\ptLen*cos(\angle+36)},{\ptLen*sin(\angle+36)},0)
                -- cycle;
        \draw[thick,gray, opacity=0.4] 
            ({3*cos(\angle)},{3*sin(\angle)},0)
                -- ({3*cos(\angle)},{3*sin(\angle)},3) 
                -- ({\ptLen*cos(\angle-36)},{\ptLen*sin(\angle-36)},3)
                -- ({\ptLen*cos(\angle-36)},{\ptLen*sin(\angle-36)},0)
                -- cycle;    
    }

    \draw[thick, dashed, capGreen]
            (0,0,1.5)
                -- ({3*cos(216)},{3*sin(216)},1.5);
    \draw[very thick, capGreen, ->- = .4 rotate 180]
            (0,0,1.5)
                -- ({3/2*cos(216)},{3/2*sin(216)},1.5);
    \draw[very thick, capGreen, ->- = .4 rotate 180]
            ({3/2*cos(216)},{3/2*sin(216)},0)
                -- ({3/2*cos(216)},{3/2*sin(216)},1.5);
    \draw[very thick, capGreen, ->- = .4 rotate 180]
            ({3/2*cos(216)},{3/2*sin(216)},3)
                -- ({3/2*cos(216)},{3/2*sin(216)},1.5)
            ;


    \foreach \angle in {0,72,...,288} {
        \draw[thick, gray, opacity=0.4] 
            (0,0,0) 
                -- ({3*cos(\angle)},{3*sin(\angle)},0) 
                -- ({3*cos(\angle)},{3*sin(\angle)},3) 
                -- (0,0,3) 
                -- cycle;
    }

    \draw[black] 
        (0,0,0) 
            -- ({3*cos(216)},{3*sin(216)},0) 
            -- ({3*cos(216)},{3*sin(216)},3) 
            -- (0,0,3) 
            -- cycle;
    
    \fill[green!50!black] ({0*cos(216)},{0*sin(216)},1.5) circle (3pt);
    \fill[green!50!black] ({3/2*cos(216)},{3/2*sin(216)},0) circle (3pt);
    \fill[green!50!black] ({3/2*cos(216)},{3/2*sin(216)},3) circle (3pt);
    
\end{tikzpicture}
}
	\end{minipage}
	\begin{minipage}{0.32\textwidth}
        {
        \centering
\tdplotsetmaincoords{75}{10}
\begin{tikzpicture}[tdplot_main_coords, baseline={(current bounding box.center)}, scale = 0.7]
    \draw[very thick, black] (0,0,0) -- (0,0,3); 
    \def\ptLen{3.5};


    \draw[thick, capGreen]
            (0,0,1.5)
                -- ({3*cos(216)},{3*sin(216)},1.5);
    \draw[very thick, capGreen]
            ({3/2*cos(216)},{3/2*sin(216)},0)
                -- ({3/2*cos(216)},{3/2*sin(216)},1.5);
    \draw[very thick, capGreen]
            ({3/2*cos(216)},{3/2*sin(216)},3)
                -- ({3/2*cos(216)},{3/2*sin(216)},1.5)
            ;
    \draw[very thick, capGreen]
        (0,0,1.5)
            -- ({3/2*cos(216)},{3/2*sin(216)},1.5);
    \foreach \angle in {0,72,...,288} {
        \draw[thick, dashed]
            (0,0,1.5)
                -- ({3*cos(\angle)},{3*sin(\angle)},1.5);
    }

    \draw[fill=gray, opacity=0.1]
            (-4,-4.5,1.5)
                -- (-4,4.5,1.5)
                -- (4,4.5,1.5)
                -- (4,-4.5,1.5)
                -- cycle;

    \draw[very thick, capGreen, ->- = .2 rotate 180]
        ({3/2*cos(72)},{3/2*sin(72)},1.5)
            -- ({0*cos(72)},{0*sin(72)},1.5)
        ;

    \foreach \angle in {0,72,...,288} {
        \draw[thick,gray, opacity=0.4] 
            (0,0,0) 
                -- ({3*cos(\angle)},{3*sin(\angle)},0) 
                -- ({3*cos(\angle)},{3*sin(\angle)},3) 
                -- (0,0,3) 
                -- cycle;
    }
    \draw[thick, black, opacity = 1] 
        (0,0,0) 
            -- ({3*cos(216)},{3*sin(216)},0) 
            -- ({3*cos(216)},{3*sin(216)},3) 
            -- (0,0,3) 
            -- cycle;
    \draw[thick, black, opacity = 1] 
        (0,0,0) 
            -- ({3*cos(72)},{3*sin(72)},0) 
            -- ({3*cos(72)},{3*sin(72)},3) 
            -- (0,0,3) 
            -- cycle;
    
    \fill[green!50!black] ({3/2*cos(72)},{3/2*sin(72)},1.5) circle (3pt);

\end{tikzpicture}
}
	\end{minipage}
	\caption{\RaggedRight{Left, a codimension-1 face in the $\{4,3,5\}$ tiling with one path already marked (\textcolor{capGreen}{\textbf{dashed green}}). A bacterium is placed at the center of the face and divides into three copies (\textcolor{green!50!black}{\textbf{dark green}}) which disperse towards the edges leaving paint (\textcolor{capGreen}{\textbf{green arrows}}) behind them. Middle, the original square face is a facet in a $\{4,3,5\}$ tiling (\textcolor{gray}{\textbf{gray cells}}). When any bacterium reaches a codimension-2 edge, there are five square faces which meet at that edge. Right, the codimension-2 edge is perpendicular to a plane (\textcolor{gray}{\textbf{gray shaded}}), and the intersections of the five square faces are naturally ordered. If the bacterium continues its journey along face $2$, then the codimension-2 junction is a $2$-edge in the tessellation.}}\label{figure:DisperseBendWalk}
\end{figure*}
We can generalize the Walk-Rotate-Walk rule geometrically to higher dimensions, but capturing it algebraically becomes slightly tricky. 

In any dimension, our goal is to describe how to carve out a ``wrinkly'' codimension-1 surface (typically with the topology of a sphere) by a similar algorithm to before. Once this can be described, it can give a clearer geometric story for our tiles, what it means to concatenate them,  various ``decomposition rules'', and the ability to perform ``bulk reconstruction.'' 

As in the 1D/2D case, our story will describe a codimension-1 surface by concatenation of codimension-1 subsurfaces which are forced to bend discretely along a hyperbolic honeycomb (actually, none of the following and preceding discussions is strictly about hyperbolic honeycombs, as the invertibility does not play a role here). In other words, we describe discrete oriented codimension-1 surfaces in $d$-dimensions. The procedure is similar to before, but slightly more care is required: we now walk along a codimension-1 surface and make a decision procedure at a codimension-2 surface. 

Let us fixate on the 2D/3D case for concreteness, and suppose we are in a $\{p,q,r\}$ tiling (in Figure \ref{figure:DisperseBendWalk}, we draw it in the case of a $\{4,3,5\}$ tiling). To build up a surface we start by picking a codimension-1 $p$-gon face in the $\{p,q,r\}$ tiling and marking a line from the center of the face out to the middle of one of its edges. Then we place a ``bacterium'' at the center of the codimension-1 $p$-gon face. Since it is a $p$-gon, it has $p$-edges, $p-1$ of which are unmarked. Then we:
\begin{enumerate}
    \item \textbf{Disperse}. We tell the bacterium to divide into $p-1$ copies and walk towards the $p-1$ unmarked codimension-2 edges of the $p$-gon face, leaving ``paint'' behind them.
    \item \textbf{Bend}. When a bacterium reaches a codimension-2 edge of the $p$-gon, this codimension-2 edge is actually a junction where $r$-many $p$-gons meet. This junction is normal to a plane, which is divided into $r$ equal sized regions by the $r$ codimension-1 surfaces that meet at the common codimension-2 junction. These surfaces are equipped with a natural ordering by $0,1,2,\dots, r-1$. The bacterium picks a number $k=0,1,2,\dots, r-1$ and bends to face along that plane.
    \item \textbf{Walk}. The bacterium walks along that line in the plane, leaving paint behind it as it walks. The trajectory walked by the bacterium is a line from the junction to the middle of a new face of a new $p$-gon.
\end{enumerate}
The trajectory of any one bacterium paints a line from the center of a $p$-gon in the surface to a neighbouring $p$-gon in the surface. By construction, a bacterium following orders as above will define a $k$-edge in the tiling.\footnote{When bacteria cross another's path, we can declare that they annihilate (because the surface has now closed locally) if we want to avoid self-intersections.} This procedure is depicted in Figure \ref{figure:DisperseBendWalk}. This gives a clear geometric analog for decomposition rules, since we have effectively turned the codimension-1 face and codimension-2 edge junction into an effective 1D problem.

The relationship of these slices to the inflation rules presented in the text is not immediately obvious, and we believe could lead to a much deeper insight into how to even better unify the inflation rules across diverse dimensions.

{
\section{Visualizing \texorpdfstring{$\{3,5,3\}$}{\{3,5,3\}} Patches with Combinatorics and Plane Projections}\label{sec:353Projections}
In this Appendix we include two large figures of inflations in the $\{3,5,3\}$ tiling to aid visualization of the inflation process and see (distorted) pictures of the prototiles, as well as refine the combinatorial growth matrix of patches described in \cite{nemeth2017growing} to half-step inflation. In particular, if we start with 1 single icosahedron, then it only has $V_1 = 12$ vertices, and under successive inflations:
\begin{equation}
    \begin{pmatrix}
        12\\ 0 \\ 0
    \end{pmatrix}
    \mapsto
    \begin{pmatrix}
        72\\ 20 \\ 0
    \end{pmatrix}
    \mapsto
    \begin{pmatrix}
        492\\ 120 \\ 30
    \end{pmatrix}
    \mapsto 
    \begin{pmatrix}
        3372\\ 800 \\ 240
    \end{pmatrix}
    \mapsto 
    \dots\,.
\end{equation}
As mentioned in the main text, this agrees with the total number of \textit{boundary} vertices, $12$, $92$, $642$, $4412$, $\dots$, described by the $\mathbf{M}$ matrix in \cite{nemeth2017growing}.

We include plane projections of this inflation process in Figures \ref{fig:353InflationProjection1} and \ref{fig:353InflationProjection2}. In both figures, the \textcolor{black}{\textbf{black layer}} shows the result of one half-step inflation on a single hyperbolic icosahedra to the dual layer with $12$ hyperbolic icosahedra meeting at a point. In the figures, only the inflation of 6 of the initial vertices is depicted. By inspection, we can verify that $6\times 6 = 36$ of the $1$-vertices are present, i.e. half of the $72$ total $1$-vertices appear in the first inflation layer since only half is depicted in the figure. Likewise, $5$ of the $3$-vertices appear internal to the diagram and $10$ additional $3$-vertices appear on the boundary of the diagram, which corresponds to $15$ of the $20$ vertices in the first inflation. In the \textcolor[RGB]{224,42,28}{\textbf{red layer}}, the figures show the result of a second inflation. The $92$ boundary vertices in the previous black layer lead to $92$ new boundary icosahedra in the subsequent layer. Note that only now in the red layer are there finally $4$-vertices appearing. The \textcolor[RGB]{45,183,112}{\textbf{green layer}} shown in Figure \ref{fig:353InflationProjection2} depicts a third inflation of a single icosahedron.

Let's expand on the construction in \cite{nemeth2017growing} slightly further. The main object computed is a matrix $\mathbf{M}$ that is essentially defined as follows. Consider a patch $S_i$ of hyperbolic honeycomb in $\mathbb{H}^d$ consisting of $w_i^0$ vertices, $w_i^1$ edges, and so on, so that $w_i^d$ is the number of cells; this defines a $d+1$-dimensional vector $\mathbf{w}_i$. Now one can consider the full-step inflated patch of tiling $S_{i+2}$ with its own $\mathbf{w}_{i+2}$. The matrix $\mathbf{M}$ is constructed so that
\begin{equation}
    \mathbf{w}_{i+2} = \mathbf{M} \mathbf{w}_{i}\,.
\end{equation}
Following the procedure described in the reference, it is a straightforward procedure to generate this matrix as a function of the Schl\"affi arguments. To obtain the number of objects added at each step is also straightforward, it is just $\mathbf{M}\mathbf{w}_i-\mathbf{w}_i$. 

The rules presented in \cite{nemeth2017growing} are ``full-step'' rules, but the generalization to ``half-step'' rules can be obtained. Work with a $\Delta:=\{p,q,\dots, r\}$ tiling and denote the matrix defined above by $\mathbf{M}(\Delta)$, denote the dual tiling $\bar\Delta := \{r,\dots,q,p\}$ tiling. We would like to find $(d+1)$-dimensional matrices $\mathbf{M}_{\Delta\mapsto\bar\Delta}$ and $\mathbf{M}_{\bar\Delta\mapsto\Delta}$ such that
\begin{equation}
    \mathbf{M}(\Delta) = \mathbf{M}_{\bar\Delta\mapsto\Delta}\mathbf{M}_{\Delta\mapsto\bar\Delta}\,.
\end{equation}
Of course, there are many such matrices, but we can find the ones which actually correspond to bulk half-step inflation by matching to a particular example. For example, in a $\{p,q\}$ tiling, we can consider the half-step inflations starting with a single $p$-gon:
\begin{equation}
    \begin{pmatrix}
        p \\ p \\ 1
    \end{pmatrix}
    \mapsto
    \begin{pmatrix}
        p(q-2)+1\\ p(q-1) \\ p
    \end{pmatrix}
    \mapsto
    \begin{pmatrix}
        p(p-2)(q-2)\\ p(p-1)(q-2) \\ p(q-2)+1
    \end{pmatrix}
    \mapsto 
    \dots\,.
\end{equation}
Then we can tune the coefficients of the matrices so that they are compatible with this particular example.

Using this method we compute the matrices and eigenvalues to be:
\onecolumngrid
\begin{alignat}{5}
    &\Delta = \{p,q\}:\qquad
        &\mathbf{M}_{\Delta\mapsto\bar{\Delta}} 
        &= \begin{pmatrix}
            q & -2 & 1 \\
            q & -1 & 0 \\
            1 & 0 & 0
        \end{pmatrix}\,,
        &&\qquad \frac{1}{2}((q-2)\pm\sqrt{(q-2)^{2}-4})\,, 1\,,\\
    &\Delta = \{3,5,3\}:\qquad
        &\mathbf{M}_{\Delta\mapsto\bar{\Delta}} 
        &= \begin{pmatrix}
            12 & -3 & 2 & -1\\
            30 & -3 & 1 & 0\\
            20 & -1 & 0 & 0\\
            1 & 0 & 0 & 0
        \end{pmatrix}\,,
        &&\qquad \frac{1}{2}(7\pm 3 \sqrt{5})\,, 1\,, 1\,,\\
    &\Delta = \{5,3,5\}:\qquad
        &\mathbf{M}_{\Delta\mapsto\bar{\Delta}} 
        &= \begin{pmatrix}
            20 & -5 & 2 & -1\\
            30 & -5 & 1 & 0\\
            12 & -1 & 0 & 0\\
            1 & 0 & 0 & 0
        \end{pmatrix}\,,
        &&\qquad \frac{1}{2}(13 + \sqrt{165})\,, 1\,, 1\,,\\
    &\Delta = \{4,3,5\}:\qquad
        &\mathbf{M}_{\Delta\mapsto\bar{\Delta}} 
        &= \begin{pmatrix}
            20 & -5 & 2 & -1\\
            30 & -5 & 1 & 0\\
            12 & -1 & 0 & 0\\
            1 & 0 & 0 & 0
        \end{pmatrix}\,,
        &&\qquad \frac{1}{2}(13 + \sqrt{165})\,, 1\,, 1\,,\\
    &\Delta = \{5,3,4\}:\qquad
        &\mathbf{M}_{\Delta\mapsto\bar{\Delta}} 
        &= \begin{pmatrix}
            8 & -4 & 2 & -1\\
            12 & -4 & 1 & 0\\
            6 & -1 & 0 & 0\\
            1 & 0 & 0 & 0
        \end{pmatrix}\,,
        &&\qquad 1\,, 1\,, 1\,, 1\,.
\end{alignat}

\clearpage
\begin{figure*}[ht]
    \centering
    \includegraphics[width=0.9\linewidth]{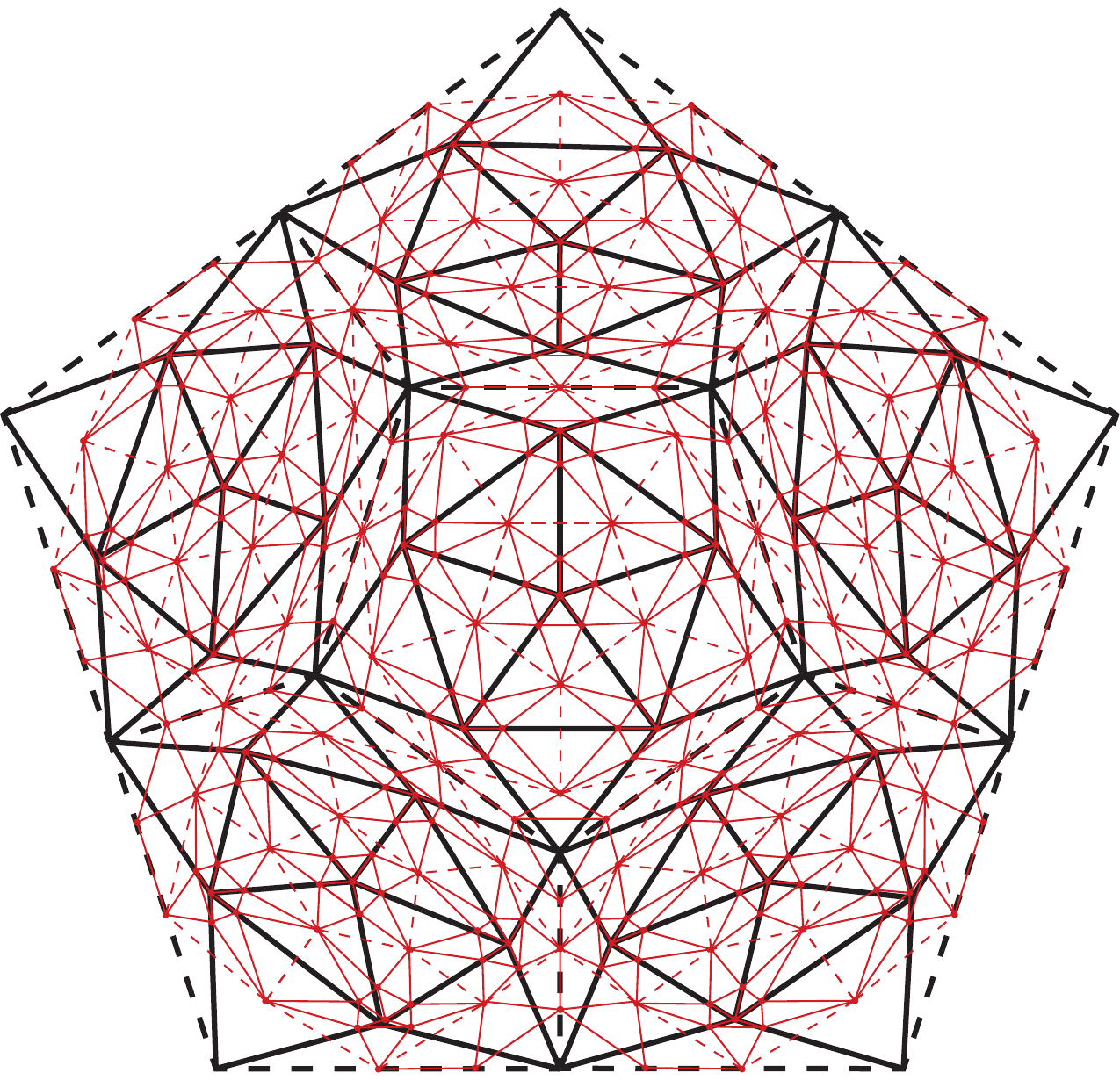}
    \caption{\RaggedRight{The \textcolor{black}{\textbf{black layer}} depicts one inflation of an icosahedron. The \textcolor[RGB]{224,42,28}{\textbf{red layer}} depicts two inflations.}}
    \label{fig:353InflationProjection1}
\end{figure*}
\clearpage
\begin{figure*}[ht]
    \centering
    \includegraphics[width=0.9\linewidth]{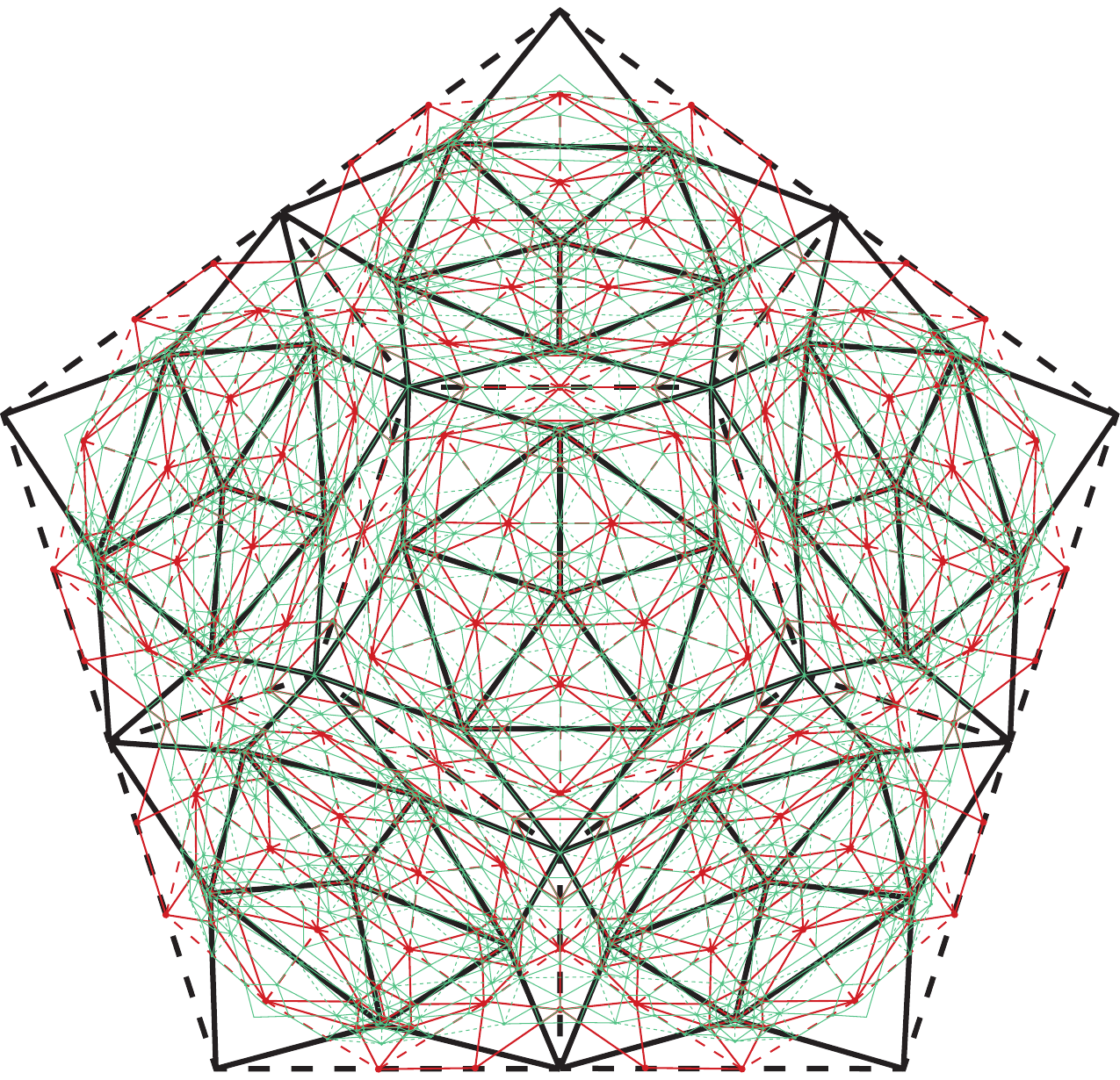}
    \caption{\RaggedRight{The \textcolor[RGB]{45,183,112}{\textbf{green layer}} depicts the third inflation of a single icosahedron.}}
    \label{fig:353InflationProjection2}
\end{figure*}
}
\clearpage
\twocolumngrid

\bibliography{HyperbolicTilings}
\end{document}